\documentclass{article}

\usepackage{authblk}
\usepackage{arxiv}
\usepackage{amsmath,amsfonts,amssymb}
\usepackage[utf8]{inputenc} 
\usepackage[T1]{fontenc}    
\usepackage{hyperref}       
\usepackage{url}            
\usepackage{booktabs}       
\usepackage{amsfonts}       
\usepackage{nicefrac}       
\usepackage{microtype}      
\usepackage{lipsum}
\usepackage{graphicx}
\graphicspath{ {./images/} }
\usepackage[super, square, comma, sort]{natbib}
\usepackage[resetlabels]{multibib}
\usepackage{array}
 \usepackage[section]{placeins}
 \usepackage{float}
 \usepackage{booktabs}
\usepackage{longtable}
\usepackage{array}
\usepackage{pdflscape}
\usepackage{amsmath, amssymb}
\title{Multi-scale measures of time-varying epidemic spread on human mobility networks}

\author{
  Cathal Mills* \\
  Department of Statistics and Pandemic Sciences Institute\\
  University of Oxford 
  \and
  Benjamin Reddy \\
  Department of Statistics and Pandemic Sciences Institute\\
  University of Oxford 
  \and
  William Hart \\
  Mathematical Institute\\
  University of Oxford 
  \and
  Robin Thompson \\
  Mathematical Institute\\
  University of Oxford  \\
  \and
  Kris V. Parag \\
  Department of Engineering, King’s College London and \\
MRC Centre for Global Infectious Disease Analysis, Imperial College London
  \and
  Moritz U. G. Kraemer \\
  Department of Biology and Pandemic Sciences Institute\\
  University of Oxford 
  \and
  Christl A. Donnelly \\
  Department of Statistics and Pandemic Sciences Institute\\
  University of Oxford 
  \and
  Ben Lambert \\
  Department of Statistics and Pandemic Sciences Institute\\
  University of Oxford 
}

\begin{document}
\maketitle
* Corresponding author: cathal.mills@stats.ox.ac.uk  \\

\begin{abstract}
Human movement drives the spatial spread and persistence of many infectious diseases, yet existing theory and real-time operational tools for inferring the instantaneous reproduction number $R(t)$ often assume static and/or homogeneously mixing populations and cannot describe how individuals generate and acquire infections heterogeneously based on their movement patterns within a day. Renewal equations underpin many such popular estimators of $R(t)$, and here, we develop a network-based modelling framework from which we derive new mechanism-led renewal equations and control indicators for outbreaks of infectious diseases. These equations directly integrate within-day human movement to rigorously define a family of instantaneous reproduction numbers; inward, outward, and type $R(t)$ for individual locations, $R(t)$ between locations, $R(t)$ at meeting locations, and $R(t)$ for the entire mobility network. These quantities correct for the unsuitability of existing location-specific $R(t)$ estimators that operate in closed, static populations. Applying our framework to epidemics on diverse types of networks alongside mobile phone data, we demonstrate how our new framework's outputs provide new, multi-scale control indicators at the network, location, and transmission corridor scales, and can be used to design targeted disease control interventions including the strength, type, and length of intervention required across space and time. We capture the biasing effects of different existing ways to measure location-specific and network-level transmission potential without capturing within-day human movements. This generalisable framework redefines reproduction numbers in real-world outbreaks that are shaped by individuals moving across connected locations, enabling more spatially and temporally precise interventions.

\end{abstract}


\newpage

\section{Introduction}
For diseases that affect humans, human movement shapes the spread of a pathogen across space and time, irrespective of the route of transmission (direct, sexual, environmental, or vector-borne). Human movement enables introductions, continued establishment, persistence, spatial spread, and reintroductions \cite{lessani_human_2024, kraemer_effect_2020,changruenngam_how_2020}.  

In recent years, there has been an increasing volume of 
human movement data available to researchers \cite{kostandova_improving_2025}, and with this, an increasing focus on the impacts of human movement on epidemic dynamics \cite{nouvellet_reduction_2021, sills_aggregated_2020, araujo_impact_2024, changruenngam_how_2020, li_association_2021}. Indeed, there is an increasingly wider variety of types of human mobility data that are commonly used by researchers, each of which has different strengths and limitations based on different research/policy questions and the local context \cite{wardle_gaps_2023, wesolowski_connecting_2016, sills_aggregated_2020}.

Despite the essential role of humans and the growing availability of data, the relative roles of human movement are rarely integrated in the most common methods for estimating real-time transmissibility (reproduction numbers $R(t)$) or epidemic speed (growth rates $r(t)$) \cite{cori_new_2013, abbott_estimating_2020, parag_are_2022, parag_improved_2021}. Instead, multi-patch models (e.g. metapopulation or network models \cite{sattenspiel_structured_1995,colizza_reactiondiffusion_2007, soriano-panos_spreading_2018,watts_multiscale_2005, kiss_mathematics_2017, keeling_networks_2005}) are often used to understand how human movement impacts outbreaks across a mobility network. Simulation studies for network models have demonstrated the difficulty in defining and estimating the basic reproduction number ($R_0$) in epidemics spreading on contact networks, and suggested the need to instead focus on real-time reproduction number ($R(t)$) estimates \cite{liu_controllability_2011}. Simpler single-patch models (e.g. renewal-equation-based models) are used for estimating $R(t)$ and $r(t)$ and therefore used to understand how fast a pathogen is spreading and inform public health policy on how best to control pathogen spread. 

While multi-patch models can offer greater expressiveness in terms of describing location connectivity and spatiotemporal dynamics in an epidemic, this can come at the expense of increased complexity, which can hinder real-time inference. This differs from renewal equations for directly transmitted diseases and vector-borne diseases, where a flexible-yet-very-general model allows for fast estimation of $R(t)$ with a few transparent assumptions. However, key limitations of the renewal equation include the homogeneous mixing of individuals and the estimation of quantities such as $R(t)$ for each location independently. These methods for estimating $R(t)$ for an entire network can average over many different spatial or demographic groups with different epidemic dynamics. 

Hierarchical models using renewal equations can borrow information/statistical strength across locations \cite{nouvellet_rtglm_2025}, yet these do not capture the true mechanistic meaning of $R(t)$. This is not just a mathematically limiting factor, but also potentially misleading as local transmissibility can be over/under-estimated, human movement effects ignored, and public health policy misinformed, or at least not fully informed \cite{birello_estimates_2024, trevisin_spatially_2023}. 
Other recent approaches \cite{roy_incorporating_2025, kim_mobility-adjusted_2025} adapt the renewal equation or similar frameworks to incorporate human movement into estimates of $R(t)$. Mathematical approaches illustrate the bias in standard estimators of $R(t)$ in spatially structured populations \cite{birello_estimates_2024, trevisin_spatially_2023}, describe early-outbreak dynamics on different networks \cite{cure_exponential_2025}, and capture epidemic dynamics under population heterogeneity \cite{roberts_new_2003, heesterbeek_type-reproduction_2007, jorge_estimating_2022}. Meanwhile, several statistical approaches allow for a background/external infection term, say $m(t)$, which creates two competing processes; one that accounts for infection importations and another that accounts for infections generated locally/internally \citep{parag_how_2024, roberts_early_2011, parag_why_2025, thompson_improved_2019}. Each of these approaches is insufficient for describing precisely how infected individuals generate new infections at differing rates across space and time based on space- and time-varying movement patterns and contact rates, the depletion of susceptible populations, and the biological infectiousness of infected individuals.  
There are many different types of human mobility data, and the suitability of each depends on the local context, the representativeness, completeness, and privacy safeguards of the data, and the overall relevance for the application \cite{kostandova_improving_2025, wardle_gaps_2023}. In our new framework, we require human mobility data that can account for time-varying movement patterns between locations. A natural source is anonymised mobile phone Global Positioning System (GPS) data that uses telemetry to measure the time spent by individuals in different locations over time \cite{oliver_mobile_2020}, thereby
differing from commonly used origin-destination flow data (e.g. using public transportation) which could misattribute infection opportunities. Although mobile phone data are subject to their own imperfections \cite{kostandova_comparing_2025, wesolowski_connecting_2016}, these allow a researcher to more accurately capture the continuous exposure and infection risks across the network over time. Examples of such anonymised mobile phone data are the SafeGraph and Google mobile phone data streams, and these generally work by tracking the GPS locations of phone users using third-party applications or a major technology company's own platform \cite{yabe_mobile_2022}.

Here, we have similar objectives to previous studies \cite{roy_incorporating_2025, kim_mobility-adjusted_2025}, yet take a different approach to address this challenge, as we return to first principles and develop a spatially explicit and mobility-informed mechanistic modelling framework. The framework is described by a set of partial differential equations (PDEs), from which we derive mechanism-led renewal equations. Our framework uses time-varying mobile phone data to more accurately capture the continuous exposure and infection risks across the network within a day. The framework differs from other approaches above in four ways; i) we derive reproduction numbers across the time-varying mobility network using mechanistic formulae that account for moving populations within a day, ii) we derive a wide taxonomy of reproduction numbers, kernels, and generation time distributions for directed pairwise combinations of locations, for inward and outward transmission, for meeting locations, for location-specific transmission lineages, and across the epidemic network as a whole, iii) we capture within-day human movement dynamics to precisely measure all infection opportunities for susceptible and infected individuals who may spend fractions of their day in one, two, or more locations, and iv) we provide a range of operational control indicators that allow for targeted risk assessments, resource planning, and control measures. The wide range of new, control-relevant quantities across space and time are enabled by our mechanistic modelling framework, and have not been available from existing $R(t)$ estimation methods. These quantities can be paired with our threshold and transience analyses to determine the strength, type, and duration of interventions required to control an outbreak.

\section{Materials and methods}
\label{sec:methods}
\subsection{Reproduction numbers, kernels, and generation times}
During an outbreak, to respond and plan appropriately, decision-makers need timely and reliable information about how fast a pathogen is spreading at any time $t$. Reproduction numbers are unitless quantities that are used to measure epidemic strength, which informally are how many infections are generated by an ``average infected individual''. These mathematical indicators have become popular and useful ways to study and quantify pathogen invasion, epidemic growth, and population immunity \cite{vegvari_commentary_2022, brouwer_why_2022, keeling_modeling_2008}. In particular, the threshold property of values greater/less than one is used to indicate whether long-term epidemic growth is possible.

At the start of an outbreak, the basic reproduction number, $R_0$, for a specific pathogen and population, defines the average number of individuals infected by an infected individual in a fully susceptible population over their infected lifetime \citep{diekmann_mathematical_2000, anderson_infectious_2010, vegvari_commentary_2022, brouwer_why_2022, delamater_complexity_2019}. As an outbreak unfolds, the susceptible population often depletes due to naturally acquired immunity, vaccination, and/or deaths, and the \textit{instantaneous reproduction number} $R(t)$ can be used to quantify how transmissibility changes over time \cite{cori_new_2013, cori_inference_2024,thompson_improved_2019}. For a single location, $R(t)$ is defined for each time $t$ as the expected number of new human infections generated by an infected individual throughout the course of their infected lifetime, should the conditions affecting human transmission at time $t$ remain the same. Mathematically, $R(t)$ can be defined for directly transmitted and vector-borne diseases alike, and can be derived by integrating the \textit{instantaneous kernel} $K(t,\tau)$ over all infection ages $\tau$:
\begin{align}\label{eq:Rt_closed_pop}
    R(t) &= \int_{0}^{\infty} K(t,\tau) d\tau,
\end{align}
where $K(t, \tau)$ represents, at time $t$, the contribution to new human infections by an individual of infection age $\tau$. Then, the probability density function for the \textit{generation time} distribution (which can vary over calendar time) is defined as:
\begin{align*}
    g(t,\tau) &= \frac{K(t, \tau)}{R(t)},
\end{align*}
thereby representing the relative/proportional contribution to new human infections at calendar time $t$ from individuals of infection age $\tau$. Then, $g(t, \tau)$ can also be thought of as the instantaneous distribution of times between primary and secondary human infections. For directly transmitted pathogens (e.g. SARS-CoV-2 for a particular variant), the generation time distribution is often assumed to be fixed over calendar time (i.e. $g(t, \tau) = g(\tau)$ for all $t$) for both practical and technical reasons, as ingredients of $g(t, \tau)$ are often difficult to measure and estimate, thereby creating identifiability issues. The time invariance can be justified by the generation time distribution being invariant to constant-strength interventions (e.g. vaccination or social distancing) \cite{park_importance_2022}. 

Formulae for $R_0$ and $R(t)$ are also often derived using a structured population model and the
spectral radius $\rho(\mathbf{K})$ of the next-generation operator (NGO) $\mathbf{K}$ \citep{diekmann_definition_1990, diekmann_mathematical_2000, arino_multi-city_2003, diekmann_construction_2010, van_den_driessche_reproduction_2002, van_den_driessche_reproduction_2017, brouwer_why_2022}:
\begin{align}    
R(t) &:= \rho(\mathbf{K}),
\end{align}
where the NGO $\mathbf{K}$ is a positive operator that represents the number of new infections in the next generation of transmission (the wave of secondary infection). If there is a finite number of states, the NGO $\mathbf{K}$ becomes the next-generation matrix (NGM) and the spectral radius, $R(t)$, is the largest real absolute eigenvalue of $\mathbf{K}$. The threshold property of $R(t)$ indicates epidemic growth if $R(t) >1$ and decay if $R(t) <1$, yet is defined instantaneously and only controls asymptotic, per-generation growth of the linearised epidemic, i.e., whether per-generation growth or decay occurs in the long run.



\subsection{Renewal equations}
The renewal equation framework is a natural way to capture how past (primary) human infections $i(t-\tau)$ generate new (secondary) human infections at time $t$ with delay $\tau$, and works by combining $R(t)$ and $g(t, \tau)$ with past human infections $i(t-\tau)$:
\begin{align}
    i(t) = R(t)\int_0^{\infty}g(t, \tau)i(t-\tau)d\tau.
\end{align}

First introduced by Euler in 1767, renewal equations are widely used throughout demography, ecology, and evolutionary biology, and more recently became popular for infectious disease modelling when estimating $R(t)$ following the seminal work of \citet{fraser_estimating_2007} and \citet{cori_new_2013}. These equations have been extended for heterogeneous populations \cite{bouros_time-dependent_2025, green_inferring_2022, jorge_estimating_2022} and
for vector-borne diseases \cite{mills_renewal_2025}.  

While there are many ways to estimate $R(t)$ from data \cite{steyn_robust_2025, steyn_primer_2025, van_den_driessche_reproduction_2017}, the renewal equation is the basis for many popular software tools that estimate $R(t)$ and additional epidemiological quantities, including EpiEstim \cite{cori_new_2013}, EpiNow2 \cite{abbott_estimating_2020}, and EpiFilter \cite{parag_improved_2021}. Practically, it is most often used in discrete time to align with daily/weekly epidemiological data and alongside state-space models that separate the observation process (for reported case incidence) and latent process (for true infection incidence).

Here, we will focus on developing the necessary mechanistic machinery for redefining renewal equations and $R(t)$ in connected, interacting populations, which then could be integrated with any of the above $R(t)$ inference software.


\subsection{Human mobility data}

Our framework uses human mobility data that can account for time-varying movement patterns between locations. Here, to demonstrate our framework (Section \ref{sec:results_part_1}), we simulated mobility data to create diverse, synthetic epidemic networks where we could isolate the true dynamics and drivers of transmission and their impacts on our framework's new metrics (Tables \ref{tab:framework_outputs} -- \ref{tab:R_operational} and Section \ref{sec:results_part_1}). We used general mobility findings from different resource-limited settings \cite{wesolowski_quantifying_2012, finger_mobile_2016, kostandova_comparing_2025} to illustrate applications in areas where infectious disease burdens continue to be most overwhelming, human movement persists (even during an outbreak) and drives complex transmission across densely populated locations, and most importantly, socioeconomic realities, limited resources, and population sizes mean that any intervention needs to be targeted in space and time. 


As human movement has a more sustained importance in urban environments, we primarily focused on simulating human movement for a densely populated megacity (Scenario A, Supplementary Material Section \ref{sec:implem_details}). For simplicity and visualisation, we assumed ten nodes/locations, consisting of two highly populated core/hub nodes, three densely populated nodes, three suburban nodes, and two low-population peripheral nodes, each of which had resident commuting fractions of $c_j$ equal to 40\%, 35\%, 28\% and 18\% respectively for each node $j$'s type (Figure \ref{fig:02_overview.pdf} A). We then allowed for these residents to most often visit their adjacent/closest nodes but with population attractiveness, day-of-week effects, and day-to-day variability (Supplementary Material Section \ref{sec:implem_details}). In two other scenarios, we considered a sparse-rural mobility network (Scenario B) and hub-amplified network (Scenario C).

\subsection{Model implementation and analysis}\label{sec:main_text_implem_details}
To implement our modelling framework (Section \ref{sec:results_part_1}) on the mobility network, we modelled a SARS-CoV-2-like pathogen, with an assumed infectiousness profile fixed across space and time using a Gamma distribution with a mean of 5.5 days and standard deviation of 1.8 days \citep{hart_inference_2022}, truncated at 25 days. We assumed $\mathcal{R}(t=0)=R_0=1.5$, a baseline contact rate parameter of 
$\lambda_W = 13.03$ contacts per person per day from the
POLYMOD contact survey \citep{mossong_social_2008} and a between-location (away) contact rate of $\lambda_B = 0.30 \times \lambda_W$. These modelling assumptions reflect reduced contact intensity outside the home residence location. Although our implementation assumes homogeneous mixing in each location, these assumptions can be extended to incorporate heterogeneous mixing (Section \ref{sec:results_part_1}), e.g. using household mixing matrices.

We performed sensitivity analyses on all epidemiological parameters. To numerically simulate our system of linear PDEs, we used a first-order upwind finite differences scheme on a uniform grid with $\Delta t = \Delta a_E = 1$ day, while verifying results with numerical convergence and stability analyses. We used the colour-blind palette of \cite{wong_points_2011} throughout to aid visualisation.

\section{Results}
\begin{figure}[H]
\centering
\includegraphics[scale = 0.25]{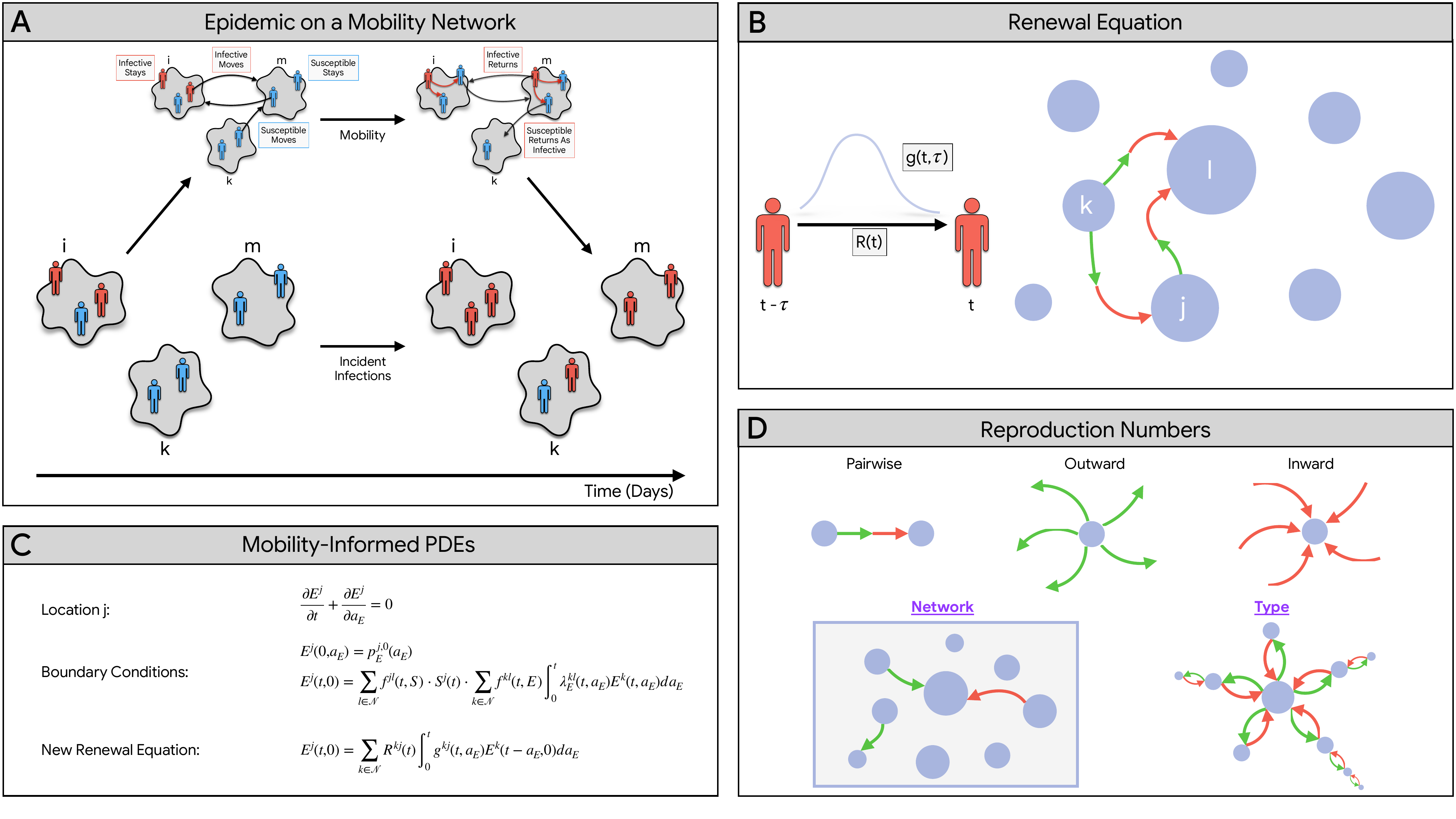}
\caption{\label{fig:figure_1} \footnotesize \textbf{Schematic of key concepts for our framework of modelling epidemics in moving populations} \textbf{A:} A snapshot of an epidemic generation on a mobility network where we describe how infections occur by infected individuals (red) infecting susceptible individuals (blue), possibly due to movement of these individuals between locations. \textbf{B:} The renewal equation (left) for directly transmitted and vector-borne diseases describes how past human infections beget new human infections at time $t$ with time delay $\tau$, controlled by the probability density function for the generation time distribution $g(t, \tau)$ and instantaneous reproduction number $R(t)$. A network (right) with nodes describing how infections propagate from residents of different locations with green outward arrows describing infections by residents of the location and red inward arrows describing inward infections to residents of the location. 
\textbf{C:} Our framework describes how infections occur in residents of location $j$ based on the movement of susceptible and infected individuals in a mobility network. It is built from first principles using partial differential equations (PDEs) with initial and boundary conditions, from which a new mobility-informed and mechanism-led renewal equation is derived. \textbf{D:} Our framework provides mechanistic definitions for a range of reproduction numbers, instantaneous kernels, and generation time distributions, as well as other control measures, each of which are described in Section \ref{sec:results_part_1} and summarised in Tables \ref{tab:framework_outputs} -- \ref{tab:R_operational}.}
\end{figure}
We now derive mechanistically a renewal equation to incorporate time-varying human  movement of susceptible and infected individuals, thereby modelling explicitly all within-day infection opportunities in an outbreak on a network. 
\subsection{Mobility-informed renewal equations and reproduction numbers for directly transmitted diseases} \label{sec:results_part_1}
\subsubsection{A new network modelling framework}
Suppose that we have a time-varying mobility network $\mathcal{N}$ with $j=1, \ldots ,\mathcal{L}$ discrete locations. For a directly transmitted pathogen such as SARS-CoV-2, we can define a mobility-informed system of age-structured, conservation PDEs as:
\begin{align}
\frac{\partial E^j}{\partial t}+\frac{\partial E^j}{\partial a_E} &= 0, 
\label{eq:direct_PDEs}
\end{align}
where $E^j(t, a_E)$ denotes the density of individual humans \textit{resident} in location $j$ at time $t$ who were exposed to a pathogen $a_E$ time units ago. $a_E$ therefore represents an individual's infection age, not biological age.

We close the system with boundary conditions, thereby introducing mobility effects via:
\begin{align}
E^j(0,a_E) &= p_E^{j, 0}(a_E) \label{eq:ic} \\
E^j(t, 0) &= \sum_{l \in \mathcal{N}} \underbrace{f^{jl}(t, S) \cdot S^j(t)}_{(a)} \cdot \sum_{k \in \mathcal{N}} \underbrace{f^{kl}(t, E) \int_{0}^{\infty}\lambda_E^{kl}(t, a_E)E^k(t, a_E)}_{(b)}da_E \label{eq:bc}
\end{align}
(a) accounts for the active population of susceptible individuals resident in location $j$ that are present in location $l$ at time $t$. (b) accounts for infected individuals (of all infection ages) that are resident in location $k$ transmitting to individuals present in location $l$. In words, the boundary condition allows susceptible residents of location $j$ to be infected at any location $l$ by infected individuals resident in any location $k$. $p_E^{j, 0}(a_E)$ denotes the initial density of infected human population resident in location $j$, $S^j(t)$ is the susceptible population at time $t$ resident in location $j$, $f^{jk}(t, S)$ and $f^{jk}(t, E)$ denote the probabilities respectively that a susceptible and exposed individual resident in location $j$ is in location $k$ at time $t$, and $\lambda_E^{kl}(t, a_E)$ is the transmission rate, inclusive of recovery probability, for an infected individual of infection age $a_E$ resident in location $k$ but (potentially) transmitting to a susceptible individual present in meeting location $l$ at time $t$.

$f^{jl}(t)$ is most naturally calculated from averaging across the daily/sub-daily time that an individual resident in location $j$ spends in location $l$. Alternatively, if using mobility co-location data, we could write the model using the probability that an individual from location $j$ meets an individual from location $k$, yet this would not account for within-day human mobility as accurately. $f^{jl}(t, S)$ and $f^{jl}(t, E)$ could be different for susceptible versus infected individuals and for different times (both within days and over several days). $f^{jl}(t, E)$ could also be a function of infection age if we suspect that disease symptoms or infection diagnosis affect human movement, and/or a demographic variable such as biological age or occupation. $f^{jl}(t, S)$ could be a function of epidemic stage, concurrent public health policy (e.g. mobility reductions), or public awareness (e.g. during information campaigns).

$\lambda_E^{kl}(t, a_E)$ could be different for different (directed) combinations of locations and possibly a function of the activity (e.g. numbers/frequency of within-location trips) experienced by a resident of location $k$ in location $l$. We could have $\lambda_E^{kk}(t, a_E) >\lambda_E^{kl}(t, a_E)$ for $k \neq l$ to reflect transmission being more likely within the home locations due to the nature of contacts experienced there (e.g. household versus workplace). Further realism could be incorporated by not just conditioning on whether the infector is in their home residence location, but also whether the potential infectee (i.e. the susceptible individual) is in their home residence location. This can remove the assumption of homogeneous mixing within each meeting location such that we have $\lambda_E^{kk}(t, a_E \mid S^k) \neq \lambda_E^{kk}(t, a_E \mid S^l)$ for $l \neq k$. 

For directly transmitted infectious diseases, we can often write the transmission rate $\lambda_E^{kl}(t, a_E)$ as: 
\begin{align}\label{eq:lambda_components}
\lambda_E^{kl}(t, a_E)= \chi^{kl}(t) \cdot \frac{1}{N_{\text{eff}}^l(t)} \cdot p^k(a_E),    
\end{align}
that is the (potentially) time-varying contact rate $\chi^{kl}(t)$ between a resident of location $k$ and an individual at location $l$ at time $t$ multiplied by the infection-age-dependent probability that a contact results in an infection $p^k(a_E)$, scaled by the total/effective population $N_{\text{eff}}^l(t) = \sum_{q \in \mathcal{N}}f^{ql}(t)N^q(t)$ in location $l$ at time $t$ (to enable frequency-dependent transmission). $\chi^{kl}(t)$ could be a function of the number of (within-location) mobility flows taken by the resident of location $k$ in location $l$, while infectiousness profile $p^{k}(a_E)$ could incorporate data on latent periods and infectiousness profiles and potentially vary based on demographic variables of residents of location $k$. Eq. \eqref{eq:lambda_components} is just one type of assumption for frequency-dependent transmission, and could be further extended, for example, by using alternative transmission types (e.g. density-dependent) or by using behavioural data on contact rates between infected and susceptible individuals of different biological ages, demographics, resident locations, and/or occupations.

\subsubsection{Deriving renewal equations and reproduction numbers}
Using the method of characteristics (Supplementary Material Section \ref{sec:renewal_eqns_derivations}) and assuming longer-term dynamics such that initial conditions (eq. \eqref{eq:ic}) have negligible effects, the boundary condition (eq. \eqref{eq:bc}) can be written as a renewal equation that describes how past human infections from all locations generate new human infections in residents of location $j$:
\begin{align}
E^j(t, 0) &= \sum_{l \in \mathcal{N}} f^{jl}(t, S) \cdot S^j(t) \cdot \sum_{k \in \mathcal{N}} f^{kl}(t, E) \int_{0}^{t}\lambda_E^{kl}(t, a_E)E^k(t -a_E, 0)da_E. 
\end{align}
As we have finite sums and the integrand is non-negative, by linearity of the integral, we can rearrange and isolate the infector's resident location $k$:
\begin{align}
E^j(t, 0) &= \sum_{k \in \mathcal{N}} \int_{0}^{t}\left[\sum_{l \in \mathcal{N}} f^{jl}(t, S) \cdot S^j(t) \cdot  f^{kl}(t, E) \lambda_E^{kl}(t, a_E)\right]E^k(t -a_E, 0) da_E. 
\end{align}
In matrix notation, we can write this as:
\begin{align}\label{eq:ngm_renewal_main_text}
\mathbf{E}(t, 0) &= \int_{0}^{t} \mathcal{K}(t, a_E)\mathbf{E}(t -a_E, 0) da_E,
\end{align}
or
\begin{align}
\left(\mathbf{E}(t, 0)\right)_j &=  \sum_{k \in \mathcal{N}}\int_{0}^{t} \mathcal{K}_{jk}(t, a_E){E}^k(t -a_E, 0) da_E,
\end{align}
where $\mathcal{K}_{jk}(t, a_E) := \mathbb{K}^{kj}(t, a_E) \geq 0$, i.e., the instantaneous kernel at each time $t$ that shapes how infected residents of location $k$ with different infection ages $a_E$ generate new infections in residents of location $j$ (which can occur in any meeting location):
\begin{align}
\mathbb{K}^{kj}(t, a_E) &:= \sum_{l \in \mathcal{N}}f^{jl}(t, S) \cdot S^j(t) \cdot f^{kl}(t, E) \lambda_E^{kl}(t, a_E), \\
\mathbb{K}_{\text{out}}^{k}(t, a_E) &:= \sum_{j \in \mathcal{N}} \sum_{l \in \mathcal{N}}f^{jl}(t, S) \cdot S^j(t) \cdot f^{kl}(t, E) \lambda_E^{kl}(t, a_E),
\label{eq:exposed_model_one}
\end{align}
while $\mathbb{K}_{\text{out}}^{k}(t, a_E)$ defines, for each time $t$, the instantaneous rate at which infected residents of location $k$ of infection age $a_E$ generate new infections in residents of any location. These instantaneous kernels will depend on day-to-day (and diurnal if using sub-daily data) movement patterns as during-week (and during-day) infection dynamics may be more strongly affected by imported infections, while during-weekend (and during-night) dynamics may be more skewed towards within-location resident-to-resident transmission. These further depend on contact rates, infectiousness profiles, and susceptible population counts (eq. \eqref{eq:lambda_components}).

By integrating instantaneous kernels over all infection ages and summing over locations, we recover different types of instantaneous reproduction numbers across space and time \cite{fraser_estimating_2007, park_importance_2022}:
\begin{align}
R^{kj}(t) &:= \int_0^t \mathbb{K}^{kj}(t, a_E) da_E \\
R_{\text{out}}^k(t) &:=  \sum_{j \in \mathcal{N}} R^{kj}(t)= \sum_{j \in \mathcal{N}} \int_0^t \mathbb{K}^{kj}(t, a_E) da_E. 
\end{align}
The \textit{between-location}/pairwise $R^{kj}(t)$ is different to instantaneous reproduction numbers in closed populations (eq. 
\eqref{eq:Rt_closed_pop}), as it is the expected number of new human infections transmitted by individuals resident in location $k$ to individuals resident in $j$ throughout the course of the individual's infected lifetime, if the conditions affecting transmission from residents of location $k$ to residents of location $j$ were to remain the same as at time $t$. $R_{\text{out}}^k(t)$ is the \textit{outward} reproduction number that defines the expected number of new human infections generated by an infected resident of location $k$ in any location throughout their infected lifetime, should conditions affecting outward transmission from location $k$ remain the same as at time $t$. $R^{kj}(t)$ tracks the pairwise transmissibility patterns for between- and within-location transmission, yet it will \textit{not} have the threshold property (for long-term epidemic growth) as for most reproduction numbers. This is because mathematically $R^{kj}(t)$ does 
not determine asymptotic/long-term behaviour of a linear system, unlike usual instantaneous reproduction numbers $R(t)$. 
Likewise, $R_{\text{out}}^k(t)$ can quantify instantaneously whether an infected resident of location $k$ generates more than one secondary infection in all locations per generation (i.e. single generation amplification), but also does \textit{not} have the familiar threshold property for the long-term growth/decay in the outbreak of secondary infections by infected residents of location $k$. The lack of threshold property is because the instantaneous measure does not capture asymptotic behaviour as in closed populations (e.g. an infected resident may generate many secondary infections in a location in which local transmission does not generate large transmission chains). Despite the lack of threshold properties, we will call both $R^{kj}(t)$ and $R_{\text{out}}^{k}(t)$ instantaneous reproduction numbers (just not in the traditional sense) for both familiarity and practical reasons as these will control the expected number of secondary infections per primary infected individual and form the backbone of the reproduction matrix/NGM from which network-level growth and threshold indicators will be derived.

As usual \cite{park_importance_2022, mills_renewal_2025, fraser_estimating_2007}, we can derive probability densities for the generation time distributions, at each calendar time, by dividing the instantaneous kernel by the corresponding reproduction number. This is similar to the results of \citet{park_importance_2022}, but with the additional spatial dimensions to allow for interacting populations. Here, we have several different types of generation time distributions, each corresponding to different types of \textit{generations} across space and time:
\begin{align}\label{eq:between_loc_out_gt_direct}
g^{kj}(t, a_E) &:= \frac{\mathbb{K}^{kj}(t, a_E)}{\int_0^t \mathbb{K}^{kj}(t, a_E) da_E}= \frac{\mathbb{K}^{kj}(t, a_E)}{{R}^{kj}(t)},\\
g_{\text{out}}^k(t, a_E) &:=  \frac{\sum_{j \in \mathcal{N}}\mathbb{K}^{kj}(t, a_E)}{\sum_{j \in \mathcal{N}} \int_0^t \mathbb{K}^{kj}(t, a_E) da_E } = \frac{\mathbb{K}_{\text{out}}^{k}(t, a_E)}{{R}_{\text{out}}^{k}(t)},
\end{align}
where $g^{kj}(t, a_E)$ defines, instantaneously at calendar time $t$, the relative contribution to the number of new infections generated in residents of location $j$ by residents of location $k$ that were infected $a_E$ time units ago. $g_{\text{out}}^k(t, a_E)$ defines, at time $t$, the relative contribution to the number of new infections generated in residents of any location caused by residents of location $k$ that were infected $a_E$ time units ago. In general, the shape of these generation time distributions can depend on biological infectiousness profiles, contact rates, susceptible and infected population counts, and importantly, the probabilities of residents being in different locations and whether this coincides with susceptible/infected individuals either staying in or moving to the meeting location. However, if we assume the separable form of eq. \eqref{eq:lambda_components} for $\lambda_E^{kl}(t, a_E)$, then $g^{kj}(t, a_E)$ will be time-invariant (a usual assumption for directly transmitted diseases) for times $t$ beyond the infectiousness support and arises as $g^{kj}(t, a_E) = \frac{p^k(a_E)}{\int_0^tp^k(a_E)da_E}$,
thereby only depending on the infectiousness profile of the infected resident of location $k$. If all locations also share the same infectiousness profile, we will have space- and time-invariant generation time distributions, i.e. the \textit{intrinsic} generation time distribution generally used in renewal equations \cite{park_importance_2022}. Note that these specific, separability and space-invariant simplifications will not hold when there are contact rates, movement patterns, or infectiousness profiles that vary over calendar time, across space, and with infection age. We therefore keep the more general notation for time- and space-varying distributions, $g^{kj}(t, a_E)$ and $g_{\text{out}}^k(t, a_E)$, throughout.

We can define our renewal equation per location in a more familiar way, capturing how previously infected individuals generate new infections in location $j$:
\begin{align}
    E^j(t, 0) &= \sum_{k \in \mathcal{N}} \int_{0}^{t}\mathbb{K}^{kj}(t, a_E) E^k(t-a_E, 0)da_E \\ &= \sum_{k \in \mathcal{N}} R^{kj}(t)\int_{0}^{t} g^{kj}(t, a_E) E^k(t-a_E, 0)da_E.
    \label{eq:exposed_model_one_b}
\end{align}
If we set $f^{jk}(t,S)=\delta_{jk}$ and $f^{jk}(t,E)=\delta_{jk}$ for all locations $j$ and $k$, then we recover a familiar single-patch renewal equation for a closed population without importations:
\begin{align}
    E^j(t, 0) &=  \int_{0}^{t}\mathbb{K}^{jj}(t, a_E) E^j(t-a_E, 0)da_E \\ &= R^{jj}(t)\int_{0}^{t} g^{jj}(t, a_E) E^j(t-a_E, 0)da_E\label{eq:exposed_model_one_c}.
\end{align}
Note that $R^{jj}(t)$ will generally differ from existing estimators of local $R(t)$ in closed populations, as $R^{jj}(t)$ accounts for outward movement of susceptible and infected residents from location $j$ and the inward movement of susceptible and infected residents from other locations which may inflate/reduce local transmissibility (e.g. in the setting of frequency-dependent transmission).

\subsubsection{Instantaneous measures of epidemic dynamics and drivers} \label{sec:instantaneous}
To derive a reproduction number that has the familiar threshold property and can measure instantaneous epidemic growth/decay, we assemble our between-location reproduction numbers into a valid (Sections \ref{sec:renewal_eqns_derivations} -- \ref{sec:tech_condns}), spatially structured  NGM $\mathbf{R(t)} \in \mathbb{R}^{\mathcal{L} \times \mathcal{L}}$, using the convention of rows as infectees and columns as infectors to align with previous work \cite{birello_estimates_2024, diekmann_definition_1990, diekmann_construction_2010, diekmann_mathematical_2000,van_den_driessche_reproduction_2002, van_den_driessche_reproduction_2017}:

\begin{align}\label{eq:reproduction_matrix}
\mathbf{R(t)} &:= 
\begin{pmatrix}
     R^{11}(t) & R^{21}(t) & \ldots & R^{\mathcal{L} 1}(t)\\
    R^{12}(t) & \ldots & \ldots & R^{\mathcal{L} 2}(t) \\
    \ldots & \ldots & \ldots & \ldots \\
    \ldots & \ldots & \ldots & R^{\mathcal{L}\mathcal{L}}(t)
\end{pmatrix},
\end{align}
from which we can derive the \textit{system-/network}-level reproduction number $\mathcal{R}(t)$ as the spectral radius of this reproduction/NGM \cite{birello_estimates_2024, diekmann_definition_1990}:
\begin{align}
\mathcal{R}(t) := \rho(\mathbf{R(t)}).
\end{align}
This quantity defines instantaneously the asymptotic growth/decay of the per-generation new infections of the (linearised) epidemic on the network. This is because we can apply the Perron-Frobenius Theorem (Supplementary Material Section \ref{sec:tech_condns}) to ensure a unique, dominant eigenvalue $\rho(\mathbf{R(t)})$ that is simple, real, and positive. We can apply the theorem as $\mathbf{R}(t)$ is a non-negative, irreducible, and primitive matrix, with the conditions satisfied because $f^{jk}(t, S), \ S^j(t), \ f^{kl}(t, E), \  \lambda_E^{kl}(t, a_E)\geq 0$ for all location pair combinations. 

Note that the spectral radius (and hence, $\mathcal{R}(t)$) will be the same if defined with rows corresponding to infectors and columns corresponding to infectees. 

$\mathcal{R}(t)$ is \textit{not} the expected/average number of secondary infections per infected human throughout their infected lifetime, should transmission conditions remain the same as at time $t$, as this will be quantified per location (by $R_{\text{out}}^j(t)$) and depend on the residence location $j$ of the infected individual. Instead, applying $\mathbf{R}(t)$ to the incidence vector $\mathbf{E}(t,0)$ repeatedly for many generations $n$ (with $\mathbf{R}(t)$ held at time $t$ value), the total incidence will grow/decay as $\left(\mathcal{R}(t)\right)^n$ because $\mathcal{R}(t)$ is the dominant eigenvalue. $\mathcal{R}(t)$ therefore, provides the asymptotic growth factor for the linearised epidemic and the threshold property at the network level, with values of $\mathcal{R}(t) >1$ indicating asymptotic growth on the network and values of $\mathcal{R}(t) < 1$ indicating asymptotic decay on the network. \citet{birello_estimates_2024} showed the bias from estimating an aggregate $\mathcal{R}(t)$ for spatially structured populations without a spatially structured approach, which arises when residents of different locations generate heterogenous totals of secondary infections.

As we have a valid NGM, by Perron-Frobenius theory \cite{horn_matrix_2017, caswell_matrix_2002}, the spectral radius $\mathcal{R}(t) = \rho(\mathbf{R(t)})$ also provides the useful property that quantifies how the network epidemic growth $\mathcal{R}(t)$ is influenced by human movement patterns:
\begin{align}\label{eq:property_spectral}
    \min_{k} R_{\text{out}}^{k}(t) \leq \mathcal{R}(t) \leq \max_{k} R_{\text{out}}^{k}(t),
\end{align}
which means that $\mathcal{R}(t)$ lies between the smallest and largest outward reproduction numbers. Equality in the upper bound occurs when all column sums of $\mathbf{R}(t)$ are equal, that is, when each location has equal instantaneous onward transmissibility (outward reproduction numbers), and more generally the location with the highest (lowest) outward reproduction numbers provide the upper (lower) bound on network epidemic growth $\mathcal{R}(t)$. 

As in population dynamics models, the eigenvalue formulation (Supplementary Material Section \ref{sec:tech_condns} and \cite{birello_estimates_2024, caswell_matrix_2002}) provides further intuition. The (normalised) right eigenvector $\mathbf{v*}$ is defined such that:
\begin{align}
\mathbf{R}\mathbf{v^*} = \mathcal{R}(t)\mathbf{v^*},    
\end{align}
and $\mathbf{v^*}$ therefore defines the long-term, equilibrium \textit{spatial distribution} of new infections \cite{caswell_matrix_2002} (i.e. resident locations where infections end up in the long term) where $v_j^*$ is the fraction of infections that would eventually occur in residents of location $j$, irrespective of the initial spatial distribution of infections. This is because repeatedly applying the NGM $\mathbf{R}$ to an input spatial distribution of infections $\mathbf{v^*}$ results in the same shape/pattern $\mathbf{v^*}$, scaled by $\mathcal{R}(t)$.

The left eigenvector $\mathbf{v}$ is defined such that:
\begin{align}
    \mathbf{v}^\top\mathbf{R} = \mathcal{R}(t)\mathbf{v}^\top.
\end{align}
Equivalently, the entry-wise equation $\sum_{j \in \mathcal{N}}{v_j}\mathbf{R}_{jk} = \mathcal{R}(t){v_k}$ captures that entries $v_k$ measure the long-term \textit{reproductive value} \cite{caswell_matrix_2002} (i.e. epidemic contributions) of a primary infected individual resident in each location $k$ because the reproductive value of one infection in location $k$ weights the number of secondary infections by the importance of the secondary resident locations $j$. Therefore, $v_k$ (known as Fisher's reproductive value in demography) summarises the long-term contribution to epidemic growth on the network from residents of location $k$ (i.e. how important locations are to growth). $v_j$ goes beyond the single-generation quantity $R_{\text{out}}^j(t)$ because it incorporates the long-term epidemic network structure to consider whether the locations of secondary infections are themselves important for amplifying in the network. For example, a location could have moderate $R_{\text{out}}^j(t)$ but high $v_j$ as it connects strongly to important amplifying locations. Collectively, $\mathbf{v^*}$ tells us the resident locations where infections end up in the long run and $\mathbf{v}$ tells us the resident locations where infections matter most for driving future epidemic growth.

We can define the network-level instantaneous kernel $\mathbb{K}_{\text{network}}(t, a_E)$ and generation time distribution $g_{\text{network}}(t, a_E)$ by weighting transmission pathways (specifically between-location instantaneous kernels) by the asymptotic stable infection distribution $\mathbf{v^*}$ and asymptotic reproductive value $\mathbf{v}$:
\begin{align}
    \mathbb{K}_{\text{network}}(t, a_E):= \sum_{j \in \mathcal{N}} \sum_{k \in \mathcal{N}}  \frac{v_jv_k^*}{v^\top v^*}  \mathbb{K}^{kj}(t, a_E), \\
    g_{\text{network}}(t, a_E):= \frac{\mathbb{K}_{\text{network}}(t, a_E)}{\mathcal{R}(t)},
\end{align}
which are the effective instantaneous kernel and generation time distribution of the network, i.e. for a single typical infection in the network after accounting for the network's dominant transmission dynamics (resident locations where infections occur and how important resident locations are).

\subsubsection{Drivers and control of epidemics}
As a scalar value, $\mathcal{R}(t)$ does not tell us which locations are driving epidemic growth, yet we can use the eigenvalue formulation to derive sensitivity and elasticity metrics. From matrix perturbation theory, the sensitivity of the spectral radius $\rho(\mathbf{R}(t)) = \mathcal{R}(t)$ with respect to between-location reproduction numbers $R^{kj}(t)=(\mathbf{R}(t))_{jk}$ is the Frechet derivative of the spectral radius (as we have a real, simple, dominant eigenvalue) or equivalently the weighting term for the network instantaneous kernel:
\begin{align}\label{eq:sens}
s^{kj}(t) &:=  \frac{\partial \mathcal{R}(t)}{\partial (\mathbf{R}(t))_{jk}} \\ &=  \frac{\partial \mathcal{R}(t)}{\partial R^{kj}(t)} \\ 
&= \frac{v_j v_k^*}{\mathbf{v}^\top \mathbf{v^*}},
\end{align}
which uses eigenvector centrality to quantify the sensitivity of the instantaneous network-level transmissibility $\mathcal{R}(t)$ to transmission from residents of locations $k$ to residents of location $j$. This tells us how much our eigenvalue $\mathcal{R}(t)$ changes when we change the strength of between-location transmission pathways/corridors $R^{kj}(t)$. The elasticity of $\mathcal{R}(t)$ is the dimensionless version of $s^{kj}(t)$:

\begin{align}
\epsilon^{kj}(t) &:=  \frac{(\mathbf{R}(t))_{jk}}{\mathcal{R}(t)} \cdot \frac{\partial \mathcal{R}(t)}{\partial (\mathbf{R}(t))_{jk}} 
\\ &= \frac{ R^{kj}(t)}{\mathcal{R}(t)} \cdot s^{kj}(t),
\end{align}
which can be interpreted as a $1\%$ change in $R^{kj}(t)$ produces an $\epsilon^{kj}(t) \%$ change in $\mathcal{R}(t)$. Elasticities $\epsilon^{kj}(t)$ are more instructive than using $R^{kj}(t)$ to target interventions at mobility corridors, as pairwise locations with high elasticities $\epsilon^{kj}(t)$ provide the rankings of directed corridors/pathways where disease control interventions produce the largest network-level benefit per unit of effort, after accounting for the network and epidemic structure (the spatial distribution of infections and per-location onward transmission potential via $\mathbf{v^*}$ and $\mathbf{v}$). For example, a pair of locations may have high $R^{kj}(t)$ yet have low eigenvector centrality (low $v_j$ and $v_k^*$) and hence, lower $\epsilon^{kj}(t)$ and lower intervention benefits/effects due to being located on the periphery of the network.

To quantify the importance of locations to onward network growth/transmissibility and rank locations by their total efficacy for control interventions, we can sum over the resident locations of secondary (primary) infections $j$ ($k$) to obtain the infector (infectee) elasticity per location: 
\begin{align}
\epsilon^{k}_{out}(t) &:= \sum_{j\in \mathcal{N}}\epsilon^{kj}(t), \\ \epsilon^{j}_{in}(t) &:= \sum_{k\in \mathcal{N}}\epsilon^{kj}(t).
\end{align}
Rankings of $\epsilon^{k}_{out}(t)$ and $\epsilon^{j}_{in}(t)$ can be used to target infection source control (exporters) and destination control measures respectively.

As $R^{kj}(t)$ is driven by the movement of individuals, we can also compute sensitivity and elasticity of $\mathcal{R}(t)$ with respect to probabilities of being in different locations and contact rates (which will often be a function of within-location activity). Note that computing sensitivity with respect to $f^{kj}(t)$ may not be particularly useful because reducing proportions of times in different locations (e.g. outside home resident location) does not account for the volume of movements in the alternative, counterfactual setting (e.g. in home residence). This problem is why sensitivity with respect to $R^{kj}(t)$ and pairwise flows $n^{kj}(t)$ between locations $k$ and $j$ may be more worthwhile as this can directly impact $\lambda^{kj}(t)$ for within- and between-location transmission.

\subsubsection{Measuring transient epidemic dynamics} \label{sec:asymptotic_transient}
Despite the PDE framework being a deterministic model of epidemic dynamics, $\mathcal{R}(t)$ only instantaneously measures the asymptotic epidemic growth/decay at time $t$, and is not predictive under rapid mobility changes, yet we can also explore the transience of epidemic dynamics using spectral properties of the NGM. This is important as, even for a deterministic model, volatile and heterogeneous mobility and transmissibility in the short term can alter the NGM quickly, render the assumption of constant transmission conditions over future generations (as in any instantaneous reproduction number) invalid, and create false-stability/growth information for decision-makers from elasticity and control indicators. 

The second largest (subdominant) eigenvalue $\lambda_2(t)$ governs the slowest-decaying transient mode, can be real or complex (if $\mathbf{R}(t)$ is asymmetric) and importantly, controls how quickly the epidemic converges to the asymptotic spatial distribution of infections, i.e. the right eigenvector $\mathbf{v}^*$. Then, the mixing time ratio $s(t)$ \cite{neubert_alternatives_1997, caswell_matrix_2002} is defined as:
\begin{align}
    s(t) := \frac{|\lambda_2(t)|}{\mathcal{R}(t)},
\end{align}
which measures instantaneously the speed of mixing of the epidemic on the network, i.e. deviations from $\mathbf{v}^*$ decay (geometrically) by a factor $s(t)$ each generation. Values of $s(t)$ lie in $(0, 1)$, with values close to 1 indicating slow mixing and the spatial configuration not changing much from generation to generation and values close to 0 indicating fast mixing and rapid convergence to $\mathbf{v}^*$. The ratio $s(t)$ therefore identifies how quickly initial perturbations/unstable spatial distributions are smoothed out by the dominant renewal equation behaviour of the system. In matrix population models \cite{caswell_matrix_2002}, the alternative damping ratio $d(t) = \frac{1}{s(t)} =\frac{\mathcal{R}(t)}{|\lambda_2(t)|}$ is often used, while the spectral gap $\zeta(t) = \mathcal{R}(t) - |\lambda_2(t)|$ also quantifies how long transience lasts before asymptotic behaviour (as implied by $\mathcal{R}(t)$) occurs, with large values of $\zeta(t)$ implying faster convergence to asymptotic behaviour (which is more likely in highly symmetric, densely connected urban networks). This spectral gap $\zeta(t)$ is important as it impacts the operational reliability of our control indicators and elasticity metrics as if the spectral gap is small (e.g. in highly disjointed, sparse national networks), eigenvectors and their inner product within the elasticity metrics can become unstable and sensitive to small amounts of noise (e.g. in the mobility data).


We can further explore transient perturbations (e.g. interventions or importations) and the possible effects of these perturbations by defining the discrete-time, \textit{reactivity} of the NGM \cite{neubert_alternatives_1997, caswell_reactivity_2005}:
\begin{align}
    \sigma(t) &:=  ||\mathbf{R}(t)||_2 = \max_{||\mathbf{E}||_2=1}||\mathbf{R}(t)\mathbf{E}||_2, 
\end{align}
which is the largest singular value or the spectral/operator/induced-2 norm of $\mathbf{R}(t)$, thereby measuring the maximum/worst-case amplification of the infection incidence after a single generation. $\sigma(t) \geq \mathcal{R}(t)$ with equality if and only if $\mathbf{R}(t)$ is a normal matrix (in which case $\mathbf{R}(t)\mathbf{R}(t)^\top = \mathbf{R}(t)^\top \mathbf{R}(t)$). $\sigma(t)$ can therefore be greater than 1 even when $\mathcal{R}(t)$ is less than 1, which implies that epidemic, per-generation perturbations can amplify before the eventual asymptotic decay. These perturbations correspond to transient growth in incident infections, despite asymptotic epidemic decline implied by $\mathcal{R}(t)$. The transience can happen when there are asymmetric mobility patterns and hence, asymmetric $\mathbf{R}(t)$, as the largest singular value $\sigma(t)$ and largest eigenvalue $\mathcal{R}(t)$ can disagree for non-normal matrices, which we anticipate for most real-world cities with distinct, directed urban mobility patterns.

The condition number $\kappa(\mathcal{R}(t))$ of the dominant eigenvalue $\mathcal{R}(t)$ can also quantify how sensitive/robust $\mathcal{R}(t)$ is to small perturbations of entries in the NGM (e.g. through misspecified or time-varying mobility or epidemiological assumptions), thereby suggesting how biases and/or fluctuations can arise in estimated $\mathcal{R}(t)$ due to small changes in the input:
\begin{align}
    \kappa\left(\mathcal{R}(t)\right) := \frac{||\mathbf{v}||_2 ||\mathbf{v^*}||_2}{|\mathbf{v}^\top \mathbf{v}^*|}.
\end{align}
Operationally, large values of $\kappa(\mathcal{R}(t))$ suggest that $\mathcal{R}(t)$ is sensitive to noisy/fluctuating human movement data, thereby suggesting caution in over-interpreting the instantaneous quantity $\mathcal{R}(t)$ when mobility data are uncertain and/or volatile. Mathematically, $\kappa\left(\mathcal{R}(t)\right)$ is related to the sensitivity quantities $s^{kj}(t)$ that define the gradient of the spectral radius (i.e. a local entry-wise, directional derivative), because the condition number provides the (operator) norm of that gradient, thereby providing a worst-case, aggregation of the sensitivities. In simple terms, it quantifies the instability of the eigenvectors, their inner product, and therefore the elasticity gradients that may be used for targeting public health measures. 

To quantify the duration of the transient growth in epidemic dynamics across $n$ generations, using $\mathbf{E}$ as short-hand for the incidence vector across locations, we can define the amplification envelope $A(n)$ \cite{trefethen_spectra_2020, neubert_alternatives_1997} as
 \begin{align}
    A(n) &:= ||\mathbf{R}(t)^n||_2 = \max_{||\mathbf{E}||_2=1}||\mathbf{R}(t)^n\mathbf{E}||_2, 
\end{align}
as this captures the maximum directional multiplication/amplification factor (i.e. the spatial pattern) after $n$ generations, irrespective of the initial incidence. The trivial settings are $A(0) = 1$ and $A(1) = \sigma(t)$ which imply respectively no amplification after zero generations and that the one-generation amplification is just the reactivity of the NGM. For non-normal matrices (i.e. heterogeneous infection, mixing and movement patterns), we can have $A(n) \gg \mathcal{R}(t)^n$ \cite{trefethen_spectra_2020}, while normal matrices will have $A(n) = \mathcal{R}(t)^n$ for all $n$. The ratio $\frac{A(n)}{\mathcal{R}(t)^n}$ quantifies how different transient dynamics are compared to asymptotic dynamics.

Alternative norms can be used, and as in any mathematical application, it is important to specify which norm is used to study transience \cite{mari_sufficient_2025}. The $\ell^1--$norm is less common but increasingly used in epidemiology. For example, \citet{mari_sufficient_2025} use the $\ell^1-$norm to define so-called \textit{epidemicity} metrics that measure transience over discrete time steps in different ways, including first-time/generation epidemicity to measure maximum growth, new thresholds for reproduction numbers to prevent transient outbreaks, and maximum epidemicity. While \citet{mari_sufficient_2025} focus on infection-age structure in a single location, we focus on transience arising from asymmetries in spatial mobility but if we use the similar $\ell^1-$norm, the amplification envelope moves from quantifying worst-case directional amplification from multiple-location seeding (i.e. the $\ell^2$ norm in $A(n)$) to a worst-case amplification from single-location seeding:
 \begin{align}
    A_1(n) &:= ||\mathbf{R}(t)^n||_1 = \max_{k \in \mathcal{N}} \sum_{j\in \mathcal{N}} \left (\mathbf{R}(t)^n \right)_{jk},
\end{align}
which means that $A_1(0) = 1$ and the first-generation epidemicity $$A_1(1) = \max_{k \in \mathcal{N}} \sum_{j\in \mathcal{N}} \left (\mathbf{R}(t) \right)_{jk} = \max_{k \in \mathcal{N}} R_{\text{out}}^{k}(t)$$ 
is the largest outward reproduction number and satisfies (by eq. \eqref{eq:property_spectral}) $A_1(1)\geq \mathcal{R}(t)$. For $n > 1, A_1(n)$ quantifies the amplification over $n$ generations from a single, worst-case seeding location for the primary infected individuals after accounting for secondary infections propagating across the network, i.e. the maximum number of total infections across all locations over $n$ generations arising from one seeding location. This differs from the amplification envelope and reactivity from the $\ell^2$ norm which can capture multi-location seeding amplifications.

Operationally during epidemics, to capture network-level transmissibility, modellers have often used averages of reproduction numbers across locations, either using an arithmetic mean or infectious-force-weighting, which can result in false indicators of a stable epidemic \cite{parag_r_2024}. Because our-network level $\mathcal{R}(t)$ accounts for how likely infections are for different resident locations and how much locations can likely contribute to epidemic growth, rather than weighting reproduction numbers by infectious force, it does not suffer from the specific false-stability, averaging issues caused by infectious-force-weighting across locations \cite{parag_r_2024}. 

However, our quantity $\mathcal{R}(t)$ has its own weaknesses because locations $j$ with high $R_{\text{out}}^j(t)$ could contribute little to $\mathcal{R}(t)$ due to low/moderate eigenvector centrality at time $t$ (low values in $\mathbf{v}$ and $\mathbf{v^*}$) and their risk for subsequent infection resurgences may be undetected. We can therefore follow a similar procedure to \citet{parag_r_2024} and define a \textit{Lehmer-mean spectrum} $X(\alpha, t)$ that weights $R_{\text{out}}^j(t)$ to balance how we weight/average transmissibility over locations:
\begin{align}
    X(\alpha, t) := \sum_{j \in \mathcal{N}} \omega_j(\alpha) R_{\text{out}}^j(t), \quad \quad \text{where } \omega_j(\alpha) = \frac{(R_{\text{out}}^j(t))^\alpha}{\sum_{k \in \mathcal{N}}(R_{\text{out}}^k(t))^\alpha}.
\end{align}
Then, for $\alpha = 0$ , $X(0, t)$ is the arithmetic mean average of outward reproduction numbers, and interestingly, for $\alpha \rightarrow \infty$, $X(\alpha, t) {\rightarrow} \max_{k \in \mathcal{N}} R_{\text{out}}^{k}(t) = A_1(1)$. This shows how the spectrum $X(\alpha, t)$ traverses from the extremes of over-averaging across locations ($X(0, t)$) and possibly blurring stability signals to over-amplifying individual locations ($A_1(1)$) and being possibly overly sensitive to fluctuations in outward reproduction numbers (e.g. in low-incidence locations). Similar to \citet{parag_r_2024}, we introduce a new \textit{spatial risk-averse reproduction number} $\mathcal{E}(t)$, defined for our system as $X(1, t)$: 
\begin{equation}
\begin{aligned}
    \mathcal{E}(t) &:= X(\alpha = 1, t) \\ &= \sum_{j \in \mathcal{N}} \omega_j(1) R_{\text{out}}^j(t) \\ &= \frac{\sum_{j \in \mathcal{N}} R_{\text{out}}^j(t)^2}{\sum_{j \in \mathcal{N}} R_{\text{out}}^j(t)}, 
\end{aligned}
\end{equation}
which measures the transmissibility of the network by weighting the relative transmissibility of individual locations. Note that $\mathcal{R}(t) = \rho(\mathbf{R}(t))$ sits outside the Lehmer-mean spectrum $X(\alpha, t)$, and therefore can disagree with all members of the continuum because $\mathcal{R}(t)$ considers whether multi-generation can sustain growth unlike $\mathcal{E}(t)$ which considers whether locations are individually transmitting strongly and upweights such locations, with weights determined by the strength of their relative transmissibility. Relatedly, a single-generation quantity, $\mathcal{E}(t)$ will not have the usual threshold property for long-term epidemic growth/decay (Supplementary Material Section \ref{sec:risk_aware_appendix}). While $\mathcal{E}(t)$ is motivated by the same $\alpha = 1$ member of the Lehmer-mean spectrum, $\mathcal{E}(t)$ differs from the risk-averse reproduction number of \citet{parag_r_2024} because it is based on $R_{\text{out}}^j(t)$ in a connected human mobility network (as opposed to local, independent reproduction numbers in closed populations), thereby further accounting for mobility-mediated transmission via inward and outward flows of susceptible and infected individuals. $\mathcal{E}(t)$ can be written as a mean-variance expression (eq. \eqref{eq:mean_variance}), thereby showing that control explicitly requires both the mean transmission to be controlled and the spatial heterogeneity to be minimised to signal stability:
 \begin{align}\label{eq:mean_variance}
\mathcal{E}(t) &:=\bar{R}_{\text{out}}(t) + \frac{\text{Var}( R_{\text{out}}^j(t))}{\bar{R}_{\text{out}}(t)}.
\end{align}

The quantity $\mathcal{E}(t)$ is a complementary spatial risk warning indicator and the advantage of using $\mathcal{E}(t)$ and $\mathcal{R}(t)$ together is that they can collectively avoid false signals for decision-making. When $\mathcal{E}(t) > 1$ and $\mathcal{R}(t) < 1$, there is a false stability signal from the network as the spectral structure of the NGM can mask resurgent infectious potential from any locations (i.e. those with high $R_{\text{out}}^j(t)$ but low eigenvector centrality within the network). $\mathcal{E}(t)$ therefore can account for spatially heterogeneous resurgence. When $\mathcal{E}(t) < 1$ and $\mathcal{R}(t) < 1$, there is no localised resurgence potential and the network cannot sustain transmission. When $\mathcal{E}(t) \approx 1$ and $\mathcal{R}(t) > 1$, there can be network-driven growth (which is captured by spectral coupling across locations in $\mathcal{R}(t))$ despite the weighted average of reproduction numbers for individual locations not looking alarming (as $\mathcal{E}(t)$ cannot handle network-mediated amplification). As the mobility network amplifies transmission across multi-step pathways in this setting, corridor-level control measures may be more effective than location-specific control measures.




\subsubsection{Between-location, inward, and meeting-location transmission}
To capture how infections are propagating across locations in the network at any given time, we can write between-location renewal equations as:
\begin{equation}
\begin{aligned}\label{eq:between_nbr_renewal_eq}
E^{k \rightarrow j}(t, 0) &= R^{kj}(t) \int_0^t g^{kj}(t, a_E) \cdot E^k(t- a_E,0)d a_E.
\end{aligned}
\end{equation}
For example, within-location incidence could be compared to incidence resulting from residents of other locations, thereby showing the importance of inward human movement to a location.

We can therefore set up an incidence/renewal equation matrix $\mathcal{E}(t)$ with individual elements ${E}(t,0)^{k \rightarrow j}$ describing the number of new infections generated by previously infected individuals from location $k$ in location $j$. Rows thereby sum to the number of new infections occurring from residents of location $j$ at time $t$, while columns sum to the number of new infections generated in residents of location $k$ at time $t$. 
$$ \mathcal{M}(t) = 
\begin{pmatrix}
     E^{1\rightarrow 1}(t, 0) & E^{1\rightarrow 2}(t, 0) & \ldots & E^{1\rightarrow \mathcal{L}}(t, 0)\\
    E^{2 \rightarrow 1}(t, 0)& E^{2\rightarrow 2}(t, 0) & \ldots & \ldots \\
    \ldots & \ldots & \ldots & \ldots \\
    \ldots & \ldots & \ldots & E^{\mathcal{L}\rightarrow \mathcal{L}}(t, 0) 
\end{pmatrix}
$$
\normalsize
The diagonal elements describe the within-location transmission from the renewal equation (similar to a traditional single-location equation, but with inward and outward human population movements accounted for), while other elements describe between-location transmission.

$\mathcal{R}(t)$ and $R_{\text{out}}^j(t)$ capture instantaneously the network-level epidemic growth and the transmission potential respectively by infected residents of location $j$, yet often decision-makers want to know if residents of location $j$ are vulnerable to imported incident infections. To capture the \textit{inward} transmissibility vulnerability/potential, we can define the expected number of new infections occurring in residents of location $j$ at time $t$ resulting from a single infected individual resident in each location after accounting for all human movements:
\begin{equation}
\begin{aligned}\label{eq:two_nbr_reprod}
R_{\text{in}}^{j}(t) &:= \sum_{k \in \mathcal{N}}  R^{kj}(t)=  \sum_{k \in \mathcal{N}} \int_0^t \mathbb{K}^{kj}(t, a_E) da_E, 
\end{aligned}
\end{equation}
which has a different meaning to the closed population setting for each location $j$ as $R_{\text{in}}^{j}(t)$ is the expected number of new human infections occurring in residents of location $j$ over the course of an infection lifetime (i.e. one generation) in a counterfactual setting of one infected individual resident seeding \textit{simultaneously in each location}, if conditions affecting transmission \textit{from each location to residents in location $j$} were to remain the same as at time $t$. Values of $R_{\text{in}}^{j}(t)$ can be compared to the result from independently estimating $R^j(t)$ for each location, as the independent approach will model the local residents' infectiousness pressure without modelling the inward and outward population movements of location $j$ throughout the day. Again, $R_{\text{in}}^{j}(t)>1$ may indicate transmission vulnerability and similar to \cite{kim_evaluating_2026, jorge_estimating_2022}, we describe it as a ``reproduction number'', but $R_{\text{in}}^{j}(t)$ does \textit{not} possess the threshold property (at time $t$ for the long-term growth/decay in the epidemic in residents of location $j$) as it does not account for where infected individuals are distributed across the network and network-mediated feedback across multiple infection generations.

To capture the heterogeneity in inward (e.g. high-burden locations) or outward (e.g. superspreading locations) transmission vulnerability/potential, and compare these heterogeneities, we can compute the coefficient of variation (CV) of the row ($\text{CV}(R_{\text{in}}^j(t))$) or column sums ($\text{CV}(R_{\text{out}}^j(t))$) respectively of $\mathbf{R}(t)$. By computing the ratio 
\begin{align}
\eta(t) := \frac{\text{CV}(R_{\text{out}}^j(t))}{\text{CV}(R_{\text{in}}^j(t))},
\end{align}
we can examine whether heterogeneity in outward transmission potential exceeds heterogeneity in inward vulnerability ($\eta(t)$) which would suggest that, based on instantaneous transmission conditions, targeting control at high-net-exporter/source locations may be more efficient than targeting high-net-importers.

We can also define the inward generation time distributions and inward instantaneous kernels per location $j$:
\begin{align}
    \mathbb{K}_{\text{in}}^{j}(t, a_E) &:=   \sum_{k \in \mathcal{N}} \mathbb{K}^{kj}(t, a_E), \\
    {g}_{\text{in}}^{j}(t, a_E) &:=    \frac{\mathbb{K}_{\text{in}}^{j}(t, a_E)}{R_{\text{in}}^{j}(t)},
\end{align}
which capture respectively, for residents of each location $j$ at time $t$, how many expected new infections are to occur, and the relative contribution to new infections, resulting from individuals of infection age $a_E$ resident in any location in the network. While $R_{\text{in}}^{j}(t)$ can inform the magnitude of interventions, ${g}_{\text{in}}^{j}(t, a_E)$ can inform how long interventions may be needed for location $j$ to experience an outbreak under control. 

Decision-makers are also often interested in targeting interventions at locations \textit{where} transmission occurs \cite{jewell_use_2022}. We can define kernels, reproduction numbers, generation times, and incidence for individual \textit{meeting} locations $l$ (i.e. indexing by location of infection and not residence):
\begin{align}
    \mathbb{K}_\text{meeting}^{l}(t, a_E) &:= \sum_{j \in \mathcal{N}} \sum_{k \in \mathcal{N}}f^{jl}(t, S) \cdot S^j(t) \cdot f^{kl}(t, E) \lambda_E^{kl}(t, a_E), \\
     {R}_\text{meeting}^{l}(t) &:= \int_{0}^{t}\sum_{j \in \mathcal{N}} \sum_{k \in \mathcal{N}}f^{jl}(t, S) \cdot S^j(t) \cdot f^{kl}(t, E) \lambda_E^{kl}(t, a_E)da_E = \int_{0}^{t}  \mathbb{K}_\text{meeting}^{l}(t, a_E) da_E, \\
     g_\text{meeting}^{l}(t, a_E) &:= \frac{\mathbb{K}_\text{meeting}^{l}(t, a_E)}{ {R}_\text{meeting}^{l}(t)}, \\
     E_\text{meeting}^l(t, 0) &=  \int_{0}^{t} \sum_{j \in \mathcal{N}} \sum_{k \in \mathcal{N}}f^{jl}(t, S) \cdot S^j(t) \cdot  f^{kl}(t, E) \lambda_E^{kl}(t, a_E)E^k(t -a_E, 0) da_E.
\end{align}
which respectively quantify instantaneously for individuals in meeting location $l$: how many infections are expected at time $t$ per infected individual of infection age $a_E$, how many infections are expected to occur in location $l$ per one infected individual drawn from the network population over their infected lifetime if conditions affecting transmission in location $l$ remain the same, the relative contributions to new infections in location $l$ from individuals of infection age $a_E$, and how many infections occur at time $t$. Again, $R_{\text{meeting}}^l(t)$ does \textit{not} have the threshold property as it cannot measure asymptotic/long-term growth of the epidemic at meeting location $l$ (e.g. a location $l$ could have high $R_{\text{meeting}}^l(t)$ yet not have the susceptible population in location $l$ over time for sustained transmission). However, $R_{\text{meeting}}^l(t)$ can measure the mixing capacity/potential for a location $l$ to act as a transmission arena/centre and motivate interventions targeted at high-footfall, transmission hubs that may be missed by residence-focused metrics. To understand effective susceptible $S_{\text{eff}}^{l}(t)$ and infected populations $E_{\text{eff}}^{l}(t)$ in any location $l$ at any time $t$ (e.g. during-day versus during-night or during-week versus weekend), we can also define
\begin{align*}
    S_{\text{eff}}^{l}(t) &:=  \sum_{k \in \mathcal{N}}f^{jl}(t, S) \cdot S^j(t), \\
     E_{\text{eff}}^{l}(t) &:= \int_0^{t}\sum_{j \in \mathcal{N}}f^{jl}(t, E) \cdot E^j(t, a_E)da_E.
\end{align*}

\subsubsection{Type reproduction numbers for locations}
To address the lack of per-location threshold properties for $R_{\text{in}}^j(t)$ and $R_{\text{out}}^j(t)$, quantify how essential resident location $j$ is to the epidemic, capture multi-step/location network feedback, and suggest more targeted interventions, we also derive an instantaneous \textit{type reproduction number} (following \cite{roberts_new_2003, heesterbeek_type-reproduction_2007}) for each location $j$ using the NGM $\mathbf{R} = \mathbf{R}(t)$ (eq. \eqref{eq:reproduction_matrix}):
\begin{align}
R_{\text{type}}^j(t) &:= {e}_j^\top \mathbf{R}\left(I - (I - P_j) \mathbf{R} \right)^{-1}{e}_j
\end{align}
which defines the expected number of eventual/return infections in residents of location $j$ resulting from an single primary infected resident of location $j$, only counting the secondary infections that occur in residents of location $j$ and treating these first passage secondary infections as counted endpoints of the transmission chains. $R_{\text{type}}^j(t)$ is not simply a local quantity as it encompasses the whole epidemic mobility network. Interpreted as a multi-step/generation transmission pathway across the network and defined instantaneously, $R_{\text{type}}^j(t)$ is the expected number of eventual infections in residents of location $j$ generated by a primary infected resident of location $j$, accounting for all possible transmission/return pathways across the mobility network and all generations, if the transmission conditions were to remain the same as at time $t$. $R_{\text{type}}^j(t)$ can be decomposed into direct ($n=1$ generations) and indirect pathways ($n > 1$ secondary infections in residents outside location $j$ that then generate onward infections in location $j$ at least once) that generate eventual infections in residents of location $j$. First, we define the set of resident locations $J:= \{1, \ldots\ \mathcal{L}\} \textbackslash \{j\} = \mathcal{N} \textbackslash \{j\}$. Then, $\mathbf{R}_{jJ}$ captures how infected residents of other locations $J$ infect residents of location $j$, $R_{JJ}$ captures within-$J$ propagation of infections, and $R_{Jj}$ captures how infected residents of location $j$ infect residents of other locations $J$.


If the Neumann series converges for $\mathbf{R}_{JJ}$, that is, if $\rho(\mathbf{R}_{JJ}) < 1$ (Supplementary Material Section \ref{sec:type_derivation}) which occurs if there is not persistent amplification within the set of locations $J$ without eventual return to location $j$, then we have:
\begin{align}
R_{\text{type}}^j(t) &= \mathbf{R}_{jj} + \mathbf{R}_{jJ}(I - \mathbf{R}_{JJ})^{-1}\mathbf{R}_{Jj},
\end{align}
which is the type reproduction number \cite{heesterbeek_type-reproduction_2007, roberts_new_2003} defined instantaneously for a mobility-informed framework with transmission pathways mediated by human movement. The feedback feature distinguishes $R_{\text{type}}^j(t)$ from $R_{\text{in}}^j(t)$ and $R_{\text{out}}^j(t)$ which do not allow such network-mediated feedback and do not measure onward/secondary transmission chains from residents of location $j$ to residents of other locations and first passage infections back to residents of location $j$ over all future generations. In contrast, $R_{\text{type}}^j(t)$ has a persistence threshold property for the transmission lineage originating in location $j$, with $R_{\text{type}}^j(t) >1$ indicating that a single infected resident of location $j$ generates more than one infection in residents of location $j$, on average, over all future generations, i.e., the type/location $j$ transmission cycle can sustain itself, at least in the frozen/instantaneous conditions of time $t$. This goes beyond $R_{jj}(t)$ and independent per-location, closed population estimators $R_{j}^{ind}(t)$ for location $j$ which only capture locally generated infections or misattribute imported infections to local dynamics. $R^j_{\text{type}}(t) > 1$ implies that the transmission lineage for residents of location $j$ can sustain itself through the network, even if local within-j transmission $R_{jj} < 1$. 



\begin{table}[t]
    \centering
    \footnotesize
    \begin{tabular}{>{\centering\arraybackslash}p{0.11\linewidth}|>{\centering\arraybackslash}p{0.118\linewidth}|>{\centering\arraybackslash}p{0.118\linewidth}|>{\centering\arraybackslash}p{0.118\linewidth}|>{\centering\arraybackslash}p{0.118\linewidth}|>{\centering\arraybackslash}p{0.118\linewidth}|>{\centering\arraybackslash}p{0.118\linewidth}}
    \toprule
         & \textbf{Between-location}  & \textbf{Inward } & \textbf{Outward}  & \textbf{Meeting location} & $\textbf{Type}^*$ & $\textbf{Network}^*$ \\\midrule
       \textbf{Instantaneous kernel} & $\mathbb{K}^{kj}(t, a_E)$ 
       
       How many infections are generated at time $t$ \textit{by} \textbf{each infected resident of location $\boldsymbol k$} of infection age $a_E$ in \textbf{residents of location $\boldsymbol j$}  &  $\mathbb{K}_{\text{in}}^{j}(t, a_E)$ 
       
       How many infections are generated at time $t$ \textit{in} \textbf{residents of location $\boldsymbol j$} from one infected individual of infection age $a_E$ \textbf{resident of each location}  & $\mathbb{K}_{\text{out}}^{j}(t, a_E)$ 
       
       How many infections are generated at time $t$ \textit{by} \textbf{each infected resident of location $\boldsymbol j$} of infection age $a_E$ in \textbf{residents of any location}  &  $\mathbb{K}_\text{meeting}^{j}(t, a_E)$ 
       
       How many infections are generated at time $t$ \textit{at} \textbf{meeting location $\boldsymbol j$} per infected individual of infection age $a_E$ that is \textbf{resident of any location}& 
       -

       
       & $\mathbb{K}_\text{network}(t, a_E)$ 

        How many infections are generated at time $t$ at infection age $a_E$ if considering \textbf{typical} individuals \textbf{in the network} and the network’s dominant \textbf{transmission dynamics} 
       
       \\\hline

      \textbf{Instantaneous reproduction number}  & $R^{kj}(t)$
       
       How many infections are generated over the infected lifetime \textit{by} \textbf{each infected resident of location $\boldsymbol k$} in \textbf{residents of location $\boldsymbol j$} & 
       $R_{\text{in}}^{j}(t)$
       
       How many infections are generated over the infected lifetime \textit{in} \textbf{residents of location $\boldsymbol j$} from one infected individual resident in \textbf{each location} & 

       $R_{\text{out}}^{j}(t)$ 
       
       How many infections are generated over the infected lifetime \textit{by} \textbf{each infected resident of location $\boldsymbol j$} in \textbf{residents of any location}
       
       & $R_\text{meeting}^{j}(t)$ 
       
       How many infections are generated over the infected lifetime \textit{at} \textbf{meeting location $\boldsymbol j$} per infected individual that is \textbf{resident in any location}
       
       & $R_{\text{type}}^j(t)$

        How many infections are generated over the infected lifetime \textit{in} \textbf{residents of location $j$ originating from a resident of location} $\boldsymbol{j}$, accounting for \textbf{all possible first-return pathways} and all generations

        & $\mathcal{R}(t)$  \textbf{Asymptotic, per-generation growth factor} at time $t$ for the linearised epidemic on the network. 
       \\ \hline
       \textbf{Generation time distribution}   &  $g^{kj}(t, a_E)$ 
       
       The proportion of new infections at time $t$ \textit{in} \textbf{residents of location $\boldsymbol j$} that are generated \textit{by} \textbf{residents of location $\boldsymbol k$} of infection age $a_E$  &  $g_{\text{in}}^{j}(t, a_E)$ 
       
       The proportion of new infections at time $t$ \textit{in} \textbf{residents of location $\boldsymbol j$} that are generated \textit{by} one infected \textbf{resident in each location} that are of infection age $a_E$  & $g_{\text{out}}^{j}(t, a_E)$ 
       
       The proportion of new infections generated in \textbf{residents of any location} \textit{by} \textbf{residents of location $\boldsymbol j$} that are of infection age $a_E$ &  $g_\text{meeting}^{j}(t, a_E)$ 
       
       The proportion of the new infections generated \textit{at} \textbf{meeting location $\boldsymbol j$} per infected individual \textbf{resident of any location} that are of infection age $a_E$ &

       - 

       & $g_\text{network}(t, a_E)$ 

    The proportion of new infections at time $t$ generated at infection age $a_E$ if considering the \textbf{typical} infected \textbf{individual in the network} and the network’s dominant \textbf{transmission dynamics} 
       
       \\\hline
    \end{tabular}
    \caption{\textbf{Mathematical outputs for our framework:} We introduce in our framework (Section \ref{sec:results_part_1}) a wide range of instantaneous reproduction numbers, kernels, and generation time distributions. We provide here simplified definitions of each mathematical quantity which quantify unfolding epidemic dynamics across space and time. It is important to note that all of these are instantaneous measures of epidemic dynamics derived from the next-generation matrix (NGM). These instantaneous measures should be complemented by our transient measures of epidemic dynamics at individual and multiple locations (Section \ref{sec:asymptotic_transient}) that can quantify short-term dynamics. We highlight (\textbf{*}) the type and network reproduction numbers as both of these have the mathematical threshold property for long-term epidemic growth/decay.}
    \label{tab:framework_outputs}
\end{table}
\FloatBarrier


As $R_{\text{type}}^j(t)$ (if defined) captures both the localised and network-mediated transmission pathways, following \cite{roberts_new_2003,heesterbeek_type-reproduction_2007}, we have for an irreducible NGM $\mathbf{R}(t)$:
\begin{align}
    R_{\text{type}}^j(t) > 1 \iff \mathcal{R}(t) =\rho(\mathbf{R}(t))>1 \quad \forall j,
\end{align}
that is, if infections from residents of location $j$ generate more than one eventual infection in residents of location $j$ (i.e. $R_{\text{type}}^j(t) > 1$), we have network-level growth ($\mathcal{R}(t) > 1$) and if the epidemic has infection growth ($\mathcal{R}(t) > 1$), then this will eventually feed back into location/type $j$. Unlike $\mathcal{R}(t)$ however, $R_{\text{type}}^j(t)$ (if defined) can pinpoint the strength/contribution of transmission pathways originating from residents of location $j$ and suggest \textit{where} to target interventions. If $R_{\text{type}}^j(t) > 1$, a transmission reduction of $c_j > 1-\frac{1}{R_{\text{type}}^j}$ is required in residents of location $j$ to bring the epidemic under control.


Note however that $R_{\text{type}}^j(t)$ can be mathematically undefined as it diverges if $\rho(\mathbf{R}_{JJ}) \geq 1$, which occurs when the background network $\mathbf{R}_{JJ}$ can sustain an epidemic on its own without infections from residents of location $j$, so control efforts targeted at residents of location $j$ will not control epidemic growth. The issue suggests the possible unsuitability of $R_{\text{type}}^j(t)$ in large and/or poorly connected networks or times (e.g. during growth stages) where the relative importance of an individual location $j$ to the epidemic elsewhere can be low, as the epidemic can sustain itself without infections from the individual location $j$. As these can be the times when public health officials require intervention targets the most, in such settings, it might make sense to quantify the contributions of a set of \textit{multiple} locations $\mathcal{P}$ (e.g. highly connected core locations or those with high $R_{\text{out}}^j(t)$ and/or $\epsilon_{\text{in}}^{j}(t)$), by regarding these as a single type. We can then follow a very similar procedure to compute a group type reproduction number:  
\begin{align}
R_{\text{type}}^\mathcal{P}(t):=  
\rho\left(\mathbf{R}_{\mathcal{P}\mathcal{P}} + \mathbf{R}_{\mathcal{P}Q}(I - \mathbf{R}_{QQ})^{-1}\mathbf{R}_{Q\mathcal{P}} \right),
\end{align}
where $Q = \mathcal{N} \backslash \mathcal{P}$. Again, if $R_{\text{type}}^\mathcal{P}(t) > 1$, a minimal transmission reduction of $c_\mathcal{P}(t) > 1-\frac{1}{R_{\text{type}}^\mathcal{P}(t)}$ is required in resident locations $\mathcal{P}$ to bring the epidemic under control.

\begin{table}[t]
\centering
\footnotesize
\setlength{\tabcolsep}{4pt}
\begin{tabular}{p{2.62cm}| p{3.6cm}| p{3.75cm}| p{5.0cm}}
\toprule
\textbf{Quantity} & \textbf{Definition} & \textbf{Measures} & \textbf{Operational use/guidance} \\
\midrule

$\mathcal{R}(t)=\rho(\mathbf{R})$
  & \textit{Network reproduction number} defined by the spectral radius of $\mathbf{R}(t)$
  & Network-level transmissibility; controls the long-run epidemic trajectory on the network
  & \textbf{Network/global stability and control}: Control measures required in the network when $\mathcal{R}>1$.  Does not identify where to target locations. \\[4pt]
\hline
$R^j_\text{out}(t)=\sum_k R^{kj}(t)$
  & \textit{Outward:} Total expected infections caused by one infected resident of $j$ in any
    location throughout the course of their infected lifetime
  & Per-location \emph{transmission potential}; potential infection burden from location on the wider network, given current epidemic network dynamics
  & Prioritise \textbf{stay-at-home / movement-restriction orders} for
    high-$R_\text{out}$ locations. Does not have threshold property for epidemic growth or account for network-mediated effects on epidemic growth. \\[4pt]
\hline
$R^j_\text{in}(t)=\sum_k R^{jk}(t)$

  & \textit{Inward:} Total expected infections received by residents of location $j$ from one
    infected resident in each location throughout the course of their infected lifetime
  & Per-location \emph{vulnerability}; risk of receiving infection from
    seeding anywhere on the network, given current transmissibility and mobility dynamics 
  & Target \textbf{travel restrictions / quarantine} for high-$R_\text{in}$
    locations. Does not have threshold property for epidemic growth or account for network-mediated effects on epidemic growth. \\[4pt]
\hline
$R^{kj}(t)$
  & Directed \textit{between-location} transmission potential
  & Strength of the $k\!\to\!j$ transmission pathway, given current transmissibility and mobility dynamics 
  & Identify \textbf{specific travel corridors} with high pairwise transmission potential. \\[4pt]
\hline
$\epsilon^{kj}(t)$
  & \textit{Elasticity between locations}: \% change in $\mathcal{R}(t)$ per \% change in $R^{kj}(t)$
  & Onward transmission potential for network epidemic growth and efficacy of control measures targeted at the $k \rightarrow j$ transmission pathway
  & Identify \textbf{specific travel corridors} with high importance for network growth and high potential control measure efficacy. \\[4pt]
\hline

$\epsilon_{\text{out}}^{k}(t) = \sum_{j} \epsilon_{kj}(t)$, $\epsilon_{\text{in}}^{j}(t) = \sum_{k} \epsilon_{kj}(t)$
  & \textit{Infector and infectee elasticity}: Sum of \% changes in $\mathcal{R}(t)$ per \% change in $R^{kj}(t)$, summed over all infected resident destination or source locations $j$ or $k$.
  & Onward/inward transmission potential contributions to epidemic growth of location $k$
  & Identify \textbf{specific infector source and infectee destination locations} with high importance for network growth and high potential control measure efficacy. \\[4pt]
\hline

$R^l_\text{meeting}(t)$
  & Reproduction number attributed to \textit{meeting} location $l$
  & Infection risk \emph{at meeting location} $l$, regardless of residents' home
    location
  & Motivates \textbf{meeting-location-level measures} (markets, stadia, transport hubs).
    High $R_{\text{meeting}}^l$, even when $R^l_\text{in}$ or $R^l_\text{out}$ are low, indicates
    high transmissibility at location $l$ driven by non-resident visitors. \\[4pt]
\hline
$R_{\text{type}}^j(t)$
  & Type reproduction number for location(s) $j$; includes all infection return pathways through the network
  & Whether location(s) $j$ alone can sustain the epidemic, accounting for direct and
    indirect routes of transmission over all future generations; $R_{\text{type}}^j>1\Leftrightarrow\mathcal{R}>1$
  & If defined, identifies the \textbf{minimal set of locations} whose control is
    sufficient: control all $j$ with $R_{\text{type}}^j>1$ to bring $\mathcal{R}<1$.
    Only defined when the background sub-network is sub-critical. \\[4pt]
\hline
$\sigma(t)$
  & \textit{Reactivity}; maximum singular value of
    $\mathbf{R}(t)$
  & Potential for \emph{transient} amplification in a single generation,
    even when $\mathcal{R}<1$
  & 
  $\sigma > 1$ indicates worst-case one-generation amplification exceeds 1 over all possible seeding locations, and can cause transient growth, even if $\mathcal{R}<1$ signals asymptotic decay, thereby suggesting \textbf{false-stability zone} and \textbf{locations to monitor in the short term}.
     \\
\hline
$A_1(n)$
  & The $\ell^1$ norm amplification envelope with $A_1(1)$ being the \textit{first generation epidemicity} \cite{mari_sufficient_2025} that bounds $\mathcal{R}(t)$ from above 
  & Maximum number of total infections across all locations over $n$ generations from one seeding location
  & Identifies worst-case transient growth per location, thereby suggesting \textbf{locations to be monitored in the short term}, though transience may be caused by noise (e.g. low-incidence locations) \\
\hline
$\mathcal{E}(t) = X(1, t)$
  & Spatial \textit{risk-averse reproduction number}
  & Transmissibility of the network with individual locations weighted by their relative transmissibility
  & \textbf{Target control measures at locations} to control localised resurgences in infections and correct for false-stability signals if $\mathcal{E}(t)>1$ and $\mathcal{R}(t) < 1$. \\
\bottomrule
\end{tabular}
\caption{\textbf{Interpretation and operational guidance for spatial reproduction numbers and related quantities:} All quantities are derived from elements of the next-generation matrix $\mathbf{R}(t)$ and are defined instantaneously at time $t$. The operational use/guidance are illustrative, non-exhaustive examples of ways to use each quantity, and there are further operationally useful quantities (e.g. the instantaneous kernels and generation time distributions, proportion of infections generated locally versus elsewhere).}
\label{tab:R_operational}
\end{table}
\FloatBarrier

\subsection{Illustrative applications of our framework}
We now apply our generalisable framework to illustrate the operational information provided by the various outputs (Tables \ref{tab:framework_outputs} -- \ref{tab:R_operational}) for epidemics on different networks.

\subsubsection{Epidemic dynamics on the network}
\begin{figure}[t]
\centering
\includegraphics[scale = 0.55]{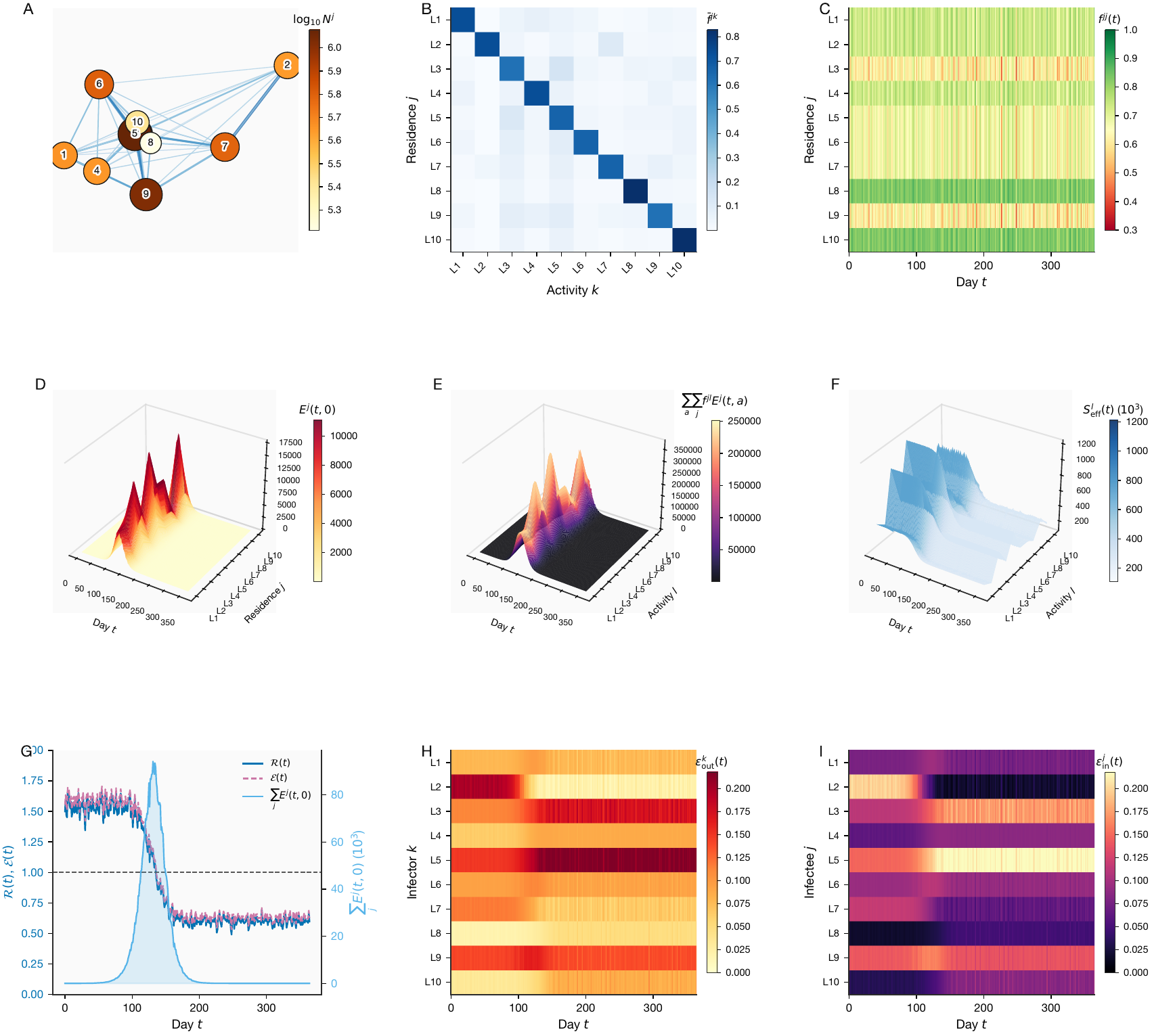}
\caption{\label{fig:02_overview.pdf} \footnotesize \textbf{Overall human mobility and epidemic dynamics across space and time:} \textbf{A)} Human mobility network in a dense-urban setting (main Scenario A) with ten nodes/locations (circles) connected by edges denoting the time-averaged probability $\overline{f^{jk}}$ that residents of location $j$ are in location $k$, with row sums equal to 1. Edge widths and colour denote $\overline{f^{jk}}$ and node size and colours denote population $N^j$ for each location $j$. \textbf{B)} The time-averaged probability $\overline{f^{jk}}$ that residents of location $j$ are in location $k$, with row sums equal to 1. \textbf{C)} Probability $f^{jj}(t)$ at each time $t$ that residents of location $j$ are in their home residence location $j$. \textbf{D)} Incidence of new infections across locations over time. \textbf{E)} Effective infected population in location $l$ at time $t$. \textbf{F)} Effective susceptible population in location $l$ at time $t$. \textbf{G)} Network-level reproduction number $\mathcal{R}(t)$, spatial risk-averse reproduction number $\mathcal{E}(t)$, and overall incidence over time. \textbf{H)} Infector elasticity $\epsilon_{\text{out}}^{k}(t)$ for resident locations $k$ of $\mathcal{R}(t)$ with respect to between-location $R^{kj}$ summed over infectee residence locations $j$ such that it summarises outward infection contributions of location $k$ to network-level epidemic growth. \textbf{I)} Infectee elasticity $\epsilon_{\text{in}}^{j}(t)$ for resident locations $j$ of $\mathcal{R}(t)$ with respect to between-location $R^{kj}$ summed over infector residence locations $k$ such that it summarises inward infection contributions of location $j$ to network-level epidemic growth. Note that $\epsilon_{\text{in}}^{j}(t)$ and $\epsilon_{\text{out}}^{j}(t)$ are equal because at the network level, every infection that leaves location $j$ must land somewhere, and every infection that lands at location $j$ must have arrived from somewhere. See Tables \ref{tab:framework_outputs} -- \ref{tab:R_operational} and main text for definitions, interpretations, and uses of our framework's quantities.}
\end{figure}

Focusing on the dense-urban setting (Scenario A from Section \ref{sec:main_text_implem_details}), the simulated movement patterns were largely dominated by within-location daily presence (Figure \ref{fig:02_overview.pdf} A -- B), with residents of locations 3 and 9 generally spending more time elsewhere and residents of locations 8 and 10 spending more time within their home location (Figure \ref{fig:02_overview.pdf} C). There was a largely synchronised epidemic across locations with a common incidence peak of approximately 131 days and an overall attack rate of 61.2\% (Figure \ref{fig:02_overview.pdf} D -- G). The network-level $\mathcal{R}(t)$ tracked the infection incidence curve and as an instantaneous measure, inherits the day-of-week movement patterns and variability, while the risk-averse $\mathcal{E}(t)$ followed a similar trajectory (Figure \ref{fig:02_overview.pdf} G). The effective susceptible and infected populations were concentrated in space and time (Figures \ref{fig:02_overview.pdf} F and \ref{fig:SI3_epi_params} E), with a distinct pattern of individuals concentrating in highly populated locations 3, 5, and 9 (Figure \ref{fig:SI0_population}), which also experienced the highest incidence in their residents (Figure \ref{fig:02_overview.pdf} B).

Early in the epidemic, as per the elasticities $\epsilon_{\text{out}}^{k}(t)$ and $\epsilon_{\text{in}}^{k}(t)$, $\mathcal{R}(t)$ was most sensitive to the secondary infections generated by and in infected residents of location 2 because this location experienced initial growth in infections fastest (Figure \ref{fig:02_overview.pdf} H -- I). As the epidemic evolved across space and time, $\mathcal{R}(t)$ became most sensitive to the secondary infections generated per infected residents of the highly populated and highly infected locations 3, 5, and 9 (Figures \ref{fig:02_overview.pdf} H -- I and \ref{fig:SI1_sensitivity}). Operationally, reducing within-location transmission, particularly involving residents of the most densely populated, dense/core nodes (locations 3, 5, and 9), was estimated to be the most efficient way to control epidemic growth on the network when at its peak incidence (Figure \ref{fig:SI1_sensitivity} E -- F).

\begin{figure}[t]
\centering
\includegraphics[scale = 0.43]{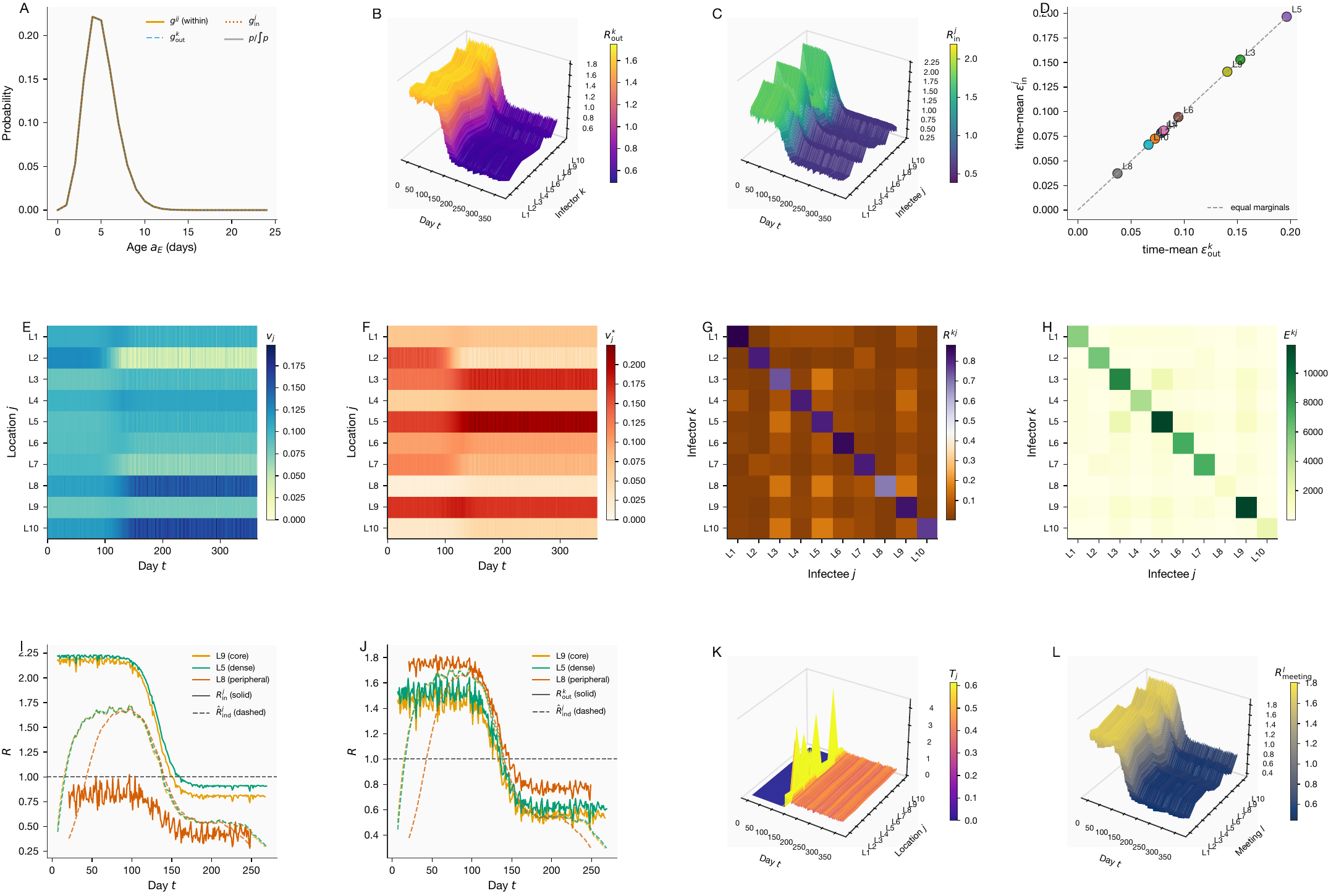}
\caption{\label{fig:03_taxonomy.pdf} \footnotesize \textbf{Taxonomy of reproduction numbers, generation time distributions, and related quantities:} \textbf{A)} Probability densities for different types of generation time distributions which overlap across space and time due to assumptions of infection age independence for contact rates and movement patterns and location-invariant infectiousness profiles. \textbf{B)} and \textbf{C)} are the outward and inward reproduction numbers respectively for each residence location over time. \textbf{D)} shows the average infector elasticity per location against their average infectee elasticity to quantify how important each location is to network epidemic growth in terms of contributing to onward infections and sinking inward infections. Note that $\epsilon_{\text{in}}^{j}(t)$ and $\epsilon_{\text{out}}^{j}(t)$ are equal because at the network level, every infection that leaves location $j$ must land somewhere, and every infection that lands at location $j$ must have arrived from somewhere. \textbf{E)} and \textbf{F)} are the left and right eigenvectors, $\mathbf{v}$ and $\mathbf{v^*}$, which reflect the long-term reproductive value of locations and long-term spatial distribution of new infections. \textbf{G)} and \textbf{H)} are the between-location reproduction numbers $R^{kj}$ and between-location incidence $E_{kj}$ respectively at the epidemic incidence peak (day 131). \textbf{F)} is the source-sink dynamics computed from  inward and outward reproduction numbers. \textbf{I)} and \textbf{J)} compare inward and outward reproduction numbers respectively to independent $R(t)$ estimates for three types of locations (hub, mid/suburb, and peripheral nodes). \textbf{K)} is the type reproduction number per location over time (when and where defined) and \textbf{L)} is the meeting location reproduction number per meeting location over time. See Tables \ref{tab:framework_outputs} -- \ref{tab:R_operational} for definitions, interpretations, and uses of our framework's quantities.}
\end{figure}

As we assumed that contact rates and movement patterns were independent of infection age and that infectiousness profiles were fixed across locations, the various types of generation time distributions (Table \ref{tab:framework_outputs}) overlapped across space and time (Figure \ref{fig:03_taxonomy.pdf} A). The largest values of $R_{\text{out}}^j(t)$ and $R_{\text{in}}^j(t)$ occurred at similar times for individual locations $j$ (Figure \ref{fig:03_taxonomy.pdf} B and C). These metrics capture inward and onward transmission vulnerability, albeit instantaneously, yet do not capture most effective location- or corridor-targeted interventions to control the network-level growth. The time-averaged infector/infectee elasticities suggest outsized reductions in network-level growth by controlling transmission involving residents of locations 3, 5, and 8 (Figure \ref{fig:03_taxonomy.pdf} D). The left eigenvectors $v_j$ over time, summarise the high reproductive value of peripheral locations 8 and 10, while the right eigenvectors $v_j^*$ show how the long-term distribution of infections concentrates in hub/dense locations (Figure \ref{fig:03_taxonomy.pdf} E -- F). The between-location reproduction numbers $R^{kj}$ and incidence $E_{kj}$ between locations at the epidemic incidence peak reveal the differential importance of human movement to incident infections across different locations (Figure \ref{fig:03_taxonomy.pdf} G and H), which further differs over time in many locations (Figure \ref{fig:04_spectral} E).

Due to the highly synchronous epidemic dynamics, the type reproduction number $R_{\text{type}}^j(t)$ is undefined for individual locations $j$ before the epidemic incidence peak (Figures \ref{fig:03_taxonomy.pdf} K, \ref{fig:type_surfaces} -- \ref{fig:type_heatmaps}), with the exceptions of hub/densely populated locations around the peak where targeted interventions would help to bring the epidemic growth under control. We recomputed the type reproduction number for groups of locations, where targeting interventions at all core/densely populated locations would bring the epidemic under control at certain epidemic stages (Figure \ref{fig:type_groups}).

\subsubsection{Comparing estimators of transmissibility}
To compare our framework's quantities with those obtained from using per-location incidence (i.e. the closed-population renewal equation, eq. \eqref{eq:Rt_closed_pop}), we computed values ${R}^\text{ind}_j(t)$ from independent, statistical estimators of reproduction numbers that are often used for a single, closed population (e.g. \citet{cori_new_2013} with a 7-day smoothing window and Gamma prior distribution):
\begin{align}
{R}^\text{ind}_j(t) \approx \frac{E^j(t, 0)}{\sum_{a_E} g^j(a_E)E^j(t-a_E, 0)},
\end{align}
that is the ratio of new local infections to the expected local infective pressure, which thereby assumes all secondary infections in residents of location $j$ are generated by primary infected residents of location $j$. 

 $R_{\text{out}}^j(t)$ and $R_{\text{in}}^j(t)$ are similar to, but differ fundamentally from, independent per-location estimators of $R(t)$. The independent estimates $R_{\text{ind}}^j(t)$ cannot describe how residents of location $j$ acquire or generate new infections at different rates based on their daily movement patterns and interactions with susceptible/infected individuals resident in different locations. Likewise, $R_{\text{in}}^j(t)$  and $R_{\text{out}}^j(t)$ are limited by not having the threshold properties for determining long-term epidemic growth/decay. The goal of $R_{\text{in}}^j(t)$ is similar to per-location, independent $R_{\text{ind}}^j(t)$, that is to track infection sink dynamics/vulnerability in residents of location $j$ (resulting from a counterfactual primary infected individual placed in each location the network). The operational use of $R_{\text{ind}}^j(t)$ is probably closer to outward $R_{\text{out}}^j(t)$ as $R_{\text{out}}^j(t)$ tracks whether an infected resident of location $j$ is generating more than one infection in the network as a whole. Independent estimators often provided different estimates compared to inward and outward $R(t)$ (Figures \ref{fig:03_taxonomy.pdf} G and \ref{fig:SI_R_comparison}). This is because the independent estimators do not account for the import and export infection dynamics in secondary infections, instead assuming all residents generate infections locally (Figures \ref{fig:03_taxonomy.pdf} G and \ref{fig:SI_R_comparison}), therefore drastically over-estimating the true local transmissibility $R^{jj}(t)$. The differences with $R_{\text{in}}^j(t)$ and $R_{\text{out}}^j(t)$ were more pronounced for more extreme movement locations, such as hub/central and peripheral/suburb locations (e.g. location 2). ${R}^\text{ind}_j$ cannot measure how infected residents of a location $j$ generate secondary infections in residents of other locations (as both inward and outward movements are not captured for location $j$), and is therefore not a true description of the onward transmissibility of infected residents of location $j$. The relative difference between ${R}^\text{ind}_j$ and $R^\text{in}_j$ was generally greater than the relative difference between ${R}^\text{ind}_j$ and $R^\text{out}_j$, likely in part because $R^\text{in}_j$ measures a different counterfactual of one infected individual in each location generating secondary infections in residents of location $j$, whereas $R^\text{out}_j$ generates the transmissibility of the average infected resident of location $j$ (similar to ${R}^\text{ind}_j$ but with additional movement structures encoded).

We also explored bias from other popular approaches for quantifying network-level epidemic dynamics that neglect spatial structure and/or human movement dynamics. We first compared estimates from two basic approaches that compute $R(t)$ for the network by weighting the incidence from each location: i) from using the aggregate incidence across locations to compute $R_{naive}$ (Figures \ref{fig:aggregate_R_bias} -- \ref{fig:aggregate_naive_vs_Et}) and ii) from using the population-weighted incidence across locations to compute $R_{pw}$, where $R_{naive}$ and $R_{pw}$ generally over-estimated and under-estimated epidemic growth of the network $\mathcal{R}(t)$ (and $\mathcal{E}(t)$) when $\mathcal{R}(t) > 1$ and $\mathcal{R}(t) < 1$ respectively.

\begin{figure}[t]
\centering
\includegraphics[scale = 0.88]{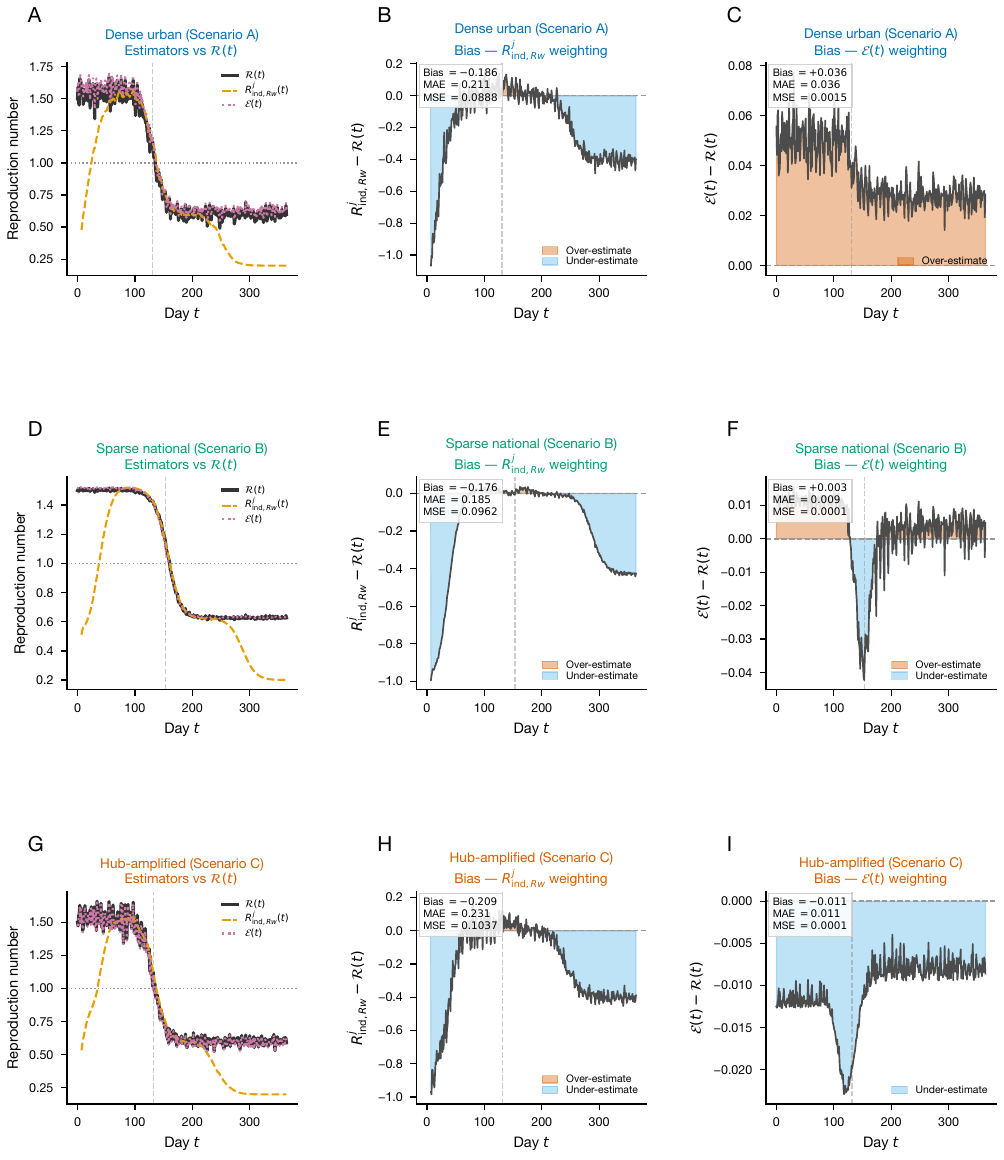}
\caption{\label{fig:main_bias_combined} \footnotesize \textbf{Bias in different estimators of network-level epidemic transmissibility:} We compare our framework's network-level reproduction number $\mathcal{R}(t)$ and risk-averse reproduction number $\mathcal{E}(t)$ alongside an approach that weights independent, per-location estimates $R_{\text{ind}}^j(t)$ by their relative transmissibility (i.e. $\frac{R_{\text{ind}}^j(t)}{\sum_{k}R_{\text{ind}}^k(t)}$). Rows denote scenarios A, B, and C, described in Materials and Methods and Results sections. The largest biases and (absolute) errors arise from the independent approach, generally under-estimating at early and late stages and over-estimating at peak stages, as the failure to capture network structure wrongly attributes the source and onward potential of new infections (panels B, E, and F). The values of $\mathcal{R}(t)$ and risk-averse reproduction number $\mathcal{E}(t)$ are generally much closer, yet $\mathcal{E}(t)$ can take larger values by weighting infection resurgences more strongly (via weighting of $R_{\text{out}}^j(t)$). See Tables \ref{tab:framework_outputs} -- \ref{tab:R_operational} for definitions, interpretations, and uses of our framework's quantities. See Figures \ref{fig:aggregate_R_bias} -- \ref{fig:SI_R_comparison} for comparing the bias from using other schemes to estimate network-level epidemic transmissibility such as population-weighted, incidence-weighted, and arithmetic-mean-weighted schemes.}
\end{figure}

Another recent approach \cite{parag_r_2024} is to instead weight the independent estimates of the reproduction number ${R}_{\text{ind}}^j(t)$ obtained from using per-location incidence, where it is generally best to weight ${R}_{\text{ind}}^j(t)$ by the corresponding relative transmissibility $\frac{{R}_{\text{ind}}^j(t)}{\sum_j {R}_{\text{ind}}^j(t)}$. Denoting these network-level estimates as ${R}_{\text{ind, RW}}(t)$, these estimates generally over-estimated and under-estimated epidemic growth of the network, $\mathcal{R}(t)$ and $\mathcal{E}(t)$, when $\mathcal{R}(t) > 1$ and $\mathcal{R}(t) < 1$ respectively (Figure \ref{fig:main_bias_combined}). We repeated this procedure using arithmetic mean weighting and incidence weighting of ${R}_{\text{ind}}^j(t)$ (Figures \ref{fig:naive_ind_am_vs_Et}), where we found similarly large biased estimates of $\mathcal{R}(t)$. The wide discrepancies from all these approaches suggest the value of human mobility data for avoiding false stability signals from estimates that simply weight or average independent estimates of transmissibility. 

If mobility data are available, an alternative is to weight $R_{\text{out}}^j(t)$ by estimates of the relative transmissibility that incorporate human movement (i.e. relative $R_{\text{out}}^j(t)$), yielding $\mathcal{E}(t)$. $\mathcal{E}(t)$ is much more closely aligned with $\mathcal{R}(t)$ and together they provide complementary information via slightly higher values of $\mathcal{E}(t)$ to account for potential resurgences in locations (Figure \ref{fig:main_bias_combined} C and F) and via slightly lower values of $\mathcal{E}(t)$ in extreme settings where network-amplified growth is responsible for epidemic growth (Figure \ref{fig:main_bias_combined} G -- I).

\subsubsection{Transience measures and time-varying mobility}
\begin{figure}[t]
\centering
\includegraphics[scale = 0.88]{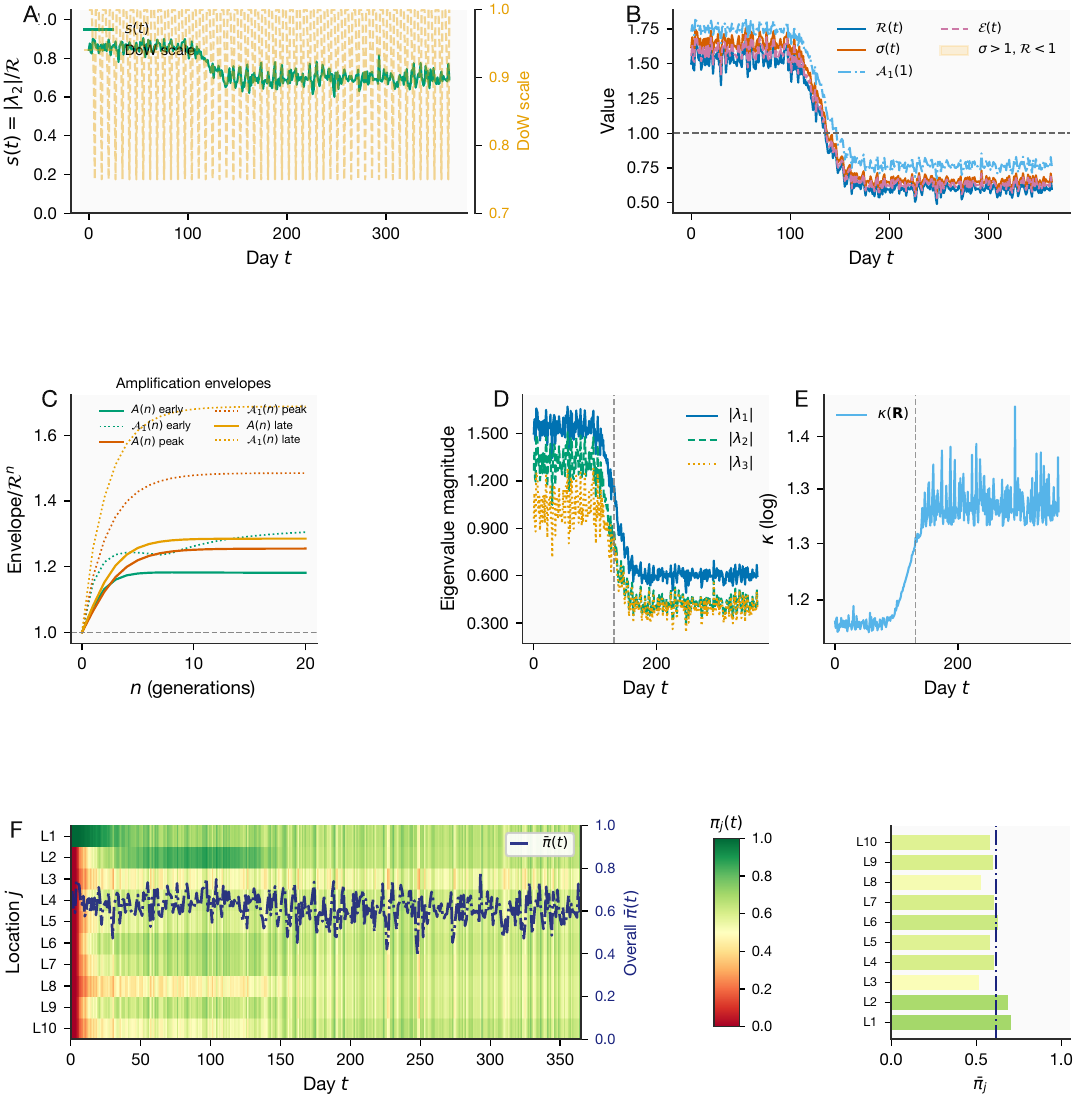}
\caption{\label{fig:04_spectral} \footnotesize \textbf{Transience and importation analyses across space and time:} \textbf{A)} Mixing time ratio $s(t)$ and day-of-week (DoW) scaling used for human movement simulation. \textbf{B)} Reactivity $\sigma(t)$ and network-level $\mathcal{R}(t)$ over time. \textbf{C)} Amplification envelope using the $\ell^1$ or $\ell^2$ norm, $A(n)$ or $A_1(n)$ respectively, divided by $\mathcal{R}^n$ for $n$ generations at early, peak, and late stages of the outbreak. \textbf{D)} Magnitudes of the top three eigenvalues of $\mathbf{R}(t)$ over time with a larger spectral gap between $\mathcal{R}(t)$ and $|\lambda_2(t)|$ indicating greater transience. \textbf{E)} Eigenvalue condition number for the dominant eigenvalue $\mathcal{R}(t)$ of $\mathbf{R}(t)$ over time. \textbf{F)} Proportion of incident infections in residents of location $j$ generated by infected individuals also resident in location $j$ over time (left) and the corresponding overall proportions (right). Lower (higher) proportion locations generally corresponded to locations where residents spent more (less) time outside of their home residence location. See Tables \ref{tab:framework_outputs} -- \ref{tab:R_operational} for definitions, interpretations, and uses of our framework's quantities.}
\end{figure}

The mixing ratio $s(t)$ generally tracked the simulated day-of-week mobility trends, while the dominance of within-location movement patterns meant that reactivity, the spectral gap, and differences between $\mathcal{R}(t)^n$ and the amplification envelope $A(n)$ were not pronounced over time (Figure \ref{fig:04_spectral} A -- D). The close correspondence between the spatial risk-averse reproduction number $\mathcal{E}(t)$ and network-level $\mathcal{R}(t)$ meant that they provided the same stability signals. On the other hand, $A_1(1) = \max_{k}R_{\text{out}}^k(t)$ often indicates larger short-term transient amplifications, even when $\mathcal{R}(t) < 1$, $\mathcal{E}(t) < 1$, and $\sigma(t) <1$, due to its focus on the worst-case amplification from a single location. $A_1(1)$ should therefore be interpreted with more caution as this can arise due to noisy signals from low-population, low-incidence settings. The amplification envelope is similarly larger for all $n$ generations when using the $\ell^1$ norm (compared to $\ell^2$ norm) for the same reason of a very localised focus (Figure \ref{fig:04_spectral} C).

There was a general trend of a sustained, relative importance of infection importations over the course of the epidemic. Residents of locations 3 and 8 most frequently acquired infection from residents of other locations (Figures \ref{fig:04_spectral} E), while residents of locations 1 and 2 more frequently acquired infections from residents of their home location. 

Due to the marked influence of the day-of-week mobility (e.g. Figure \ref{fig:04_spectral} A--B and D--E) on various outputs, we performed a simple reanalysis of a simulation where we removed the day-of-week scaling (i.e. a static mobility graph with time-invariant $f^{jk}$, Figure \ref{fig:SI_static_02_overview}). The static reanalysis had the expected effect of reducing oscillations in $\mathcal{R}(t)$ (Figure \ref{fig:SI_static_02_overview} C) as populations in different locations were stable over time, while the attack rate was lower (58.0\%) and epidemic incidence peak occurred later (Figure \ref{fig:SI_static_03_taxonomy} D -- E). Likewise, smoother values of $R_{\text{out}}^j(t)$ and $R_{\text{in}}^j(t)$ capture more general trends, including in the over- and under-estimation from independent estimators (Figure \ref{fig:SI_static_03_taxonomy} B -- C and G -- H). We can also see clearer, smoother trends in the mixing ratio, eigenvalues, condition number, and imported-versus-local infections (Figure \ref{fig:SI_static_04_spectral}). The transience plots also more clearly show the generally small gap between $\mathcal{E}(t)$ and $\mathcal{R}(t)$ and between $\sigma(t)$ and $\mathcal{R}(t)$ and the larger gap between $A_1(1)$ and $\mathcal{R}(t)$ (Figure \ref{fig:SI_static_04_spectral} B).

\subsubsection{Counterfactual settings}
We considered several counterfactual settings, including a sparse national scenario (Scenario B) which was dominated by within-location movement (Figure \ref{fig:05_comparison} A) and within-location transmission (Figure \ref{fig:05_comparison} B, C, and F). Imported infections in the sparse-national setting accounted for smaller proportions (compared to the dense urban setting) and were of greatest importance for infection seeding at the start and end of an outbreak that occurred over a longer time period (Figure \ref{fig:05_comparison} E -- F and Figure \ref{fig_SI8_3d_earlypeak}).

We also amplified the role of a single hub in human movement for a different urban setting (Scenario C) by creating a high-population, highly attractive hub, thereby creating a human mobility matrix with heightened non-normality (Figures \ref{fig:scenario_C_transience} -- \ref{fig_counterfactual_nonnormal}). The hub-amplified setting resulted in a more peaked epidemic which had a greater importance of between-location (i.e. imported) infections and slower mixing (Figure \ref{fig_counterfactual_nonnormal} j). The amplified role of the hub, (denoted as location 1) was clear as its residents experienced the largest number of infections (Figure \ref{fig_counterfactual_nonnormal}) and had a strong influence on other locations (Figure \ref{fig_counterfactual_nonnormal}). The highly asymmetric movement patterns meant that reactivity $\sigma(t)$ (the largest singular value of $\mathbf{R}(t)$) far exceeded $\mathcal{R}(t)$ (the dominant eigenvalue of $\mathbf{R}(t)$), that is the one-step directional amplification far exceeded the asymptotic network-level growth. Operationally, this could lead to a possible false-action and/or false-stability zone where transience means that $\sigma(t) \gg \mathcal{R}(t)$ at several time points and $\sigma(t) >1$ for 31 days even when $\mathcal{R}(t) < 1$ (Figures \ref{fig:scenario_C_transience} -- \ref{fig_counterfactual_nonnormal}). The network-level reactivity $\sigma(t)$ also exceeded $A_1(1)$ because amplification from seeding in multiple locations, in the worst case, here exceeded the amplification from seeding infections in the location with largest $R_{\text{out}}^k(t)$ alone, again highlighting that network-level growth/decay is very different to local epidemic growth/decay. Values of $\sigma(t)$ far exceeded $\mathcal{E}(t)$. This is because $\mathcal{E}(t)$ focuses only on columns of the NGM $\mathbf{R}(t)$ via the weighted average $R_{\text{out}}^k(t)$ and cannot capture the correlations in rows and columns of $\mathbf{R}(t)$ that correspond to spatially structured transmission patterns (i.e. spectral feedback driving network amplification). So, there were times with $\sigma(t) > 1$ even when $\mathcal{R}(t) < 1$, $A_1(1) <1$, and $\mathcal{E}(t) < 1$, which means that despite there being no single location infection resurgence, the network being asymptotically stable, and single-location amplifications being all less than one, a correlated multi-location seeding/importation event can still be transiently amplified in one generation.

In other sensitivity analyses, higher $R_0$ and shorter infectiousness profiles also produced sharper, faster epidemic trajectories, while initial infection seeding in space and different between-location versus within-location transmission rates had little effect on the network's incidence and $\mathcal{R}(t)$ trajectories (Figure \ref{fig:SI7_sensitivity}).

\section{Discussion}
We have introduced a new mechanistic framework to enhance the targeting of interventions in real time across space and time. The framework addresses a fundamental limitation of the most common ways to estimate and use $R(t)$ in moving populations as we explicitly model continuous movements of infected and susceptible individuals across locations. We redefine measures of epidemic spread in spatially structured interacting populations and inform how epidemics could be brought under control by targeted interventions of varying types, durations, and scales. Unlike existing approaches for $R(t)$ estimation, our framework precisely describes how infections occur across space and time by accounting for the wide array of within-day infection opportunities, that is by the time that individuals spend in their home residence location and other locations.

Our framework introduces many control-relevant quantities (Tables \ref{tab:framework_outputs} -- \ref{tab:R_operational}) that are not obtainable from renewal equations for single-location, closed populations. This poses the question; which reproduction number or control indicator should be used in practice to inform epidemic control strategies? We argue that this depends on the policy question and local context and no one scalar measure can capture all aspects of epidemic control for a network, but together, our family of indicators can answer these questions. These indicators provide network-level, location-level, mobility-corridor-level, and venue-level diagnoses: i) \textit{between-location} and within-location reproduction numbers and elasticity metrics can inform temporary, targeted reductions for specific between-location mobility corridors/pathways, ii) \textit{inward} reproduction numbers and infectee elasticity can inform localised interventions/protection (e.g. border restrictions) to reduce inward flow for importer/destination locations that are vulnerable to transmission, iii) \textit{outward} reproduction numbers and infector elasticity can inform targeted interventions (e.g. stay-at-home or movement restrictions for residents) for the outward flow of residents from exporter/source locations that are contributing, on average, more than one secondary infection, iv) \textit{meeting}-location reproduction numbers can inform targeted interventions for locations where growing totals of infections can occur, v) \textit{network}-level reproduction numbers and transience measures can most rigorously inform necessary interventions targeted at driving the long-term stability and control of the network as a whole, and finally vi) the \textit{type}-reproduction number (if defined) can quantify locations that sustain transmission (through inter-location feedback) and inform interventions targeted at these locations with self-sustaining transmission lineages.

These quantities provide a principled targeting methodology as they can inform both the timing and magnitude of interventions for controlling different aspects of the outbreak, and isolate different locations and pathways for which interventions could have outsized impacts in reducing pathogen transmission. The various metrics (Tables \ref{tab:framework_outputs} -- \ref{tab:R_operational}) should be applied and interpreted collectively, because for example, reducing specific outward or between-location flows could increase within-location flows of local residents, and within-location transmission vulnerability should therefore be considered. The generation time distribution is also important, not only for reliably estimating $R(t)$, but also for knowing \textit{when} and \textit{how fast} infections will occur (i.e. time-to-control, not just magnitude), the type of intervention that might be suitable, and how long that intervention might need to be sustained. 


Recognising that such instantaneous measures can be misleading \cite{neubert_alternatives_1997, mari_sufficient_2025, parag_r_2024} as they depend on current movement and infection dynamics, we encourage the application of the various transience measures (Section \ref{sec:asymptotic_transient}). These capture how short-term growth in the incidence of new infections can be generated from importation/seeding events from individual and multiple locations and suggest where short-term interventions may be needed to control localised resurgences in infections, even when $\mathcal{R}(t)$ indicates asymptotic equilibrium decay. Overall, this complementary approach is crucial as, for example, $\mathcal{R}(t)$ can miss localised infection resurgence in low-centrality locations, whereas $\mathcal{E}(t)$ can miss network-mediated amplification that only emerges through multi-step spectral feedback on the network. Together, $\mathcal{R}(t)$ and $\mathcal{E}(t)$ can answer relevant stability questions for real-time decision-making in an epidemic on a network.

Several statistical methods have been developed to estimate $R(t)$ across space \cite{roy_incorporating_2025, kim_mobility-adjusted_2025, nouvellet_rtglm_2025}. These per-location $R(t)$ estimators differ from our framework above as the estimates often do not have a mechanistic meaning due to either smoothing across space or assuming that inward and outward transmission occurs according to the same reproduction process (e.g. identical rates). For example, these approaches can often allow residents of location $k$ to generate infections in residents of a location $j$, yet the estimate of $R(t)$ for location $j$ is used for the transmission/reproduction process from residents of other locations $k$ into residents of location $j$. Likewise, many metapopulation models exist (e.g. \cite{sattenspiel_structured_1995,colizza_reactiondiffusion_2007, soriano-panos_spreading_2018,watts_multiscale_2005, kiss_mathematics_2017, keeling_networks_2005, cosner_effects_2009, zeng_dengue_2023}), but these are less often linked to instantaneous and transient location-specific measures of human-to-human transmissibility that account for within-day human mobility. We are also not the first to illustrate a network-level $\mathcal{R}(t)$ in a spatially structured population (see \cite{birello_estimates_2024} for rigorous quantification of spatial biases in regular, real-time estimators of $R(t)$), and there are spatially explicit reproduction numbers \cite{trevisin_spatially_2023} and a well-established literature on epidemic heterogeneity in mathematical epidemiology \cite{heesterbeek_type-reproduction_2007, inaba_state-reproduction_2008, bouros_time-dependent_2025, roberts_early_2011, pellis_estimation_2022}. Recent work has also connected $R_0$ to the exponential growth of complex networks and quantified early-outbreak dynamics under static networks \cite{cure_exponential_2025}. Other applications of compartmental models \cite{kim_evaluating_2026, jorge_estimating_2022} use commuting-volume mobility (rather than within-day fractional time) to produce per-location inward and between-location reproduction numbers which do not have the threshold property for long-term epidemic growth. We extend the research above as we derive equations and control indicators that mechanistically describe \textit{how} the movement of interacting humans (and possibly vectors) within a day in different locations results in network-level and location-level epidemic growth (Tables \ref{tab:framework_outputs} -- \ref{tab:R_operational}). Starting from human mobility and infection data, our framework can distinguish where infections come from, where they are generated, where they are felt, whether they can feed back into sustained transmission, and how they can be best controlled.

Our new framework also provides a causally explicit platform to perform causal inference using the gold-standard of counterfactual predictions \cite{pearl_causality_2022, runge_inferring_2019} (e.g. estimating effects of varying stay durations, contact rates, types of movement etc.). The quantities of our framework (Tables \ref{tab:framework_outputs} -- \ref{tab:R_operational}) enable the effects of specific interventions that target specific reproduction numbers to be predicted more accurately, enabling the effects of different interventions to be compared. There have been many modelling analyses where the effects of mobility and/or non-pharmaceutical interventions on transmission are inferred/deduced using association or correlation with $R(t)$ \cite{jewell_use_2022,nouvellet_reduction_2021, bracher_evaluating_2021, flaxman_estimating_2020}. Such approaches can neglect the true causal mechanisms by which human movement shapes $R(t)$ in spatiotemporally evolving epidemics and more generally, estimates are subject to unobserved/unmodelled confounding, e.g. as multiple changes happen concurrently in reality, leading to incorrect attribution of the change in $R(t)$ to the single modelled event. This is not just a statistical issue, but also a trust issue for modelling and public health policy as inferred effects often inform decision-makers faced with difficult and uncertain decisions \cite{hadley_how_2024, anderson_reproduction_2020}. We suggest that by capturing the causal mechanisms of the effects of human movement on $R(t)$, our framework provides a platform to elevate from a correlation- or- association-based study to a causal one. 

There are many future directions for our framework. The unveiled biasing effects of human movement is especially relevant for the task of inferring $R(t)$ and real-time analyses for informing decision-making during an epidemic \cite{anderson_reproduction_2020, gostic_practical_2020, steyn_primer_2025, steyn_robust_2025}. Uncertainty will inevitably arise from imperfections of epidemiological, biological, demographic, environmental, and human mobility data, and should ideally be propagated through the modelling process. Inference for our new framework could follow this workflow. First, conditional on having case incidence data (that matches the spatial resolution of the network) and biological data (e.g. infectiousness profiles), in its simplest form our framework requires only one additional data stream compared to popular closed-population methods, that is high-quality, relevant data/estimates for human mobility that capture within- and across-day movements of individuals. Requiring such additional input data is the key limitation of our framework for inference. We have illustrated the potentially misleading results from estimating $R(t)$ for locations and for the network from independent, closed-population renewal equations, yet to incorporate our framework alongside existing inference methods, we will need to address identifiability issues. For example, different combinations of $\lambda_E^{kl}(t, a_E)$ and $f^{kl}(t, a_E)$ could produce the same incidence patterns, which means that careful, setting-specific prior regularisation of key parameters (e.g. biological infectiousness or contact rates) may be important. Data availability has long been an issue for accurate, representative human mobility data and this will be amplified in resource-limited settings where the burden of infectious diseases is highest yet human movement patterns are poorly observed \cite{oliver_mobile_2020,wardle_gaps_2023, kostandova_improving_2025, lessani_human_2024}. These challenges motivate high-quality contact tracing data (e.g. the COVID-19 mobile application of the UK National Health Service \citealp{ferretti_quantifying_2020, ferretti_digital_2024}), transfer learning (e.g. to borrow data from other settings using a local calibration model \cite{kraemer_utilizing_2019, kraemer_artificial_2025}), and the next generation of human mobility models \cite{kostandova_improving_2025, cabanas-tirapu_human_2025}. Conditional on addressing identifiability, these heterogeneous data could be embedded into our mechanistic formula alongside existing joint inference techniques (e.g. Bayesian state-space models, akin to existing software such as EpiFilter or EpiNow2 \cite{abbott_estimating_2020, parag_improved_2021}) that correctly quantify the various sources of uncertainty \cite{steyn_primer_2025}. Of course, there will be an additional complexity of multivariate latent states across space and time, particularly for vector-borne diseases, that will require careful considerations of scalability, stability, and accuracy. All inference approaches will also face the same challenges that existing $R(t)$ estimators face, including noisy, delayed, truncated, and generally incomplete observed data \cite{gostic_practical_2020, park_estimating_2024, steyn_primer_2025, bajaj_renewal-equation_2025}, and carefully developing the observational model that accounts for these imperfections is crucial for the success of the latent mechanistic model. 

We largely discussed mobility-focused interventions yet the range of available interventions may be wider or narrower for different settings and different pathogens, and their success will depend on biological, social and public health factors that ultimately determine the controllability of an outbreak \cite{thompson_infectious_2026, fraser_factors_2004}. Recent work on epidemic control has shown the importance of controlling minor outbreaks \cite{kucharski_controlling_2024}, the differential impacts of noisy surveillance data on decisions for control strategies at different epidemic stages \cite{parag_asymmetric_2025}, the limitations of $R(t)$ and $r(t)$ when policies are implemented with delay 
\cite{parag_how_2024}, and the utility of dynamic optimal control algorithms for noisy, imperfect data \cite{beregi_optimal_2025}. We therefore recommend blending our framework with existing epidemic control theory, more extensive perturbation analyses \cite{neubert_alternatives_1997, alshanskiy_model_2018}, additional network analyses (including alternative network models) \cite{liu_controllability_2011,pastor-satorras_epidemic_2015}, further type and target reproduction numbers and related control strategies \cite{heesterbeek_type-reproduction_2007, inaba_new_2012}, and importantly the local context \cite{eggo_importance_2021} to better inform the control of complex epidemic dynamics across space and time.

The transient measures defined in Section \ref{sec:asymptotic_transient} apply to any non-negative matrix produced by our framework or other structured models of epidemic dynamics, not only the residence-location NGM $\mathbf{R}(t)$. For example, we could define a meeting-location reproduction number matrix $\mathbf{R}_{\text{meeting}}(t)$ to capture how infections propagate across meeting locations and apply transience measures to study which locations could experience transient infection surges. Likewise, applying transience measures to to the background sub-network matrices $\mathbf{R}_{JJ}(t)$ could quantify how long transient epidemicity could persist after targeted control of location $j$, thereby complementing $R^j_\text{type(}t)$ which measures asymptotic control of location-j lineage but not the short-term transience that could impact intervention efficacy.


Our framework's suitability is limited by the deterministic assumption in the mechanism, thereby relying on the assumption of a sufficiently large population in each location for deterministic outbreak dynamics to be a valid assumption. We mentioned how uncertainty from the observation process and human mobility data might impact conclusions. Stochasticity could also be embedded in the latent infection process using stochastic PDEs \cite{ponosov_stochastic_2020} or time-varying branching processes
\cite{pakkanen_unifying_2023}, and potentially compared to probabilistic, agent-based models (e.g. OpenABM \cite{hinch_openabm-covid19agent-based_2021}) that track the movements and infection exposures of individuals probabilistically.

Our applications are limited by the real-time availability of human movement data that captures the time spent by individuals in different locations over time. In this study, we did not focus heavily on modelling real-world mobility patterns of cities. Instead, we illustrated in diverse mobility networks that if we had access to such data, we have a framework to provide a wide range of new outputs that inform targeted and timely actions for decision-makers. Future work would be worthwhile to capture when and where mobility information is most valuable, perhaps focusing on ill-specified or unrepresentative mobility data and their impacts on reliability of outputs (e.g. via further sensitivity metrics). Carefully tested mobility approximations may be valuable here, possibly using informed assumptions and multiple mobility data streams (e.g. from censuses or public transport operators) on commuter stay durations, mobile phone ownership and usage while travelling, and/or within-day and within-location movement activity \cite{kostandova_comparing_2025, kostandova_improving_2025}. To address false stability signals from fluctuating human movement (e.g. day-of-week effects), it may be important to define a periodic NGM, use Floquet analysis, or implement other periodicity analyses \cite{inaba_new_2012, mitchell_comparison_2017}). While not modelled here, our framework allows for long-term/long-distance migration, which may be crucial to consider for seeding events or environmentally driven displacement. More generally, understanding when model approximations and more readily available origin-flow data (which are Eulerian rather than Lagrangian like mobile phone data) succeed and fail will be vital to quantify the setting- and task-specific value of information of making often already collected mobility data available. Different mobility patterns will likely give rise to different epidemic patterns across space and time \cite{balcan_multiscale_2009, mills_time-_2025, kostandova_comparing_2025}, and our framework could then integrate heterogeneous data streams to suggest the different types, durations, and strengths of interventions necessary for reducing an epidemic's social, economic, and health impacts.

To conclude, our framework redefines reproduction numbers into mobility-informed, mathematically-derived measures of when, where, and why interventions are necessary to control epidemics. We hope that these different multi-scale measures of epidemic spread will be useful to inform targeted public health measures during future outbreaks of a range of pathogens.

\subsection*{Author contributions}
C.M.\ conceived the idea and designed the methodology for the project. All authors approved the final submitted drafts. For the purpose of Open Access, the author has applied a CC BY public copyright licence to any Author Accepted Manuscript version arising from this submission. 
\subsection*{Acknowledgements}

\subsection*{Conflict of interests statement}
The authors declare no competing interests.
\subsection*{Data and code availability}
 All code and data used in this article are available at \url{https://github.com/cathalmills/mobility_Rt}.

\bibliographystyle{unsrtnat}
\bibliography{references}


\appendix

\newpage

\section*{Appendix}
\renewcommand{\thefigure}{SI~\arabic{figure}}
\setcounter{figure}{0}
\renewcommand{\thetable}{SI~\arabic{table}}
\setcounter{table}{0}

\begin{appendix}
\section{Further implementation details} \label{sec:implem_details}
As human movement has a more sustained importance in urban environments, we primarily focused on simulating human movement for densely populated megacity. For simplicity and visualisation, we assumed ten nodes/locations, consisting of two highly populated core/hub nodes, three densely populated nodes, three suburban nodes, and two low-population peripheral nodes, each of which had resident commuting fractions of $c_j$ equal to 40\%, 35\%, 28\% and 18\% respectively for each node $j$'s type. We then allowed for these residents to most often visit their adjacent/closest nodes but with population attractiveness.  

Specifically, we placed nodes using polar coordinates $(r_i, \theta_i)$ about a common
centroid. Radii $r_i$ were drawn from $\mathrm{Exponential}(\lambda_r)$ with $\lambda_r=8$ km
(dense-urban) or $\lambda_r=120$ km (coarse-national) and angles $\theta_i\sim\mathrm{Uniform}[0,2\pi)$. The type of each node determines its base population ${N^j}_{\mathrm{base}}$ and node resident populations $N^j$ follow $N^j={N^j}_{\mathrm{base}}\exp(\epsilon^j)$ with
$\epsilon^j\!\sim\!\mathcal{N}(0,\sigma_\epsilon^2)$,
$\sigma_\epsilon=0.25$ (dense-urban) or $0.20$ (coarse-national). 

To compute the probability $f^{jk}(t)$ that a resident of location $j$ is in location $k$ at time $t$, we first set up the time-invariant base matrix that encodes fraction of time at home (diagonal) and distance–population-weighted off-diagonal flows. We assumed $$\bar{f}_{jj} = 1 - c_j,$$ and for location $k \neq j$, we computed the unnormalised weight as:
$$w_{jk} = e^{-d_{jk}/\delta} \times \sqrt{\frac{N_k}{\bar{N}}},$$
where $d_{jk}$ is the Euclidean distance (km) between nodes $j$ and $k$, $\delta$ is the distance decay scale (7km for the dense-urban and 200km for the coarse-national), and $\sqrt{N_k / \bar{N}}$ is the square-root population attractiveness term ($\bar{N}$ is the mean population) that favours larger population destinations \cite{simini_universal_2012}, creating a compromise between the radiation model and gravity models to avoid over-concentration at the largest population hub (as in gravity model) while preserving realistic flow gradients. 

Off-diagonal entries are obtained by row-normalising and rescaling to the
commuting fraction:
$$\bar{f}_{jk} = c_j \cdot \frac{w_{jk}}{\sum_{k' \neq j} w_{jk'}}, \quad k \neq j,$$
ensuring row-stochasticity, $\sum_k \bar{f}_{jk} = 1$.

To create time-varying mobility networks, we applied to $\bar{f}_{jk}$ i) day-of-week scaling $[s_{\text{Mon}}, s_{\text{Tue}}, s_{\text{Wed}}, s_{\text{Thu}},
   s_{\text{Fri}}, s_{\text{Sat}}, s_{\text{Sun}}]
 = [1.00,\;1.00,\;1.00,\;1.00,\;0.95,\;0.90,\;0.80].$ and ii) day-to-day idiosyncratic variability $\xi_t \sim \operatorname{LogNormal}(0, \sigma_\xi)$. These capture respectively the regular weekly commuting patterns and the irregular day-to-day movement variability (e.g. due to weather or events), and we combined them to form time-varying scaling factor $s(t) = s_{\text{DoW}(t)} \cdot \xi_t$.
The away fraction of each origin location $j$ is scaled by $s(t)$, capped at 0.95,
and the home fraction set to $f_{jj}(t) = 1 - \sum_{k \neq j} f_{jk}(t)$
before renormalising each row to sum to 1. For simplicity, we simulated movement of individuals irrespective of infection status.

\newpage

\section{Further derivation details for renewal equations}\label{sec:renewal_eqns_derivations}
We can write the solution for the infected human population as follows:
\begin{equation}
    E^j(t,a_E) =
    \begin{cases}
    E(t-a_E,0), & \text{if $a_E\leq t$},\\
    p_E^{j, 0}(a_E-t), & \text{if $a_E> t$}.
  \end{cases}
\end{equation}
Then, focusing on long-term behaviour such that $t > a_E$ ($t\rightarrow \infty$ and $p_E^{j, 0}(a_E-t) = 0$) , we can rewrite the boundary condition eq. \eqref{eq:bc} as:
\begin{align}
E^j(t, 0) &= \sum_{l \in \mathcal{N}} f^{jl}(t, S) \cdot S^j(t) \cdot \sum_{k \in \mathcal{N}} f^{kl}(t, E) \int_{0}^{t}\lambda_E^{kl}(t, a_E)E^k(t -a_E, 0)da_E. 
\end{align}

Recalling the matrix equation of eq. \eqref{eq:ngm_renewal_main_text}, we have:
\begin{align}\label{eq:ngm_renewal_si}
\mathbf{E}(t, 0) &= \int_{0}^{t} \mathcal{K}(t, a_E)\mathbf{E}(t -a_E, 0) da_E.
\end{align}
Then, as in previous works \cite{murray_mathematical_2002, bouros_time-dependent_2025, mills_renewal_2025, jorge_estimating_2022}, we can suppose there is a separable, long-term solution of the form:
\begin{align}
    \mathbf{E}(t, a_E) &= e^{\gamma_t t} \Phi,
\end{align}
where $\Phi$ is a constant, non-negative vector. This equation is solved for each time $t$ (i.e. freezing human movement patterns as at time $t$) and states that the age distribution of infected humans is altered by a factor which grows or decays with time depending on whether $\gamma_t > 0$ or $\gamma_t < 0$, respectively. Simplifying both sides, we have:
\begin{align}
    \Phi &= \int_0^{t}\mathcal{K}(t, a_E)e^{-\gamma_t a_E}da_E \Phi =: \bar{K}(\gamma_t)\Phi,
\end{align}
which means that $\Phi$ is the eigenvector of $\bar{K}(\gamma_t):= \int_0^{t}\mathcal{K}(t, a_E)e^{-\gamma_t a_E}da_E$ with eigenvalue 1. Note that $\bar{K}(\gamma_t)$ is the Laplace integral transform of the positive matrix $\mathcal{K}(t, a_E)$ with upper integral limit $t \rightarrow \infty$. 

Using the Perron Frobenius Theorem (see Section \ref{sec:tech_condns}), we have that the spectral radius of $\rho(\bar{K}(\gamma_t))$ is one:
\begin{align}\label{eq:threshold_spectral_ngo}
    1 &= \rho(\bar{K}(\gamma_t)),
\end{align}
and now, as $\bar{K}(0) = \int_0^t\mathcal{K}(t,a_E)da_E$ gives the expected number of new infections generated by a single resident of a given location, say $k$, in residents of another given location, say $j$, this allows us to define the NGM (i.e. reproduction number matrix) $\mathbf{R}(t)$ as in eq. \eqref{eq:reproduction_matrix}. 

As in \citet{mills_renewal_2025, bouros_time-dependent_2025}, because $\bar{K}(\gamma_t)$ is monotonically decreasing in $\gamma_t$ and the spectral radius of a non-negative matrix is monotonically decreasing, we also have that $\rho(\bar{K}(\gamma_t))$ in eq. \eqref{eq:threshold_spectral_ngo} is monotonically decreasing in $\gamma_t$ and long-term epidemic growth occurs if and only if the network growth rate $\gamma_t >0$, which occurs if and only if $\mathcal{R}(t)=\rho(\bar{K}(0))= \rho(\mathbf{R}(t))>\rho(\bar{K}(\gamma_t))=1$. This recovers the threshold property of the network-level reproduction number $\mathcal{R}(t)$ and the instantaneous growth property of the growth rate $\gamma_t$ (i.e. the sign of the growth rate $\gamma_t$ is determined by whether $\mathcal{R}(t)=\rho(\bar{K}(0))= \rho(\mathbf{R}(t))> 1$).

\newpage
\normalsize
\newpage
\section{The eigenvalue problem and network-level reproduction number} \label{sec:tech_condns}
The Perron-Frobenius theorem guarantees that:
\begin{enumerate}
    \item $\mathcal{R}(t) = \rho(\mathbf{R}(t))$ is a simple, real, positive eigenvalue that is strictly larger than all other eigenvalues
    \item There is a unique, right eigenvector $\mathbf{v^*}$ such that $\mathbf{R}(t)\mathbf{v^*} = \mathcal{R}(t)\mathbf{v^*}$.
    \item There is also a unique left eigenvector $\mathbf{v}$ such that $\mathbf{v}^\top\mathbf{R} = \mathcal{R}(t)\mathbf{v}^\top$.
\end{enumerate}

By normalising such that $\sum_{j=1}^\mathcal{L}v_j^*=1$, the right eigenvector defines a probability distribution that describes the stable/equilibrium, long-term spatial distribution of infections across locations. In words, this means that irrespective of the initial conditions for the spatial distribution of incidence, after many generations with a counterfactually constant $\mathbf{R}(t)$, the fraction of all infections occurring in residents of location $k$ converges to $v_k^*$. Meanwhile, the left eigenvector entries $v_j$ measure the contribution of a single infected individual in location $j$, and therefore summarises the onward transmission potential (asymptotically) for residents of location $j$.

\section{Type reproduction number derivation}\label{sec:type_derivation}
Recall that from \citet{heesterbeek_type-reproduction_2007}, we have:
\begin{align}
R_{\text{type}}^j(t) &:= {e}_j^\top \mathbf{R}\left(I - (I - P_j) \mathbf{R} \right)^{-1}{e}_j,
\end{align}
where $P_j = {e}_j{e}_j^\top$ is the rank-one/single-entry projection matrix onto the j-th coordinate direction ($P_{jj} = 1$, $P_{ij} = 0$ otherwise). $\mathbf{R} {e}_j$ is the number of infected individuals in the second generation starting from one infected individual of type $j$ (i.e. residents of location $j$).

Decomposing and using block matrix notation with $j$ written as the first row, we have that $(I - P_j)\mathbf{R}$ is $\mathbf{R}$ with its $j$-th row zeroed out and describes all transmission except new infections in residents of location $j$:
\begin{align*}
(I - P_j) \mathbf{R} &= \begin{pmatrix}
 0 & 0 \\
 \mathbf{R}_{Jj} & \mathbf{R}_{JJ}
\end{pmatrix},
\end{align*}
which means that 

\begin{align*}
    \left(I - (I - P_j) \mathbf{R} \right)^{-1} &= \begin{pmatrix}
 1 & 0 \\
 -\mathbf{R}_{Jj} & (I - \mathbf{R}_{JJ})
\end{pmatrix}^{-1}, 
\end{align*}
and using the formula for the inverse of a block lower triangular matrix
\begin{align*}
    \left(I - (I - P_j) \mathbf{R} \right)^{-1} &= \begin{pmatrix}
 1 & 0 \\
  \left(I -  \mathbf{R}_{JJ} \right)^{-1} \mathbf{R}_{Jj} & (I - \mathbf{R}_{JJ})^{-1}
\end{pmatrix}, 
\end{align*}
and left-multiplying by $\mathbf{R}$
\begin{align*}
    \mathbf{R}\cdot\left(I - (I - P_j) \mathbf{R} \right)^{-1} &= \begin{pmatrix}
 \mathbf{R}_{jj} +  \mathbf{R}_{jJ}(I -  \mathbf{R}_{JJ})^{-1}\mathbf{R}_{Jj} & \mathbf{R}_{jJ}(I -  \mathbf{R}_{JJ})^{-1} \\
 \mathbf{R}_{Jj} +  \mathbf{R}_{JJ}(I -  \mathbf{R}_{JJ})^{-1}\mathbf{R}_{Jj}  & \mathbf{R}_{JJ}(I - \mathbf{R}_{JJ})^{-1}
\end{pmatrix}, 
\end{align*}
and finally left-multiplying by $e_j^\top$ and right-multiplying by $e_j$ to extract the $(j,j)$ entry of this matrix:
\begin{align*}
R_{\text{type}}^j(t) &= {e}_j^\top \mathbf{R}\left(I - (I - P_j) \mathbf{R} \right)^{-1}{e}_j \\ &= \mathbf{R}_{jj} + \mathbf{R}_{jJ}(I - \mathbf{R}_{JJ})^{-1}\mathbf{R}_{Jj}.
\end{align*}
\section{Vector-borne diseases}
\subsection{Mobility-informed renewal equations and reproduction numbers for vector-borne diseases}\label{sec:vbds_supp}

Vector-borne diseases, such as dengue, Zika, or lyme disease, require an intermediate vector to spread the pathogen from a primary infected human to a secondary susceptible human. Similar to eq. \eqref{eq:direct_PDEs} and \citet{mills_renewal_2025}, for a mosquito-borne disease, we can set up a similar age-structured system of PDEs across human and mosquito populations in locations $j$ 
\begin{align}
\frac{\partial E^j}{\partial t}+\frac{\partial E^j}{\partial a_E} &= -b_{E I}^j\left(t, a_{E}\right) E^j\left(t, a_{E}\right), \label{eq:vbd_1_app} \\
\frac{\partial I^j}{\partial t}+\frac{\partial I^j}{\partial a_I} &= 0, \label{eq:vbd_2_app} \\
\frac{\partial W^j}{\partial t}+\frac{\partial W^j}{\partial a_W} &= -b_{W V}^j\left(t, a_{W}\right) W^j\left(t, a_{W}\right) -\mu^j W^j\left(t, a_{W}\right), \label{eq:vbd_3_app} \\
\frac{\partial V^j}{\partial t}+\frac{\partial V^j}{\partial a_V} &= 0, \label{eq:direct_simple_app}
\end{align}
where $E$ and $I$ denote infected and infectious humans respectively and $W$ and $V$ denote infected and infectious mosquitoes respectively. We close the system with the following boundary conditions that are informed by human mobility : 
\begin{align}
E^j(t, 0) &= \sum_{k \in \mathcal{N}}\int_0^\infty b_{VE}^{kj}\left(t, a_V\right) V^k\left(t, a_V\right) d a_V, \label{eq:boundary_on_exposed_model_four}\\
&= \sum_{k \in \mathcal{N}}\int_0^\infty b^k(t)f^{jk}(t, S_H)\frac{S_{H}^{j}(t)}{\sum_{l \in \mathcal{N}}f^{lk}(t, S_H)N_{H}^{l}(t)}V^k\left(t, a_V\right) d a_V, \\
E^j\left(0, a_E\right) &= p_E^j\left(a_E\right), \\
I^j(t, 0) &=\int_0^{\infty} b_{E I}^j\left(t, a_E\right) E^j\left(t, a_E\right) d a_E, \label{eq:boundary_on_infectious_model_four} \\
I^j\left(0, a_I\right) &= p_I^j\left(a_I\right), \\
W^j(t, 0) &= \sum_{k \in \mathcal{N}}\int_0^{\infty} b_{IW}^{kj}\left(t, a_I\right) I^k\left(t, a_I\right) d a_I, \label{eq:boundary_on_exposed_mosquitoes_model_four} \\
&= \sum_{k \in \mathcal{N}}\int_0^{\infty} \frac{b^k(t) S_{M}^{k}(t)}{\sum_{l \in \mathcal{N}}f^{lk}(t, I)N_{H}^{l}(t)} f^{kj}(t, I)I^k\left(t, a_I\right) p_{WI}(a_I) d a_I \\
W^j\left(0, a_W\right) &=p_W^j\left(a_W\right),  \\
V^j(t, 0) &=\int_0^{\infty} b_{WV}^j\left(t, a_W\right) W^j\left(t, a_W\right) d a_W, \label{eq:boundary_on_infectious_mosquitoes_model_four} \\
V^j\left(0, a_V\right) &=p_V^j\left(a_V\right).  \label{eq:initial_on_infectious_mosquitoes_model_four}
\end{align}
where $b_\text{VE}^{kj}(t, a_ V)$ and $b_\text{IW}^{kj}(t, a_ I)$ represent the infection processes by which a susceptible human and mosquito resident in location $j$ respectively are infected by an infectious human and mosquito resident in location $k$ respectively. The development processes $b_\text{EI}^j(t, a_ I)$ and $b_\text{WV}^j(t, a_ V)$ represent the rate at which infected humans and mosquitoes resident in location $j$ become infectious. $p_E^j(a_E), p_I^j(a_I), p_W^j(a_W),$ and $p_V^j(a_V)$ are the initial densities for the resident exposed human, infectious human, exposed mosquito, and infectious mosquito populations respectively. We have assumed that locations are sufficiently large in area such that mosquitoes can only infect and be infected within their home location. This is the case for most real-world applications to surveillance data, but to account for between-location movement of mosquitoes, this would simply require an analogous term to $f^{jk}(t, S_H)$ for mosquitoes that represents the probability that each mosquito resident in location $j$ is in location $k$ at any time $t$ and then modified effective mosquito population terms in renewal equations for the possible meeting locations with susceptible and infected humans.

For each locations' birth/development processes, we can follow \citet{mills_renewal_2025} and write these as:
\begin{equation}
\begin{aligned}\label{eq:birth_process_examples}
b_{EI}^j\left(t, a_E\right) &=  \lambda_{E, \text{Latent}}^j(a_E) \\
b_{IW}^j\left(t, a_I\right) &=   \frac{b^j(t)}{N_{H}^{j}(t)}\cdot S_{M}^{j}(t) \cdot p_{WI}^j(a_I) \\
b_{WV}^j\left(t, a_W\right) &=   \lambda_{\text{W, Latent}}^j(t, a_W)  \\
b_{VE}^j\left(t, a_V\right) &=     b^j(t) \cdot \frac{S_{H}^{j}(t)}{{N_{H}^{j}(t)}} \cdot p_{EV}^j(a_V)
\end{aligned}
\end{equation}
\normalsize
where $\lambda_{E,\text{Latent}}^j(a_E)$ and $\lambda_{W,\text{Latent}}^j(t, a_W)$ denote the rates at which latent periods end for infected humans and mosquitoes respectively, $N_{H}^{j}(t)$ and $N_{M}^{j}(t)$ denote the total human and mosquito populations,
$S_{H}^{j}(t)$ and $S_{M}^{j}(t)$ denote the susceptible human and mosquito population counts, $p_{WI}$ and $p_{EV}$ denote the probability of mosquito and human infection upon infectious bite respectively, and $b^j(t)$ denotes the rate at which a mosquito bites a human. 

Hazard rates for the human latent periods $b_\text{EI}^j(t, a_ E)$  may differ by population demographic variables (e.g. older populations), while the mosquito latent period $b_\text{WV}^j(t, a_ W)$  may differ by location/space (e.g. urban landscape and mosquito adaptation) in addition to time. 

Following the method of characteristics and seeking a separable solution, the renewal equation for new human infections in residents of location $j$ can be written as:
\begin{equation}
\begin{aligned}\label{eq:birth_process_examples_b}
E^j(t, 0) &= \sum_{k \in \mathcal{N}}\int_0^t \int_0^{t-a_V}  \int_0^{t-a_V-a_W}  \int_0^{t-a_V - a_W - a_I}  b^k(t) f^{jk}(t) \frac{S_{H}^{j}(t)}{\sum_{l \in \mathcal{N}}f^{lk}(t)N_{H}^{k}(t)}p_{EV}^k(a_V)   b_{WV}^k\left(t-a_V, a_W\right)   \\ & \quad\cdot \sum_{c \in \mathcal{N}} \frac{b^c(t-a_V - a_W) S_{M}^{c}(t-a_V - a_W)}{\sum_{d \in \mathcal{N}}f^{dc}(t-a_V - a_W)N_{H}^{d}(t-a_V - a_W)} f^{kj}(t-a_V - a_W) p_{WI}^c(a_I) \\ & \quad \quad \cdot b_{E I}^c\left(t-a_V - a_W - a_I, a_E\right) e^{-\mu a_V} e^{-\int_0^{a_W} b_{WV}^c(t-a_V-a_W+r, r)d r - \mu a_W}    \\ & \quad \quad \quad \cdot E^c(t-a_V - a_W - a_I - a_E,0) e^{-\int_{0}^{a_E} b_{EI}^c(t-a_V - a_W - a_I-a_E+s,s) ds} d a_E  d a_I  d a_W   d a_V  
\end{aligned}
\end{equation}

or compactly:
\small
\begin{equation}
\begin{aligned}\label{eq:compact}
E^j(t, 0) &= \sum_{k \in \mathcal{N}} \int_0^t \int_0^{t-a_V}  \int_0^{t-a_V-a_W}  \int_0^{t-a_V - a_W - a_I} \sum_{c \in \mathcal{N}} \mathbb{K}^{ckj}(t, a_E, a_I, a_W, a_V) \\ & \quad \quad \quad \quad \quad \quad \quad \quad \quad \quad \quad \quad \quad \quad  \cdot E^c(t-a_V - a_W - a_I - a_E,0) d a_E  d a_I  d a_W   d a_V,  
\end{aligned}
\end{equation}
\normalsize
where we have defined:
\begin{align}
    \mathbb{K}^{ckj}(t, a_E, a_W, a_I, a_V) &:= b^k(t) f^{jk}(t) \frac{S_{H}^{j}(t)}{\sum_{l \in \mathcal{N}}f^{lk}(t)N_{H}^{k}(t)}p_{EV}^k(a_V)   b_{WV}^k\left(t-a_V, a_W\right)   \\ & \quad\cdot \sum_{c \in \mathcal{N}} \frac{b^c(t-a_V - a_W) S_{M}^{c}(t-a_V - a_W)}{\sum_{d \in \mathcal{N}}f^{dc}(t-a_V - a_W)N_{H}^{d}(t-a_V - a_W)} f^{kj}(t-a_V - a_W) p_{WI}^c(a_I) \\ & \quad \quad \cdot b_{E I}^c\left(t-a_V - a_W - a_I, a_E\right) e^{-\mu a_V} e^{-\int_0^{a_W} b_{WV}^c(t-a_V-a_W+r, r)d r -\mu a_W}    \\ & \quad \quad \quad \cdot  e^{-\int_{0}^{a_E} b_{EI}^c(t-a_V - a_W - a_I-a_E+s,s) ds}
\end{align}

$\mathbb{K}^{ckj}(t, a_E, a_W, a_I, a_V)$ is the instantaneous rate at which previously infected resident humans of location $c$ produce secondary infections that occur at meeting location $k$ in residents of location $j$. In further detail, this is the expected number of secondary infections occurring in residents of location $j$ at calendar time $t$ from the bite of infectious mosquitoes resident in location $k$ which became infectious at calendar time $t-a_V$ and were infected at calendar time $t-a_V-a_W$ in location $k$ by an infectious human resident in location $c$ who was originally infected at time $t-a_V-a_W-a_I-a_E$. The ordering of notation $(c, k, j)$ reflects the resident locations of the ordered infection events (primary human infection of resident of location $c$, infection of mosquito resident of location $k$, and secondary human infection of resident of location $j$). As in \citet{mills_renewal_2025}, the stage-specific ages in $\mathbb{K}^{ckj}(t, a_E, a_W, a_I, a_V)$ reflect the ages of the humans/mosquitoes at the times when they play their role in the multi-stage transmission cycle.

As before, we can sum over meeting (/mosquito resident) locations to write instantaneous kernels that describes instantaneous transmission between humans resident in two locations:
\begin{equation}
\begin{aligned}\label{eq:two_nbr_kernel}
\mathbb{K}^{cj}(t, a_E,a_I, a_W, a_V) &:= \sum_{k \in \mathcal{N}}  \mathbb{K}^{ckj}(t, a_E, a_I, a_W, a_V),  
\end{aligned}
\end{equation}
where the expression sums over all meeting locations $k$ where the susceptible human resident in location $j$ can be infected by an infectious mosquito resident in any location $k$ who was themselves infected by the infected human resident in location $c$.

Different types of human-to-human instantaneous reproduction numbers can again be formed by integrating over all ages. We define the between-location $R^{cj}(t)$, the expected number of secondary human infections in location $j$ generated by a previously infected human resident in location $c$ throughout their infected lifetime if conditions affecting this human transmission route were to remain the same at time $t$:
\begin{equation}
\begin{aligned}\label{eq:two_nbr_reprod_vbd1}
R^{cj}(t) &:=  \int_0^t \int_0^{t-a_V}  \int_0^{t-a_V-a_W}  \int_0^{t-a_V - a_W - a_I} \sum_{k \in \mathcal{N}} \mathbb{K}^{ckj}(t, a_E, a_I, a_W, a_V).  
\end{aligned}
\end{equation}

This suggests probability densities for the between-location generation time distributions can be formed as:
\begin{equation}
\begin{aligned}\label{eq:two_nbr_kernel_b}
g^{cj}(t, a_E,a_I, a_W, a_V) &:= \frac{\mathbb{K}^{cj}(t, a_E, a_I, a_W, a_V)}{R^{cj}(t)},   
\end{aligned}
\end{equation}
which describes, at each time $t$, the relative contributions of different stage-specific ages to the expected number of new human infections for residents of location $j$ arising from the original human infection from a resident of location $c$. Ages are evaluated at the calendar times when each stage-specific individual played their role in the transmission cycle. Unlike the directly transmitted setting (eq. \eqref{eq:between_loc_out_gt_direct}), this $g^{cj}(t, a_E,a_I, a_W, a_V)$ will not generally be fixed across space and time, due to the time- and age-varying birth processes across different locations, creating different environmental and biological conditions for different individual vectors.

Our renewal equation can then be written as:
\begin{equation}
\begin{aligned}\label{eq:renewal_eq_vbds_final_form_app}
E^j(t, 0) &= 
\sum_{c \in \mathcal{N}} \int_0^t \int_0^{t-a_V}  \int_0^{t-a_V-a_W}  \int_0^{t-a_V - a_W - a_I} \mathbb{K}^{cj}(t, a_E, a_I, a_W, a_V) \\ & \quad \quad \quad \cdot E^c(t-a_V - a_W - a_I - a_E,0)d a_E  d a_I  d a_W   d a_V \\  
&=\sum_{c \in \mathcal{N}}R^{cj}(t) \int_0^t \int_0^{t-a_V}  \int_0^{t-a_V-a_W}  \int_0^{t-a_V - a_W - a_I} g^{cj}(t, a_E, a_I, a_W, a_V) \\ & \quad \quad \quad \cdot E^c(t-a_V - a_W - a_I - a_E,0)d a_E  d a_I  d a_W   d a_V.  
\end{aligned}
\end{equation}

Again, we also have per-location \textit{outward} instantaneous kernels for each residence location $c$:
\begin{equation}
\begin{aligned}\label{eq:overallkernel}
\mathbb{K}_{\text{out}}^{c}(t, a_E,a_I, a_W, a_V) &:= \sum_{j \in \mathcal{N}} \sum_{k \in \mathcal{N}}  \mathbb{K}^{ckj}(t, a_E, a_I, a_W, a_V),  
\end{aligned}
\end{equation}
which suggests defining an \textit{outward} instantaneous reproduction number $R_{\text{out}}^{c}(t)$ as:
\begin{equation}
\begin{aligned}\label{eq:two_nbr_reprod_vbd2}
R_{\text{out}}^{c}(t) &:= \int_0^t \int_0^{t-a_V}  \int_0^{t-a_V-a_W}  \int_0^{t-a_V - a_W - a_I} \mathbb{K}_{\text{out}}^{c}(t, a_E,a_I, a_W, a_V)d a_E  d a_I  d a_W   d a_V \\ &=
\int_0^t \int_0^{t-a_V}  \int_0^{t-a_V-a_W}  \int_0^{t-a_V - a_W - a_I} \sum_{j \in \mathcal{N}}  \sum_{k \in \mathcal{N}} \mathbb{K}^{ckj}(t, a_E, a_I, a_W, a_V) d a_E  d a_I  d a_W   d a_V.  
\end{aligned}
\end{equation}
This transmission potential indicator sums over all locations $j$ where a secondary infected human could be resident to capture, at time $t$, how many expected infections are generated by the infected individual resident in location $c$, if conditions affecting human transmission were to remain the same as at time $t$. Note that we have also summed all possible locations where the intermediate mosquito was infected, or equivalently where the infected human resident in location $j$ was infected. 

We then define the probability densities for the outward generation time distribution as:
\begin{equation}
\begin{aligned}\label{eq:overallkernel_b}
g_{\text{out}}^{c}(t, a_E,a_I, a_W, a_V) &= \frac{\mathbb{K}_{\text{out}}^{c}(t, a_E, a_I, a_W, a_V)}{R_{\text{out}}^{c}(t)},  
\end{aligned}
\end{equation}
which describes, at each time $t$, the relative contributions of different stage-specific ages to the expected number of new human infections arising from the original human infection in a resident of location $c$. 


We can also define \textit{inward} reproduction numbers as the expected number of new infections occurring in residents of location $j$ by an average infected individual resident in any location throughout their infected lifetime if transmission conditions were to remain the same as at time $t$.
\begin{equation}
\begin{aligned}\label{eq:two_nbr_reprod_vbd3}
R_{\text{in}}^{j}(t) &= \sum_{k \in \mathcal{N}} \sum_{c \in \mathcal{N}} \int_0^t \int_0^{t-a_V}  \int_0^{t-a_V-a_W}  \int_0^{t-a_V - a_W - a_I} \mathbb{K}^{ckj}(t, a_E, a_I, a_W, a_V) d a_E  d a_I  d a_W   d a_V,  
\end{aligned}
\end{equation}
which indicates whether a location is experiencing net instantaneous growth/decay in the incidence of resident infections.


We can also compute between-location renewal equations: 
\begin{equation}
\begin{aligned}\label{eq:between_nbr_vbd_renewal_eq}
E^{c \rightarrow j}(t, 0) &= R^{cj}(t) \int_0^t \int_0^{t-a_V}  \int_0^{t-a_V-a_W}  \int_0^{t-a_V - a_W - a_I} g^{cj}(t, a_E, a_I, a_W, a_V) \\ &  \quad \quad \cdot E^c(t-a_V - a_W - a_I - a_E,0)d a_E  d a_I  d a_W   d a_V,  
\end{aligned}
\end{equation}
which capture how infections propagate from resident humans of locations $c$ to resident humans of locations $j$.

As before, the incidence/renewal equation matrix $\mathcal{M}(t)$ has individual elements $E^{c \rightarrow j}(t, 0)$ describing the number of new human infections generated in location $j$ by previously infected resident humans of location $c$ . 

In matrix form,
$$ \mathcal{M}(t) = 
\begin{pmatrix}
 E^{1 \rightarrow 1}(t, 0) &E^{1 \rightarrow 2}(t, 0) &\ldots  & E^{1 \rightarrow \mathcal{L}}(t, 0) \\
E^{2 \rightarrow 1}(t, 0)  & E^{2 \rightarrow 2}(t, 0) & \ldots & E^{2 \rightarrow \mathcal{L}}(t, 0) \\
    \ldots & \ldots & \ldots & \ldots \\
    \ldots & \ldots & \ldots & E^{\mathcal{L} \rightarrow \mathcal{L}}(t, 0)
\end{pmatrix}
$$

\normalsize
where for simplicity, we have dropped the limits of integration from eq. \eqref{eq:between_nbr_vbd_renewal_eq}. Rows sum to the number of new infections occurring in location $j$ at time $t$. This formulation makes explicit the relative contributions of each location $c$ to the new human infections in location $j$ at time $t$. As before, the diagonal elements describe the  within-location transmission from the renewal equation.



\section{Further derived quantities}
\subsection{Risk-averse reproduction number reformulation}\label{sec:risk_aware_appendix}
The spatial risk-averse reproduction number can be rewritten as:
\begin{align}
    \mathcal{E}(t) &= X(\alpha = 1, t)   \\ &= \frac{\sum_{j \in \mathcal{N}} R_{\text{out}}^j(t)^2}{\sum_{j \in \mathcal{N}} R_{\text{out}}^j(t)} \\
    &= \bar{R}_{\text{out}}(t) + \frac{\text{Var}( R_{\text{out}}^j(t))}{\bar{R}_{\text{out}}(t)},
\end{align}
which recasts $\mathcal{E}(t)$ from the ratio of the second moment to the first moment of the outward reproduction numbers to the arithmetic mean of outward reproduction numbers plus a heterogeneity/variance-to-mean penalty. 

This form makes explicit that $\mathcal{E}(t) < 1$ requires both the mean outward reproduction to be controlled and spatial heterogeneity to be small.

Note that $\mathcal{E}(t)$ cannot determine epidemic growth/decay with the usual threshold property (like $\mathcal{R}(t)$) as it is simply a weighted average of single-generation quantities $R_{\text{out}}(t)$ and therefore does not handle the full, multi-generation structure of the network obtained from computing $\mathcal{R}(t)$ as the spectral radius of the NGM $\mathbf{R}(t)$. It is still, however, a useful spatial risk warning indicator for detecting potential heterogeneous infection resurgences from locations.

\newpage
\section{Further application figures}
\subsection{Incidence across space and time}

\newpage
\begin{figure}[H]
\centering
\includegraphics[scale = 0.88]{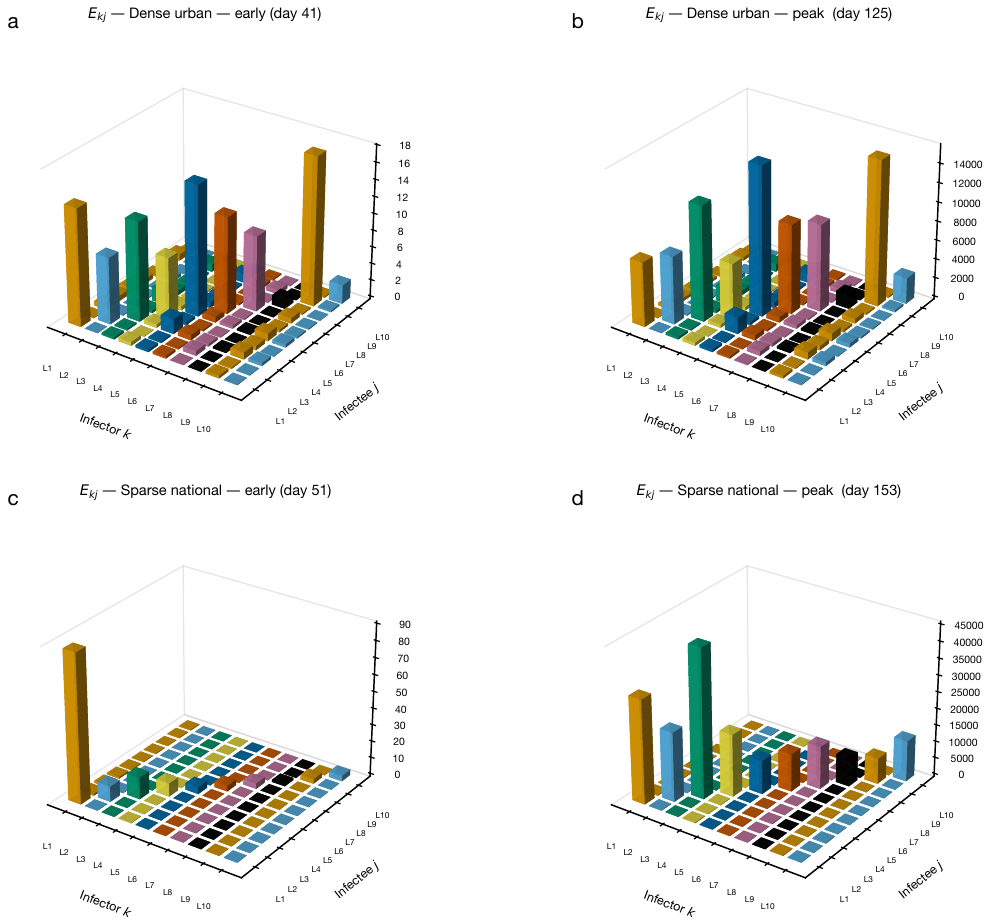}
\caption{\label{fig_SI8_3d_earlypeak} \footnotesize \textbf{Early-stage and peak-stage epidemic dynamics:} For our dense-urban (Scenario A, top) and sparse-national (Scenario B, bottom) settings, we visualise between-location incidence at early (left) and peak (right) stages of the epidemic. $E_{kj}(t, 0)$ represents the number of infections transmitted from residents of location $k$ to residents of location $j$ at time $t$.}
\end{figure}

\subsection{Modelling assumptions}
\begin{figure}[H]
\centering
\includegraphics[scale = 0.88]{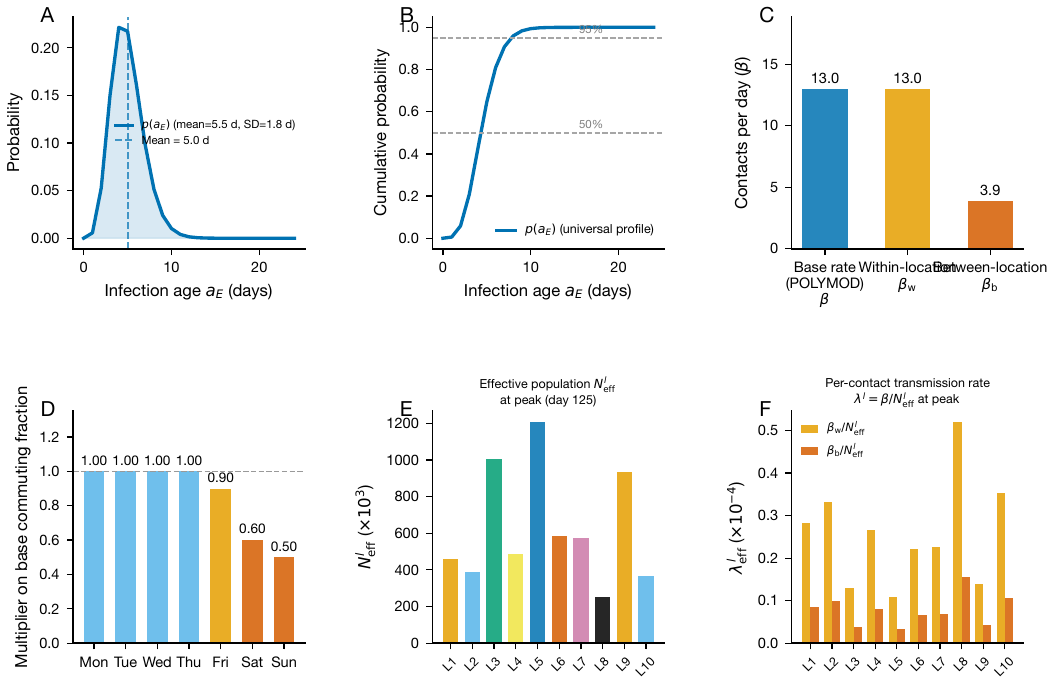}
\caption{\label{fig:SI3_epi_params} \footnotesize \textbf{Epidemiological parameters for applications of our framework for dense-urban scenario:} \textbf{A)} and \textbf{B)} show the assumed probability mass function and cumulative distribution function for the infectiousness profile across all time points and locations. \textbf{C)} The assumed contact profiles for within-location and between-location transmission. \textbf{D)} The day-of-week-scaling, commuting multiplier fractions, per day of week. \textbf{E)} The effective population in each location at the peak stage of the epidemic. \textbf{F)} Per-contact transmission rates for between- and within-location transmission in each location at the peak stage of the epidemic.}
\end{figure}
\FloatBarrier
\newpage
\begin{figure}[H]
\centering
\includegraphics[scale = 0.88]{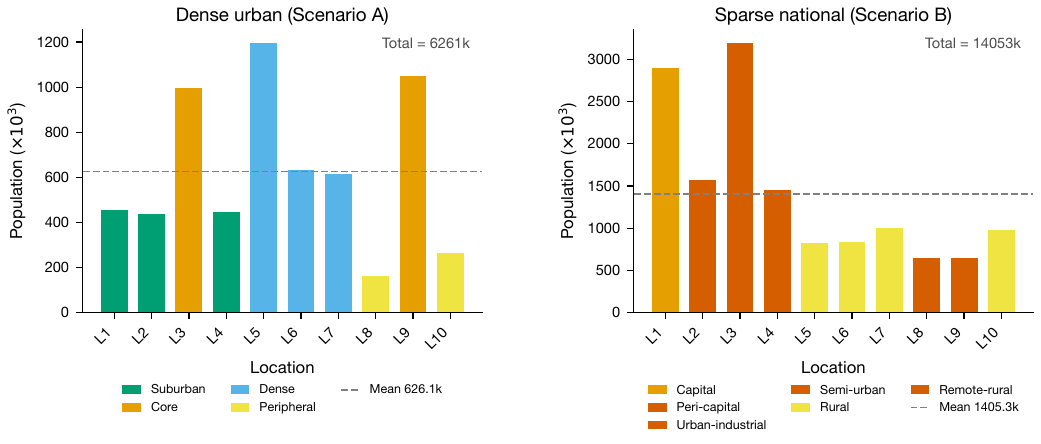}
\caption{\label{fig:SI0_population} \footnotesize \textbf{Population structure in dense-urban (Scenario A) and sparse-national scenarios (Scenario B):} Ten location nodes were used in each simulated scenario, each with differing population counts and belonging to a particular class/type of location, each of which are visualised here.}
\end{figure}

\subsection{Numerical convergence and stability analyses}
\begin{figure}[H]
\centering
\includegraphics[scale = 0.88]{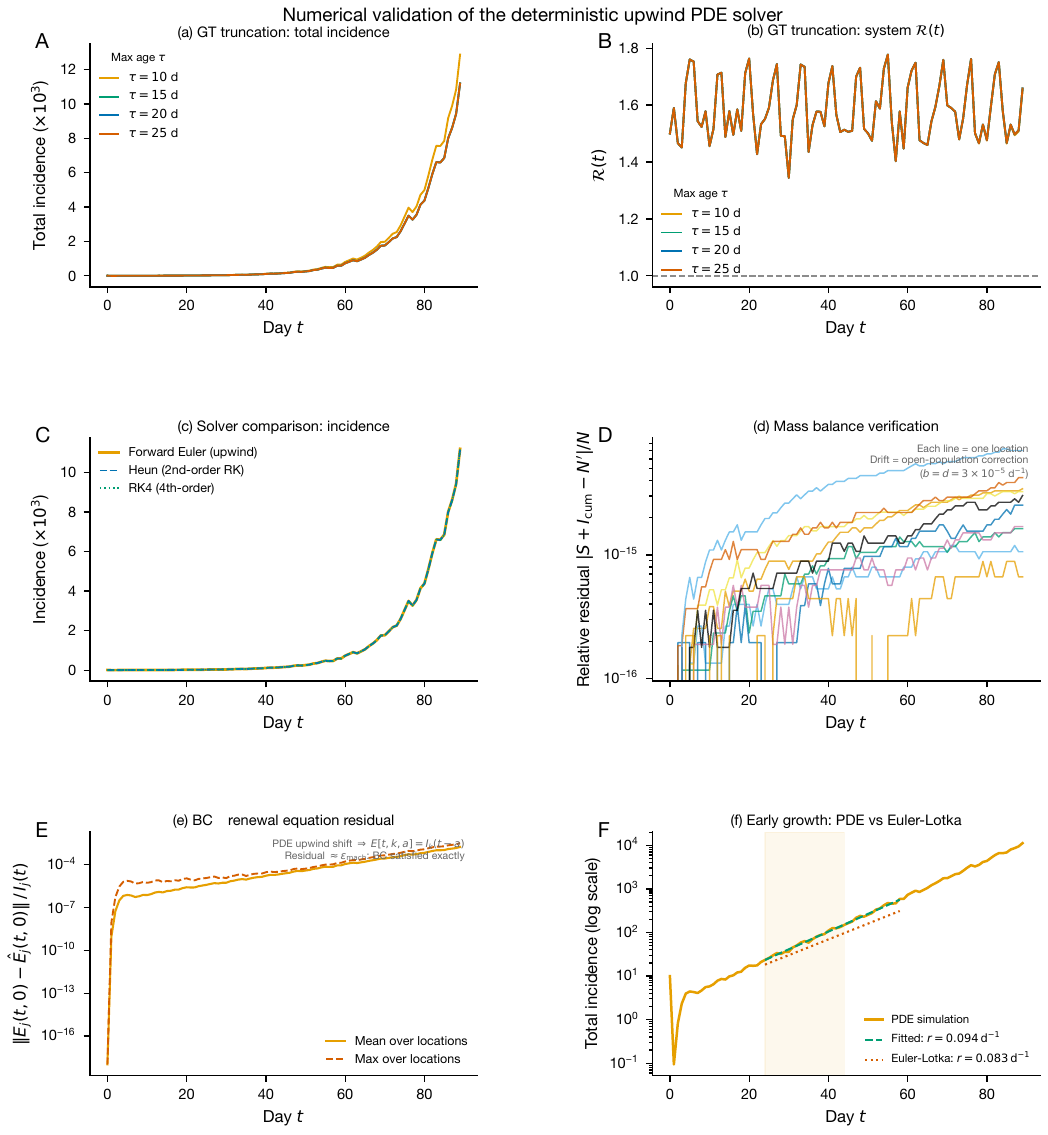}
\caption{\label{fig:SI4_convergence} \footnotesize \textbf{Diagnostic checks for numerical simulations of partial differential equations (PDEs) for Scenario A:} \textbf{A)} Total incidence and \textbf{B)} $\mathcal{R}(t)$ under different truncations for the generation time distribution. \textbf{C)} visualises the incidence using different numerical PDE solvers. \textbf{D)} visualises the verification of mass balance. \textbf{E)} visualises the boundary condition for infection incidence $E(t, 0)$ minus renewal equation residuals over time and across locations. \textbf{F)} displays the infection incidence $E(t, 0)$ over time from the PDE simulations, alongside the exponential growth rate.}
\end{figure}
\FloatBarrier
\newpage
\subsection{Sensitivity and elasticity analyses}
\begin{figure}[H]
\centering
\includegraphics[scale = 0.88]{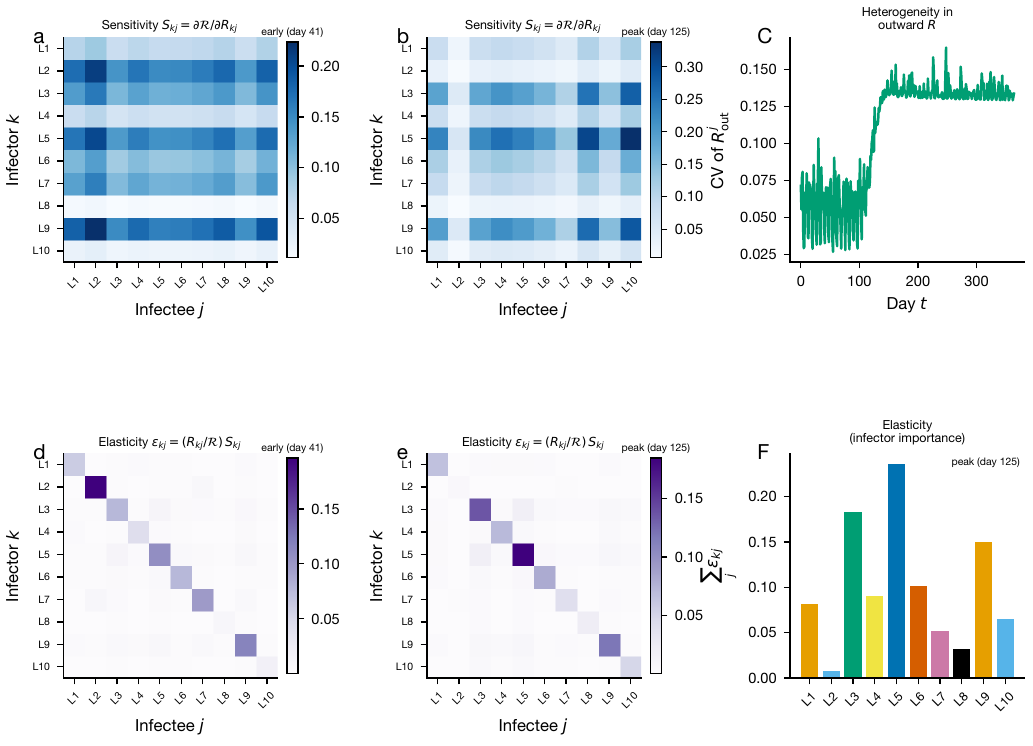}
\caption{\label{fig:SI1_sensitivity} \footnotesize \textbf{Sensitivity and elasticity of $\mathcal{R}(t)$}: \textbf{A)} and \textbf{B)} describe the sensitivity $S_{kj}$ of $\mathcal{R}(t)$ with respect to between-location reproduction numbers at an early and peak epidemic stage respectively, with higher values indicating greater importance of that between-location transmissibility to network-level transmissibility. \textbf{C)} is the coefficient of variation (CV) in $R_{\text{out}}^j(t)$ across locations $j$ over time, with higher values indicating grater heterogeneity. \textbf{D)} and \textbf{E)} describe the elasticity (i.e. normalised sensitivity) $\epsilon_{kj}$ of $\mathcal{R}(t)$ with respect to between-location reproduction numbers at an early and peak epidemic stage respectively, with higher values indicating greater importance of that between-location transmissibility to network-level transmissibility. \textbf{F)} is the infector elasticity $\epsilon_{\text{in}}^{k}(t) = \sum_{j \in \mathcal{N}}\epsilon_{kj}$ that sums over destination location for secondary infections to quantify the relative importance each primary infection location $k$. We visualise this elasticity at the epidemic incidence peak.}
\end{figure}
\FloatBarrier

\subsection{Further counterfactual sensitivity simulation analyses}
\newpage
\begin{figure}[H]
\centering
\includegraphics[scale = 0.88]{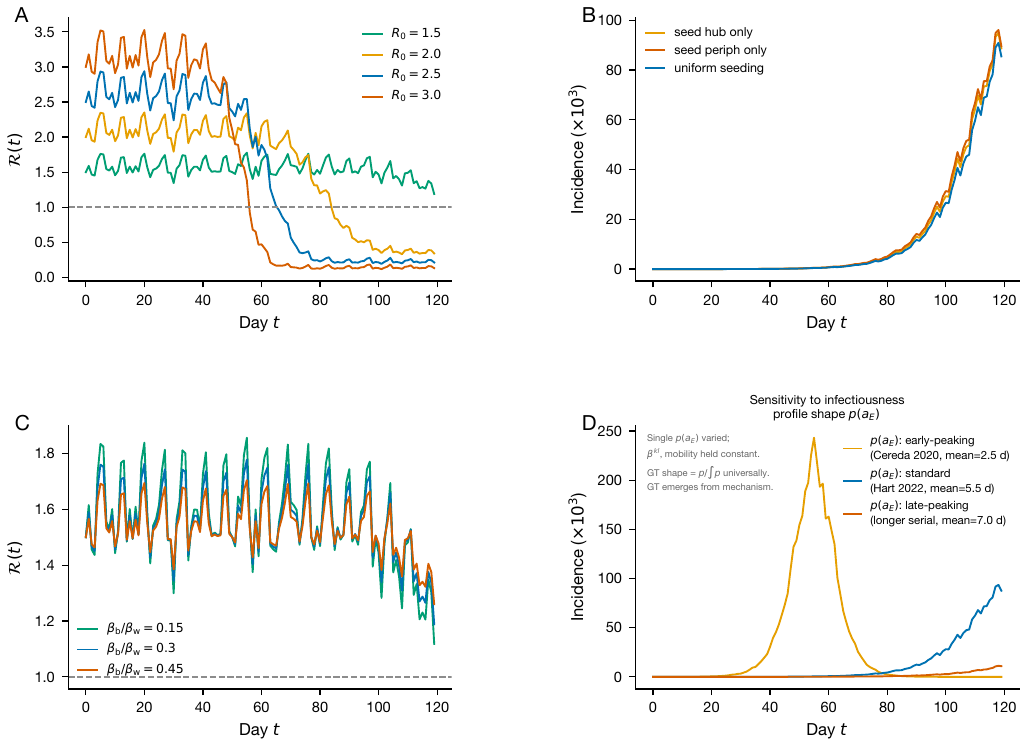}
\caption{\label{fig:SI7_sensitivity} \footnotesize \textbf{Sensitivity analysis for main text Scenario A application:} \textbf{A)} visualises network-level $\mathcal{R}(t)$ for different basic reproduction numbers $R_0$, with higher values resulting in faster epidemics. \textbf{B)} visualises the effects on daily infection incidence of different locations for initial condition seeding. \textbf{C)} captures $\mathcal{R}(t)$ for different assumed levels of between-location transmissibility versus within-location transmissibility. \textbf{D)} visualises the effects on daily infection incidence of different assumptions for infectiousness profiles with shorter infectivity profiles producing faster epidemics.}
\end{figure}
\FloatBarrier

\newpage
\subsection{Static network analyses}
\begin{figure}[H]
\centering
\includegraphics[scale = 0.88]{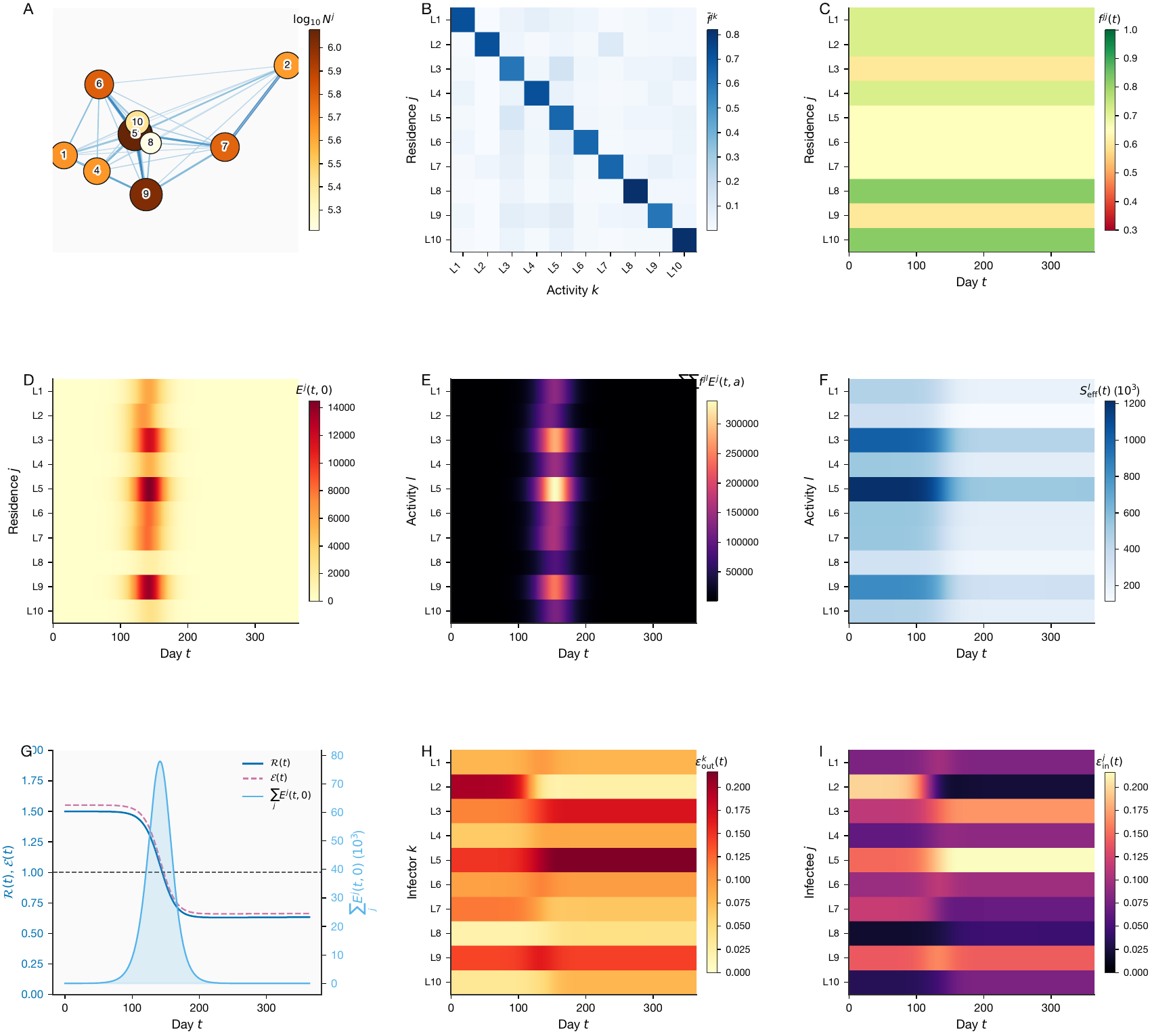}
\caption{\label{fig:SI_static_02_overview} \footnotesize \textbf{Overall human mobility and epidemic dynamics across space and time for Scenario A under static mobility patterns:} \textbf{A)} Time-averaged probability that residents of location $j$ are in location $k$, with row sums equal to 1. \textbf{B)} Incidence of new infections across locations (y-axis) over time (x-axis). \textbf{C)} Network-level reproduction number $\mathcal{R}(t)$ and overall incidence over time. \textbf{D)} Probability at each time t that residents of location $j$ are in their home residence location $j$. \textbf{E)} Effective susceptible population in location $l$ at time $t$. \textbf{F)} Infector elasticity of $\mathcal{R}(t)$ with respect to pairwise $R^{kj}$ summed over infectee residence locations $j$ such that it summarises infection contributions to network-level reproduction. Figure \ref{fig:02_overview.pdf} is the analogous visualisation for Scenario A when assuming time-varying movement patterns.}
\end{figure}

\newpage
\begin{figure}[H]
\centering
\includegraphics[scale = 0.5]{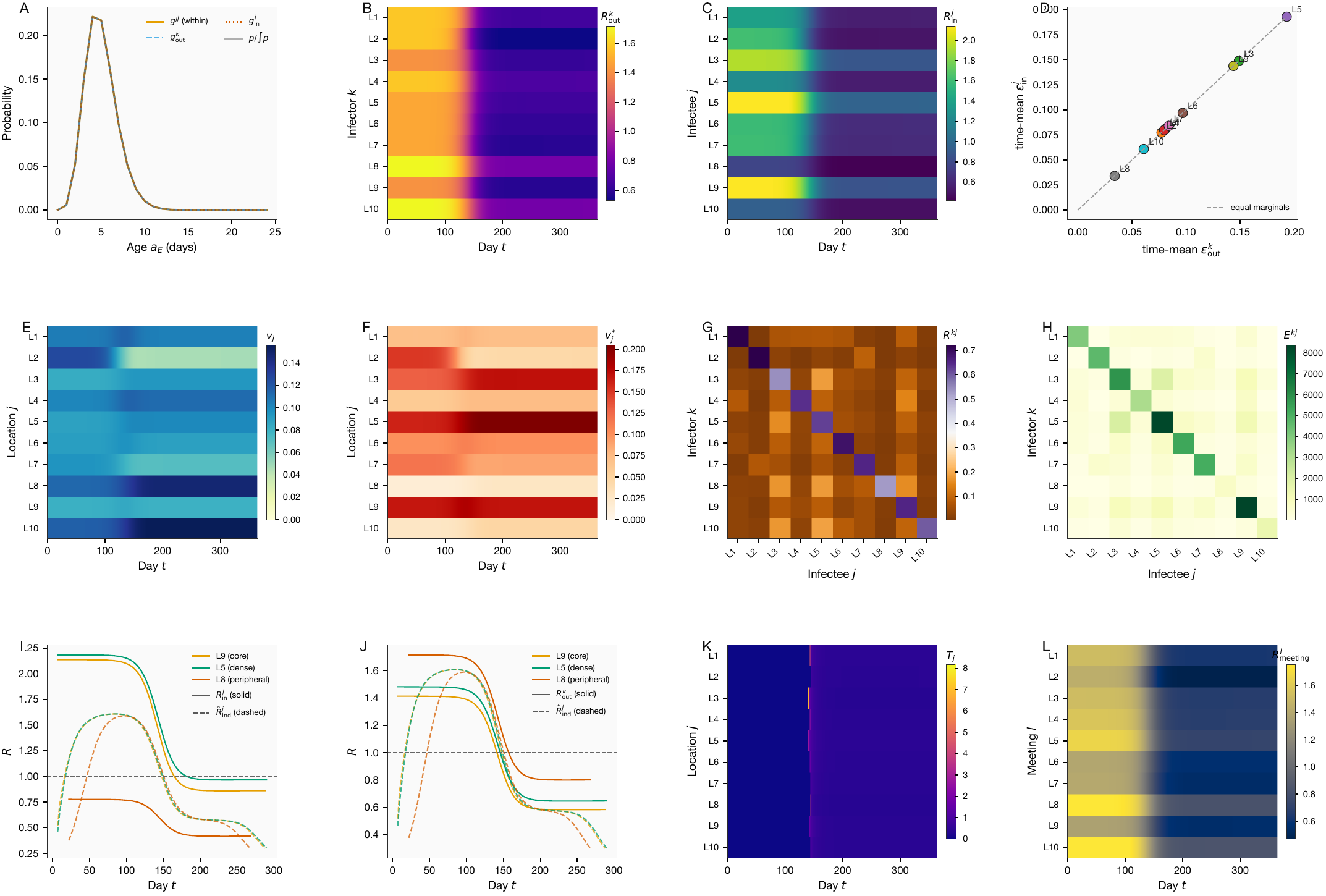}
\caption{\label{fig:SI_static_03_taxonomy} \footnotesize \textbf{Taxonomy of reproduction numbers and generation time distributions for Scenario A under static movement patterns:} \textbf{A)} Probability densities for different types of generation time distributions which closely overlap across space and time due to assumptions of infection age independence for contact rates and movement patterns and location-invariant infectiousness profiles. \textbf{B)} and \textbf{C)} are the inward and outward reproduction numbers respectively for each location over time. \textbf{D)} and \textbf{E)} are the between-location reproduction numbers $R^{kj}$ and incidence $E_{kj}$ respectively at at the epidemic incidence peak (day 141). F) is the source-sink dynamics. \textbf{G)} and \textbf{H)} compare inward and outward reproduction numbers respectively to independent $R(t)$ estimates for three types of locations (hub, mid/suburb, and peripheral nodes). \textbf{I)} is the type reproduction number per location over time (where defined) and \textbf{J)} is the meeting location reproduction number per meeting location over time. See Tables \ref{tab:framework_outputs} -- \ref{tab:R_operational} for definitions, interpretations, and uses of our framework's quantities. Figure \ref{fig:03_taxonomy.pdf} is the analogous visualisation for Scenario A when assuming time-varying movement patterns.}
\end{figure}
\FloatBarrier
\newpage

\begin{figure}[H]
\centering
\includegraphics[scale = 0.88]{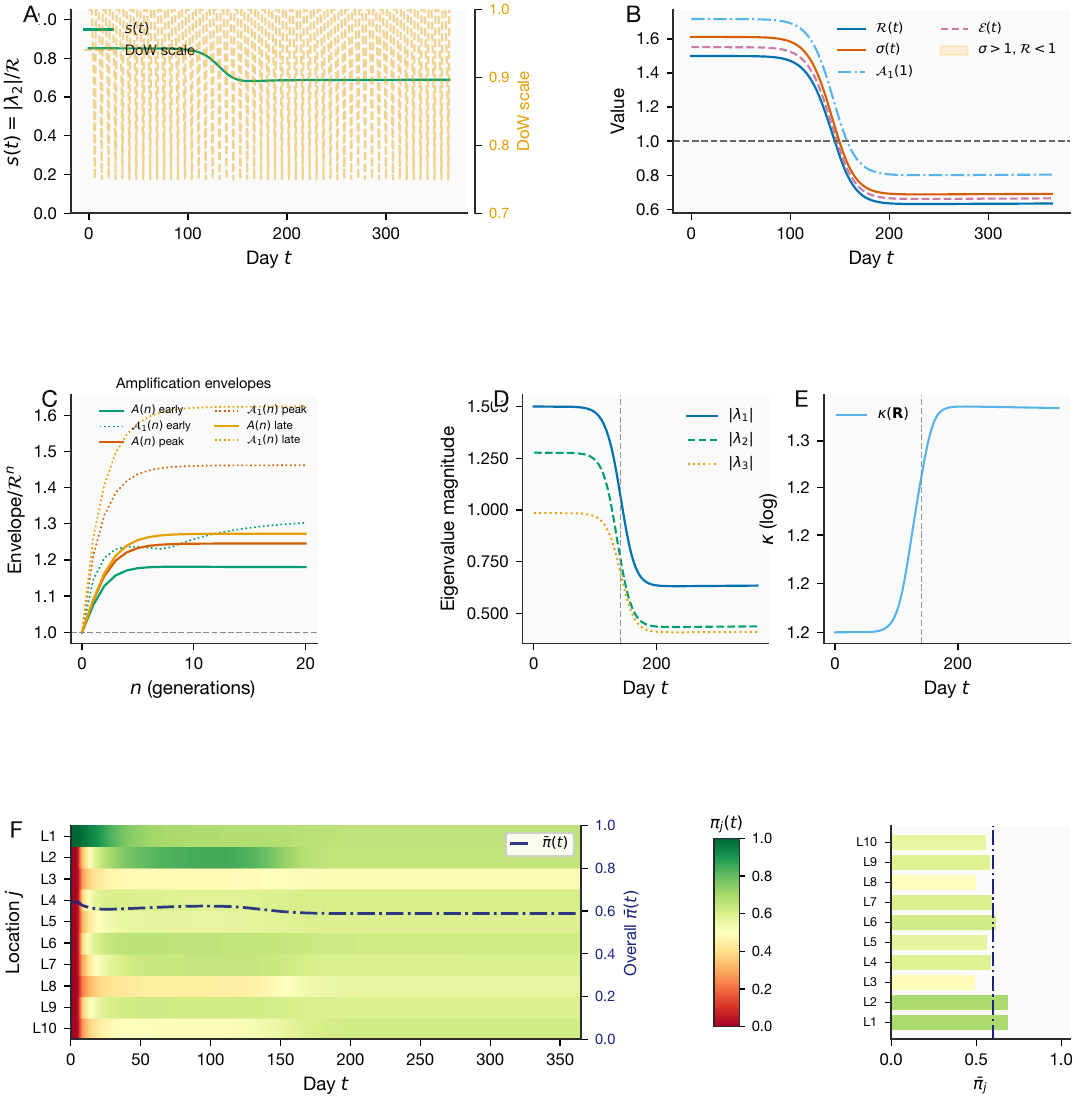}
\caption{\label{fig:SI_static_04_spectral}\footnotesize \textbf{Transience and importation analyses across space and time:} \textbf{A)} Mixing time ratio $s(t)$ and day-of-week (DoW) scaling used for human movement simulation. \textbf{B)} Reactivity $\sigma(t)$ and network-level $\mathcal{R}(t)$ over time. \textbf{C)} Amplification envelope $A(n)$ divided by $\mathcal{R}^n$ for $n$ generations at early, peak, and late stages of the outbreak. \textbf{D)} Magnitudes of the top three eigenvalues of $\mathbf{R}(t)$ over time with a larger spectral gap between $\mathcal{R}(t)$ and $|\lambda_2(t)|$ indicating greater transience. \textbf{E)} Eigenvalue condition number for the dominant eigenvalue $\mathcal{R}(t)$ of $\mathbf{R}(t)$ over time. \textbf{F)} Proportion of incident infections in residents of location $j$ generated by infected individuals also resident in location $j$ over time (left) and the corresponding overall proportions (right). See Tables \ref{tab:framework_outputs} -- \ref{tab:R_operational} for definitions, interpretations, and uses of our framework's quantities. Figure \ref{fig:04_spectral} is the analogous visualisation for Scenario A when assuming time-varying movement patterns.}
\end{figure}
\newpage

\subsection{Bias/error analyses for reproduction numbers}

\subsubsection{Aggregate incidence reproduction numbers}

\begin{figure}[H]
\centering
\includegraphics[scale = 0.88]{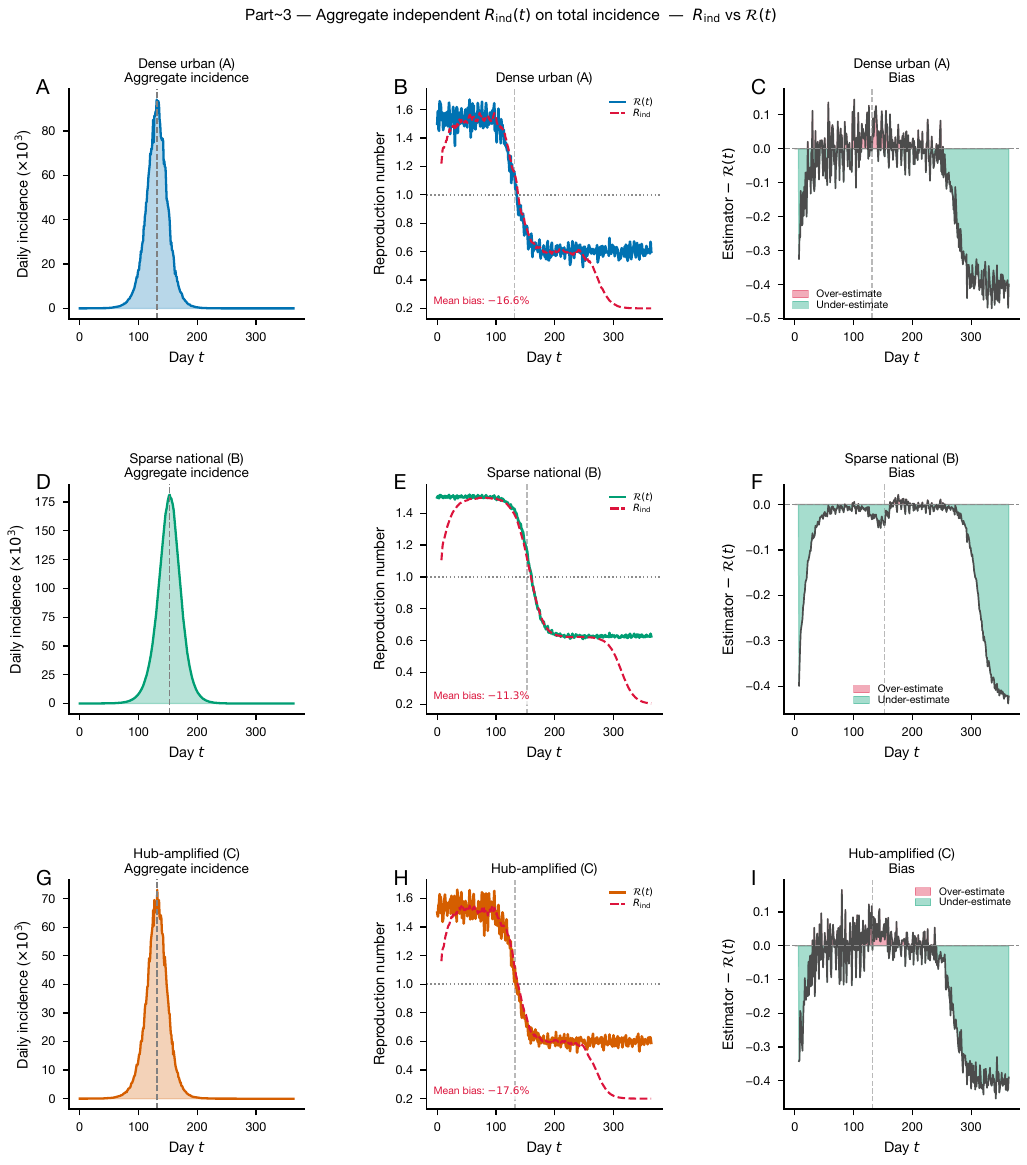}
\caption{\label{fig:aggregate_R_bias} \footnotesize \textbf{Aggregate $R(t)$ versus $\mathcal{R}(t)$} We compare the reproduction number $R_{\text{naive}}(t)$ estimated from the network-aggregated infection incidence (i.e. treating all locations as a single, well-mixed population) against our framework's network-level reproduction number $\mathcal{R}(t)$ for the dense-urban setting (Scenario A), sparse-national setting (Scenario B), and hub-amplified setting (Scenario C).  Aggregating incidence across locations neglects spatial structure and human movement, generally over-estimating network-level transmissibility during epidemic growth ($\mathcal{R}(t)>1$) and under-estimating it during decline ($\mathcal{R}(t)<1$).}
\end{figure}
\FloatBarrier

\begin{figure}[H]
\centering
\includegraphics[scale = 0.88]{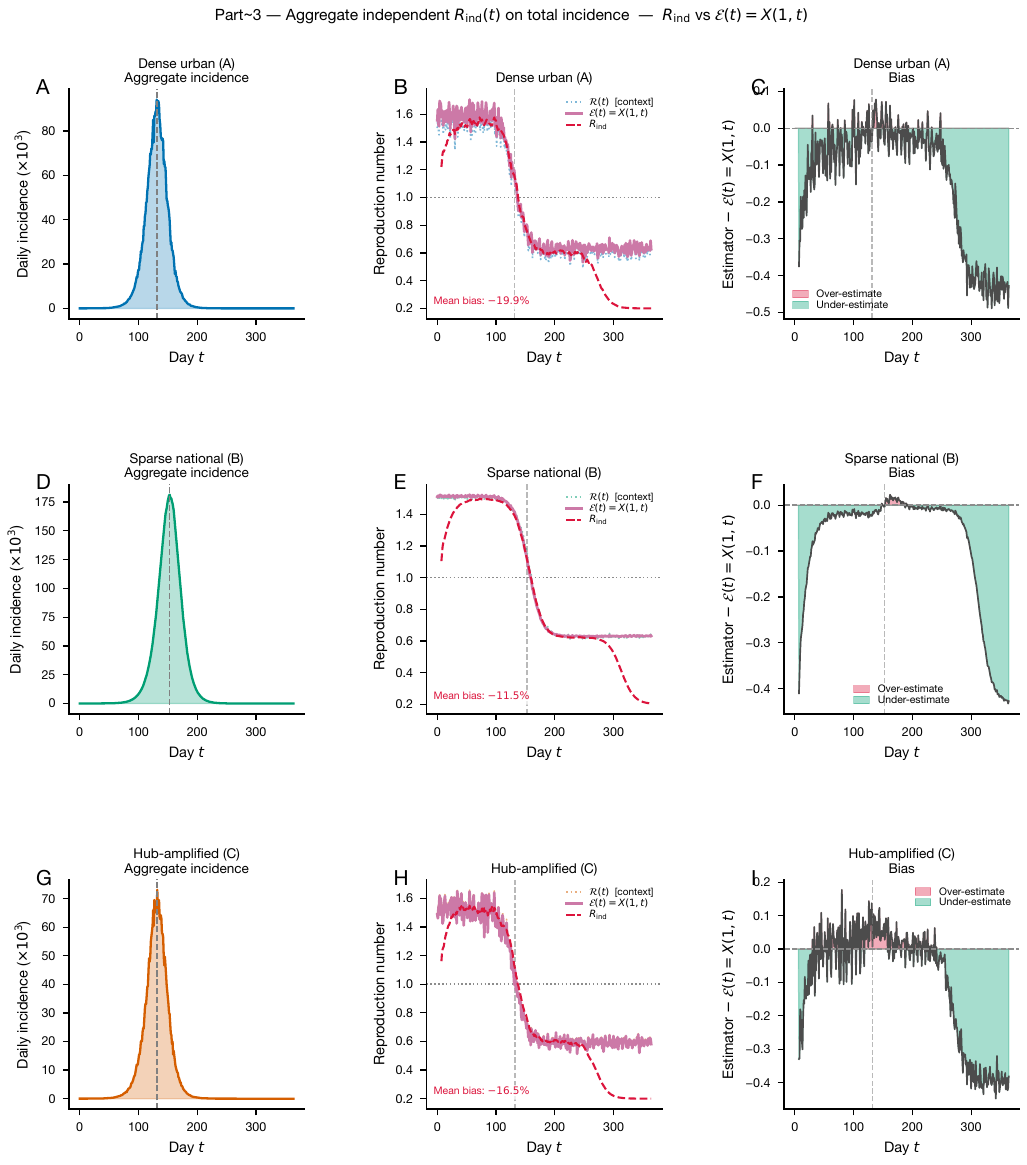}
\caption{\label{fig:aggregate_R_bias_Et} \footnotesize \textbf{Aggregate incidence $R(t)$ versus $\mathcal{E}(t)$} As in Figure \ref{fig:aggregate_R_bias}, but comparing the aggregate-incidence reproduction number $R_{\text{naive}}(t)$ against our framework's spatial risk-averse reproduction number $\mathcal{E}(t)$ for the dense-urban setting (Scenario A).}
\end{figure}
\FloatBarrier

\subsubsection{Population-weighted mean of independent, closed-population reproduction numbers}

\begin{figure}[H]
\centering
\includegraphics[scale = 0.88]{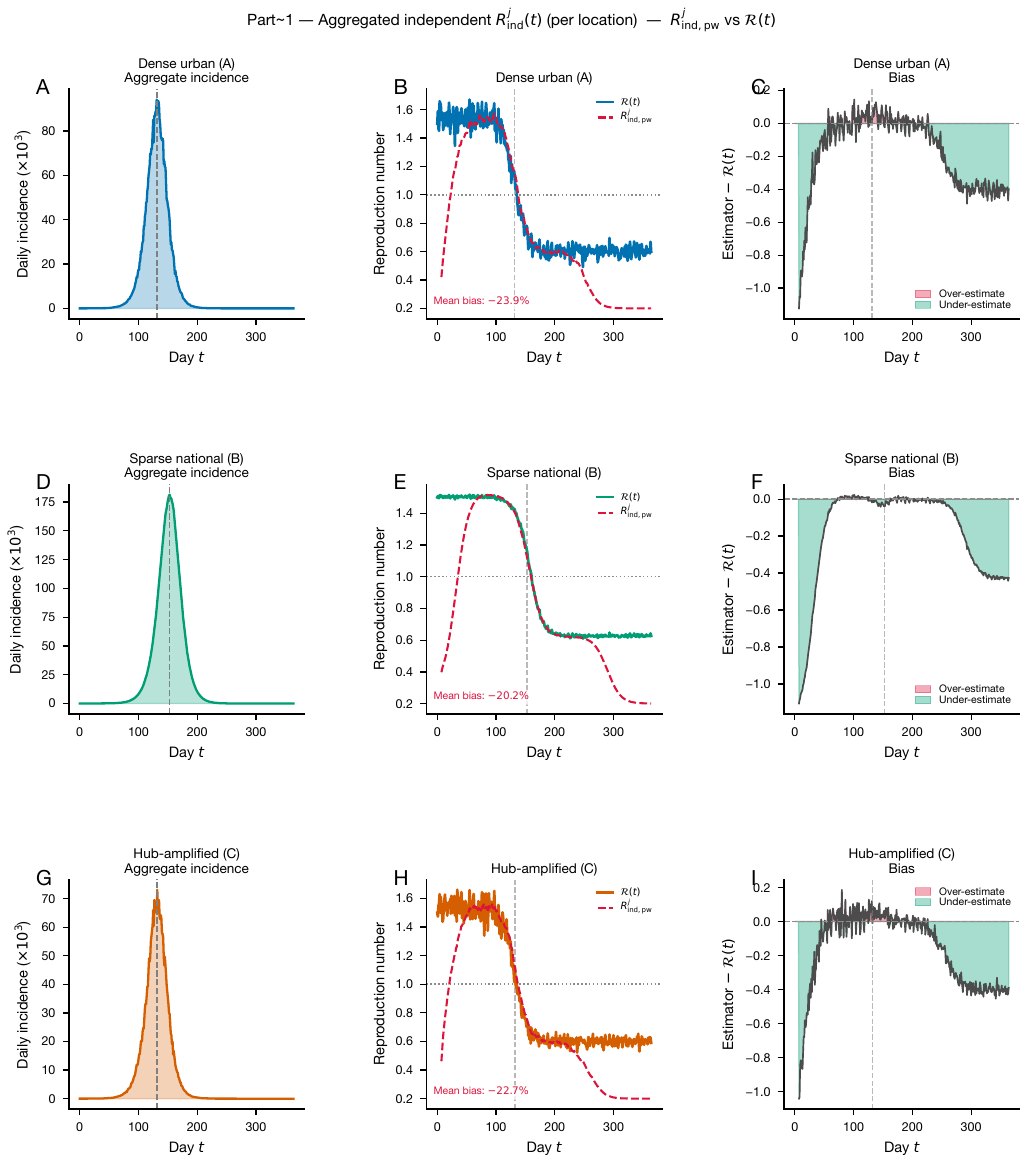}
\caption{\label{fig:pw_vs_Rt} \footnotesize \textbf{Aggregate $R(t)$ versus $\mathcal{R}(t)$} Comparison of the population-weighted reproduction number $R_{\text{pw}}(t)$ (the incidence-weighted average of independent, per-location estimates $R_{\text{ind}}^j(t)$, weighted by resident population) against our framework's network-level reproduction number $\mathcal{R}(t)$ for the dense-urban setting (Scenario A), sparse-national setting (Scenario B), and hub-amplified setting (Scenario C). }
\end{figure}

\begin{figure}[H]
\centering
\includegraphics[scale = 0.88]{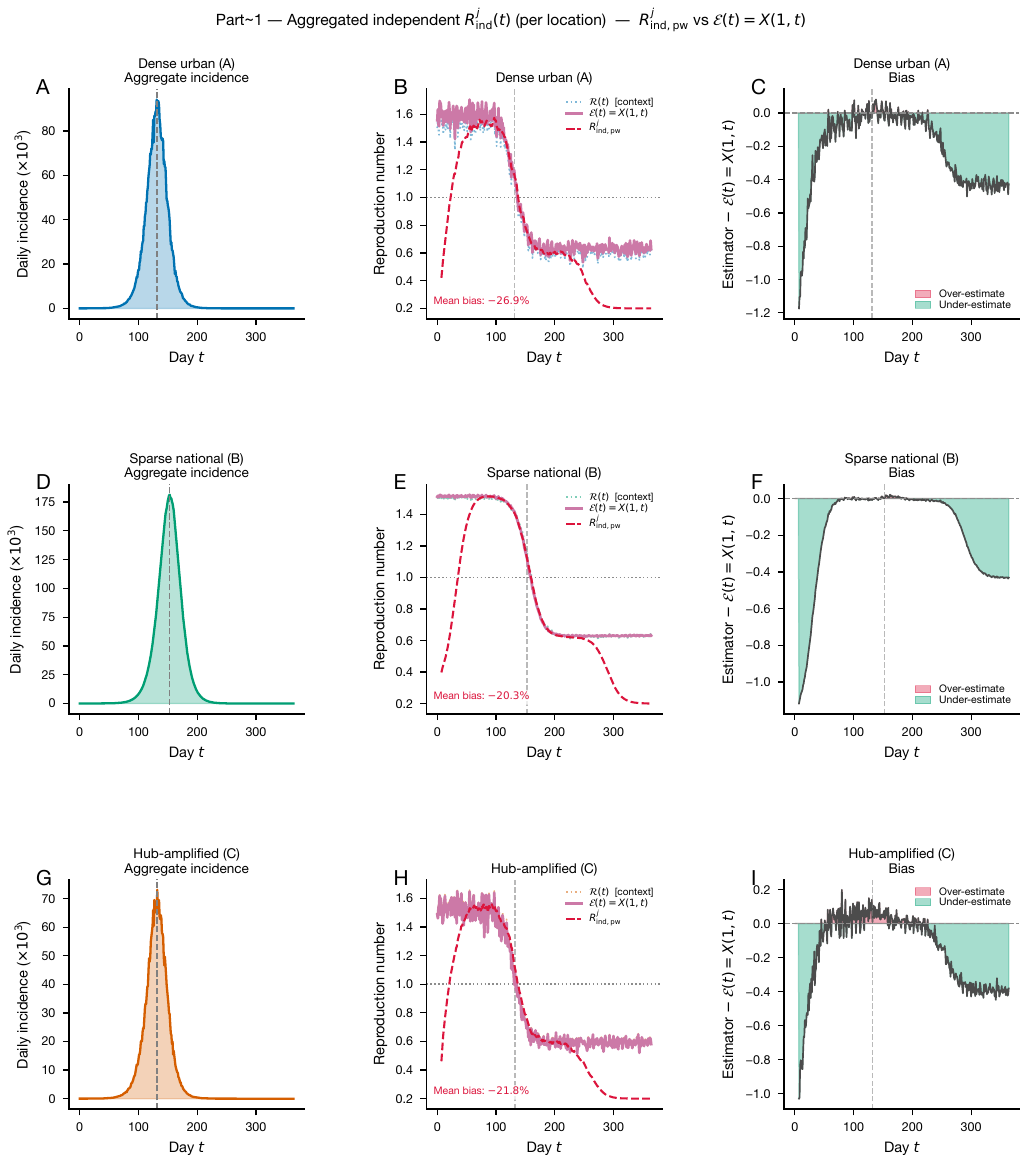}
\caption{\label{fig:pw_vs_Et} \footnotesize \textbf{Aggregate $R(t)$ versus $\mathcal{R}(t)$} Comparison of the population-weighted reproduction number $R_{\text{pw}}(t)$ against our framework's spatial risk-averse reproduction number $\mathcal{E}(t)$ for the dense-urban setting (Scenario A).}
\end{figure}
\FloatBarrier


\newpage
\begin{figure}[H]
\centering
\includegraphics[scale = 0.88]{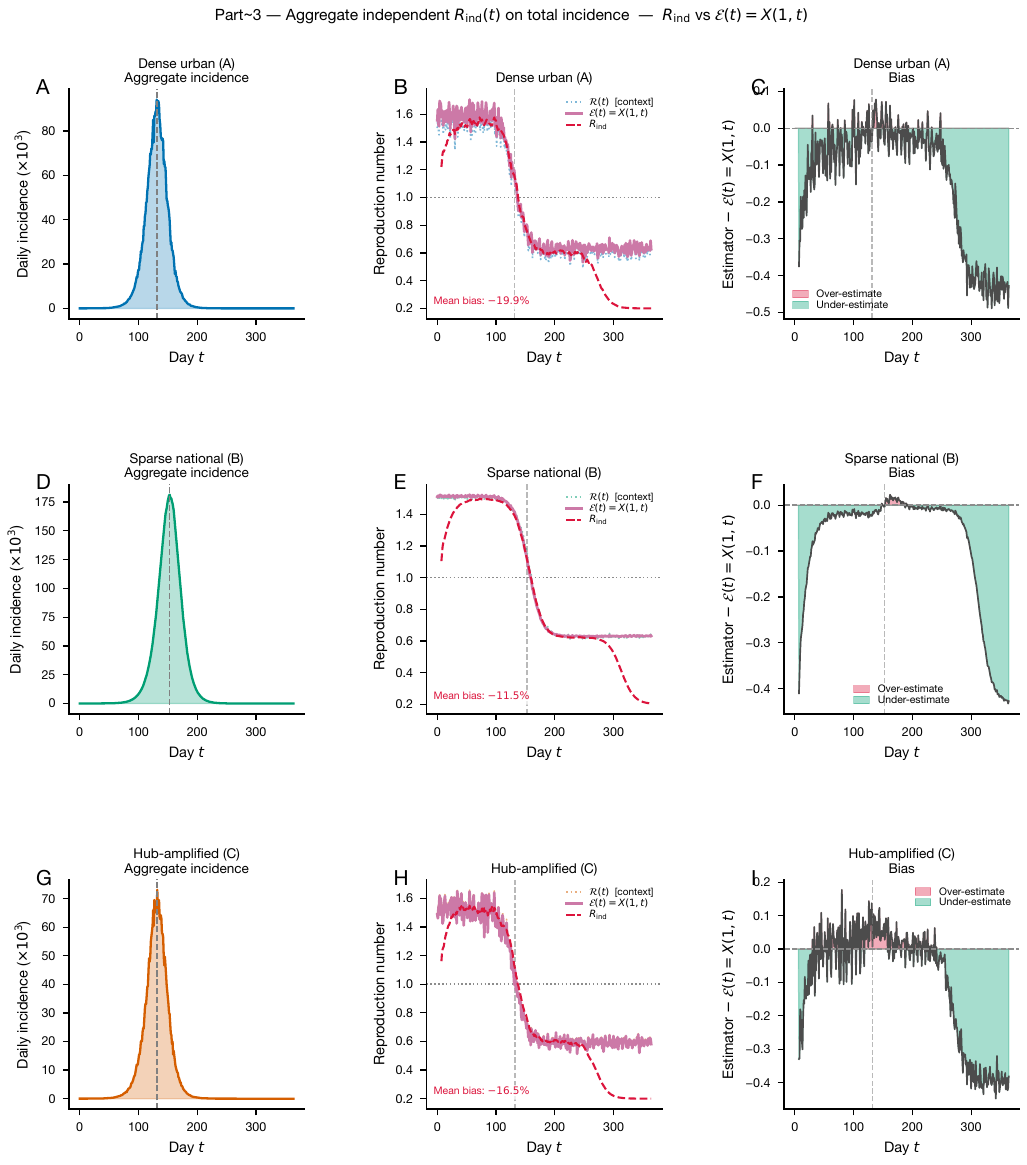}
\caption{\label{fig:aggregate_naive_vs_Et} \footnotesize \textbf{$R(t)$ from aggregate incidence versus $\mathcal{E}(t)$} Comparison of the reproduction number estimated from the network-aggregated infection incidence against our framework's spatial risk-averse reproduction number $\mathcal{E}(t)$ for the dense-urban setting (Scenario A).}
\end{figure}
\FloatBarrier

\newpage

\subsubsection{Arithmetic mean of independent, closed-population reproduction numbers}

\begin{figure}[H]
\centering
\includegraphics[scale = 0.88]{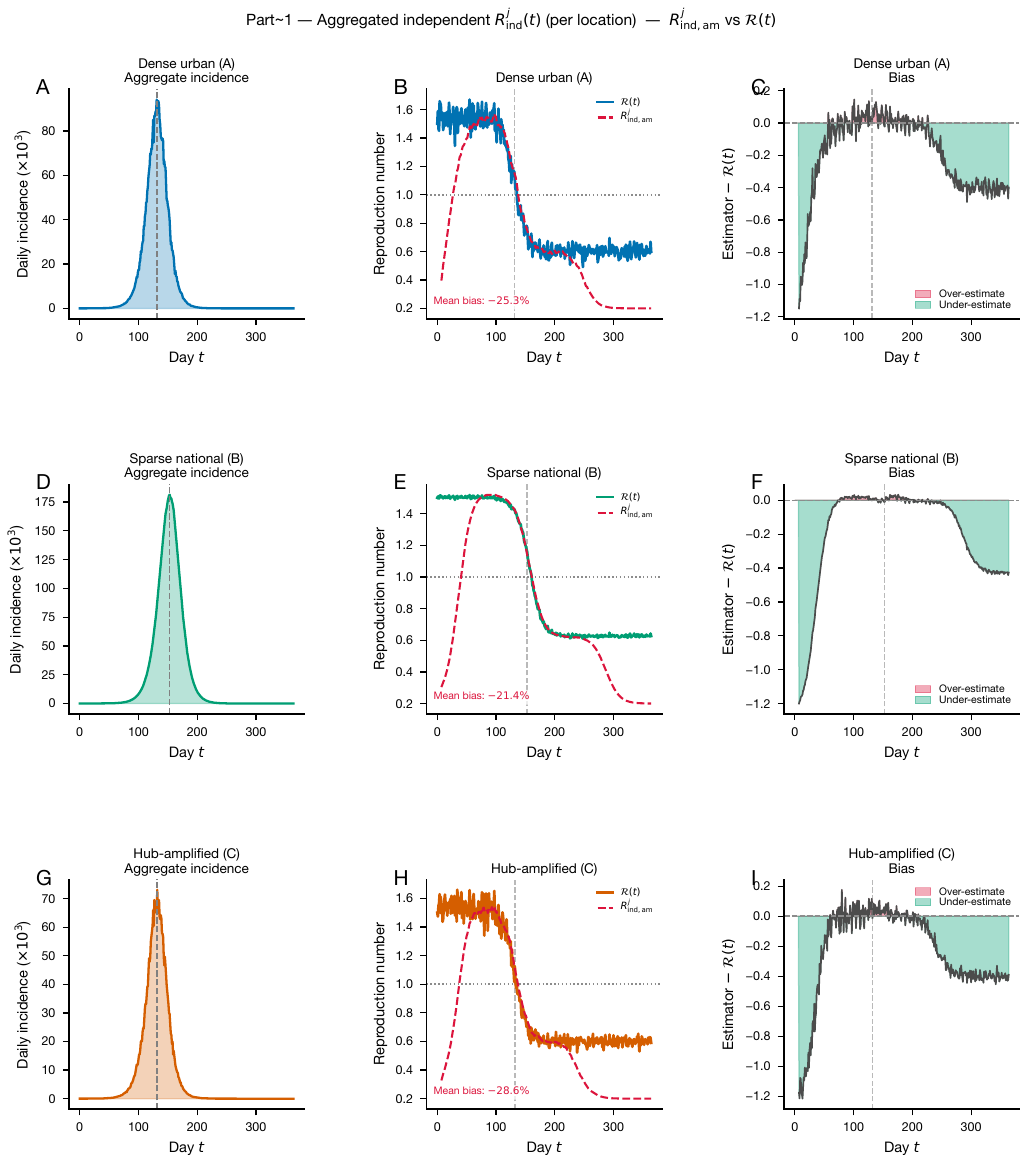}
\caption{\label{fig:naive_ind_am_vs_Rt} \footnotesize \textbf{Arithmetic mean of $R_{\text{ind}}(t)$ versus $\mathcal{R}(t)$} Comparison of the arithmetic mean across locations of the independent, closed-population reproduction numbers $R_{\text{ind}}^j(t)$ against our framework's network-level reproduction number $\mathcal{R}(t)$ for the dense-urban setting (Scenario A), sparse-national setting (Scenario B), and hub-amplified setting (Scenario C).  Simple averaging neglects spatial structure and human movement and generally yields biased estimates of $\mathcal{R}(t)$.}
\end{figure}
\FloatBarrier
\newpage

\begin{figure}[H]
\centering
\includegraphics[scale = 0.88]{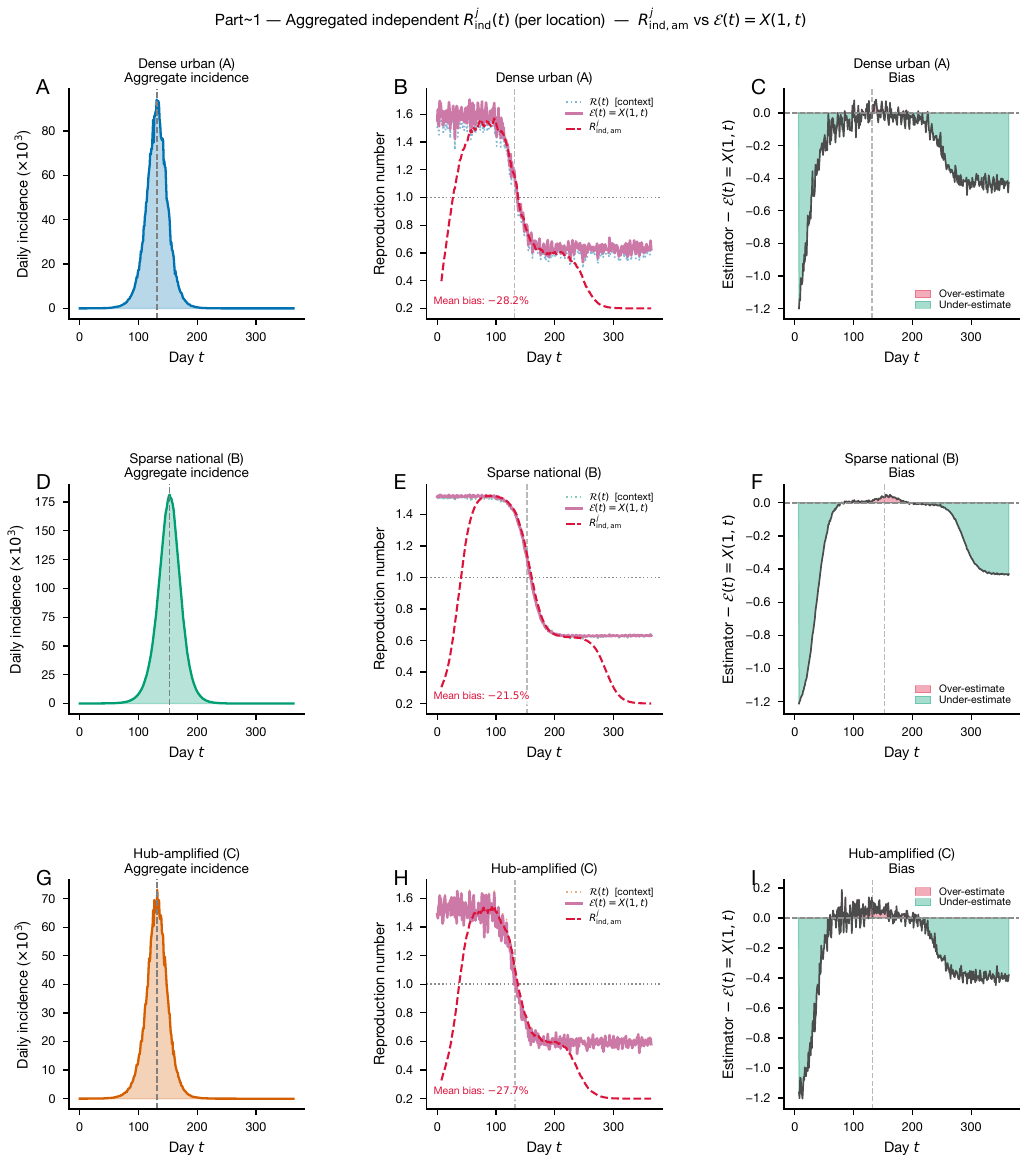}
\caption{\label{fig:naive_ind_am_vs_Et} \footnotesize \textbf{Arithmetic mean of $R_{\text{ind}}(t)$ versus $\mathcal{E}(t)$} As in Figure \ref{fig:naive_ind_am_vs_Rt}, but comparing the arithmetic mean of independent reproduction numbers $R_{\text{ind}}^j(t)$ against the spatial risk-averse reproduction number $\mathcal{E}(t)$ for the dense-urban setting (Scenario A).}
\end{figure}
\FloatBarrier
\newpage
\subsubsection{Incidence-weighted mean of independent, closed-population reproduction numbers}
\begin{figure}[H]
\centering
\includegraphics[scale = 0.88]{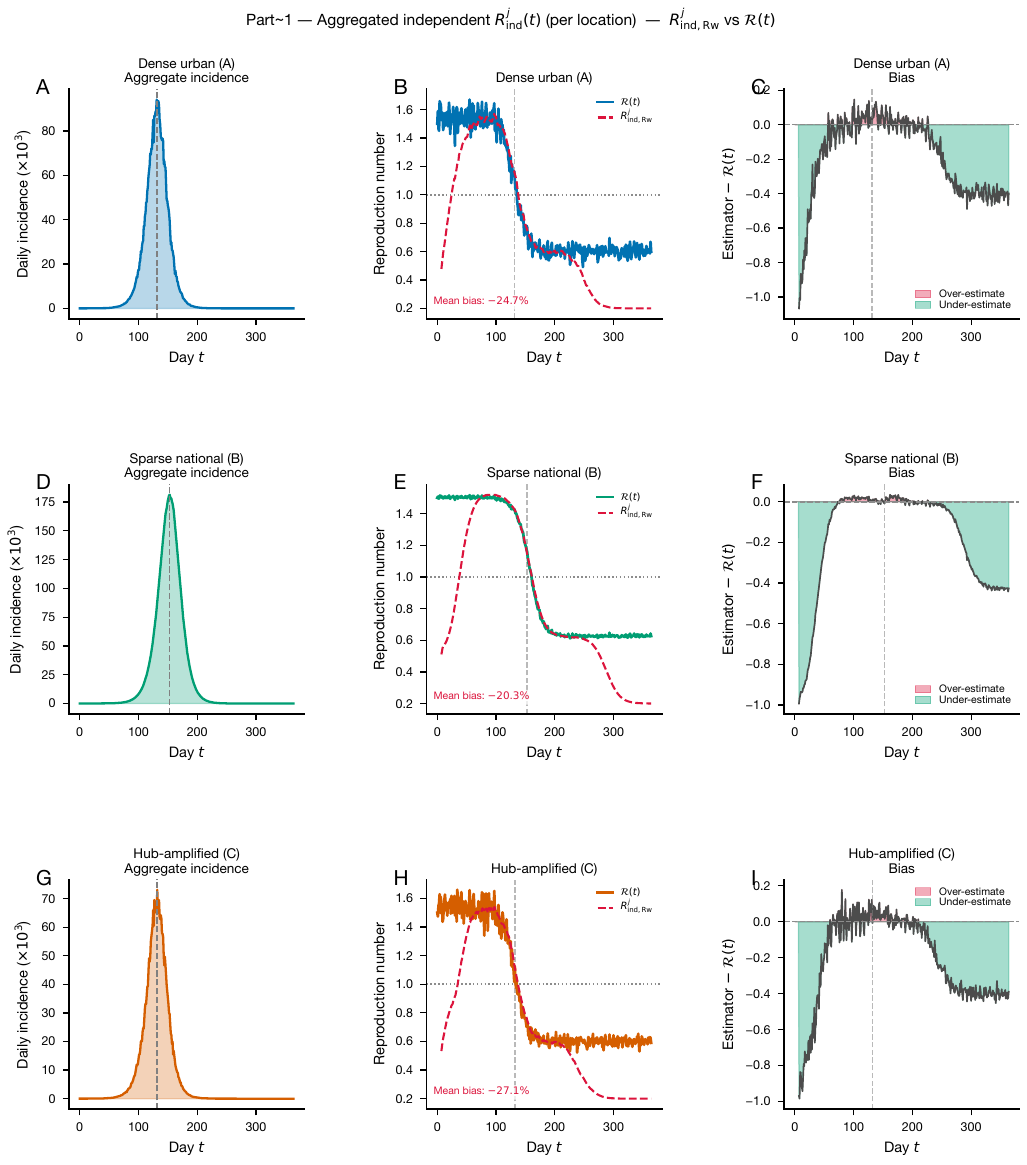}
\caption{\label{fig:naive_ind_Rw_vs_Rt} \footnotesize \textbf{Incidence-weighted $R_{\text{ind}}(t)$ versus $\mathcal{R}(t)$} Comparison of the incidence-weighted mean across locations of the independent, closed-population reproduction numbers $R_{\text{ind}}^j(t)$ against our framework's network-level reproduction number $\mathcal{R}(t)$ for the dense-urban setting (Scenario A), sparse-national setting (Scenario B), and hub-amplified setting (Scenario C). }
\end{figure}
\FloatBarrier
\newpage
\begin{figure}[H]
\centering
\includegraphics[scale = 0.88]{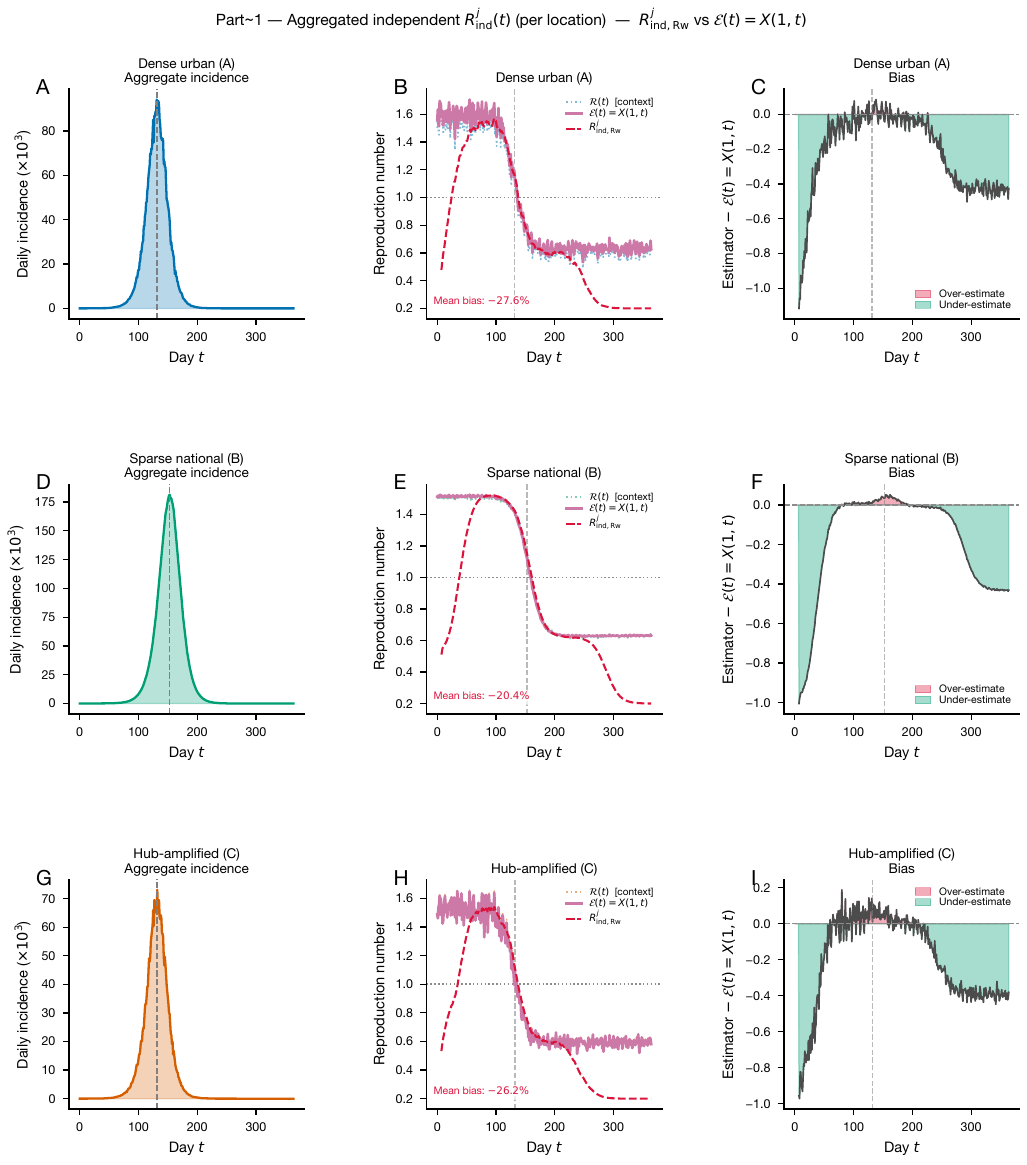}
\caption{\label{fig:naive_ind_Rw_vs_Et} \footnotesize \textbf{Incidence-weighted $R_{\text{ind}}(t)$ versus $\mathcal{E}(t)$} As in Figure \ref{fig:naive_ind_Rw_vs_Rt}, but comparing the incidence-weighted mean of independent reproduction numbers $R_{\text{ind}}^j(t)$ against the spatial risk-averse reproduction number $\mathcal{E}(t)$ for the dense-urban setting (Scenario A).}
\end{figure}
\FloatBarrier
\newpage
\subsubsection{Incidence-weighted mean of mobility-informed outward reproduction numbers}
\newpage
\begin{figure}[H]
\centering
\includegraphics[scale = 0.88]{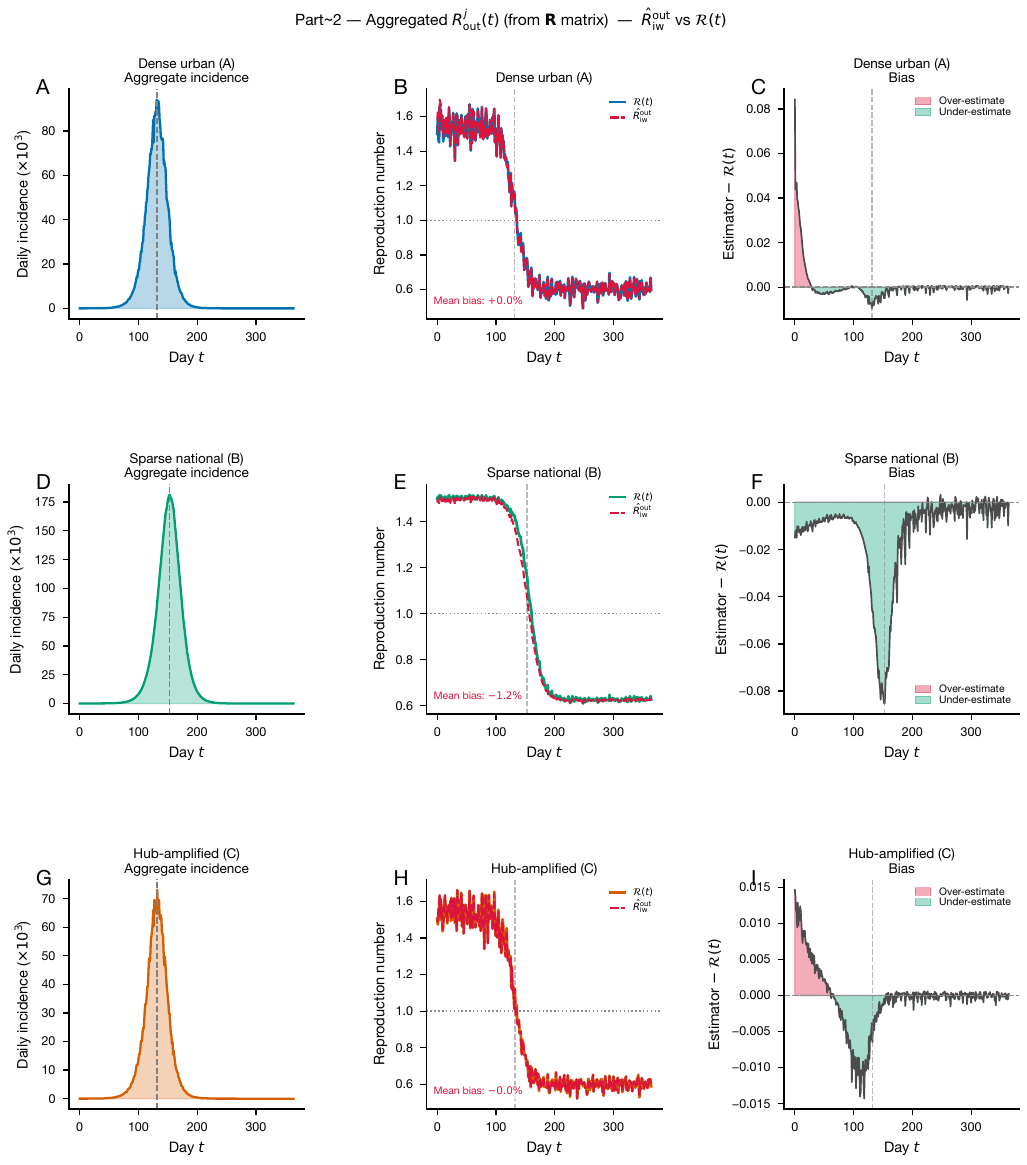}
\caption{\label{fig:naive_out_iw_vs_Rt} \footnotesize \textbf{Incidence-weighted $R_{\text{out}}(t)$ versus $\mathcal{R}(t)$} Comparison of the incidence-weighted mean across locations of the mobility-informed outward reproduction numbers $R_{\text{out}}^j(t)$ against our framework's network-level reproduction number $\mathcal{R}(t)$ for the dense-urban setting (Scenario A), sparse-national setting (Scenario B), and hub-amplified setting (Scenario C). }
\end{figure}
\FloatBarrier

\begin{figure}[H]
\centering
\includegraphics[scale = 0.88]{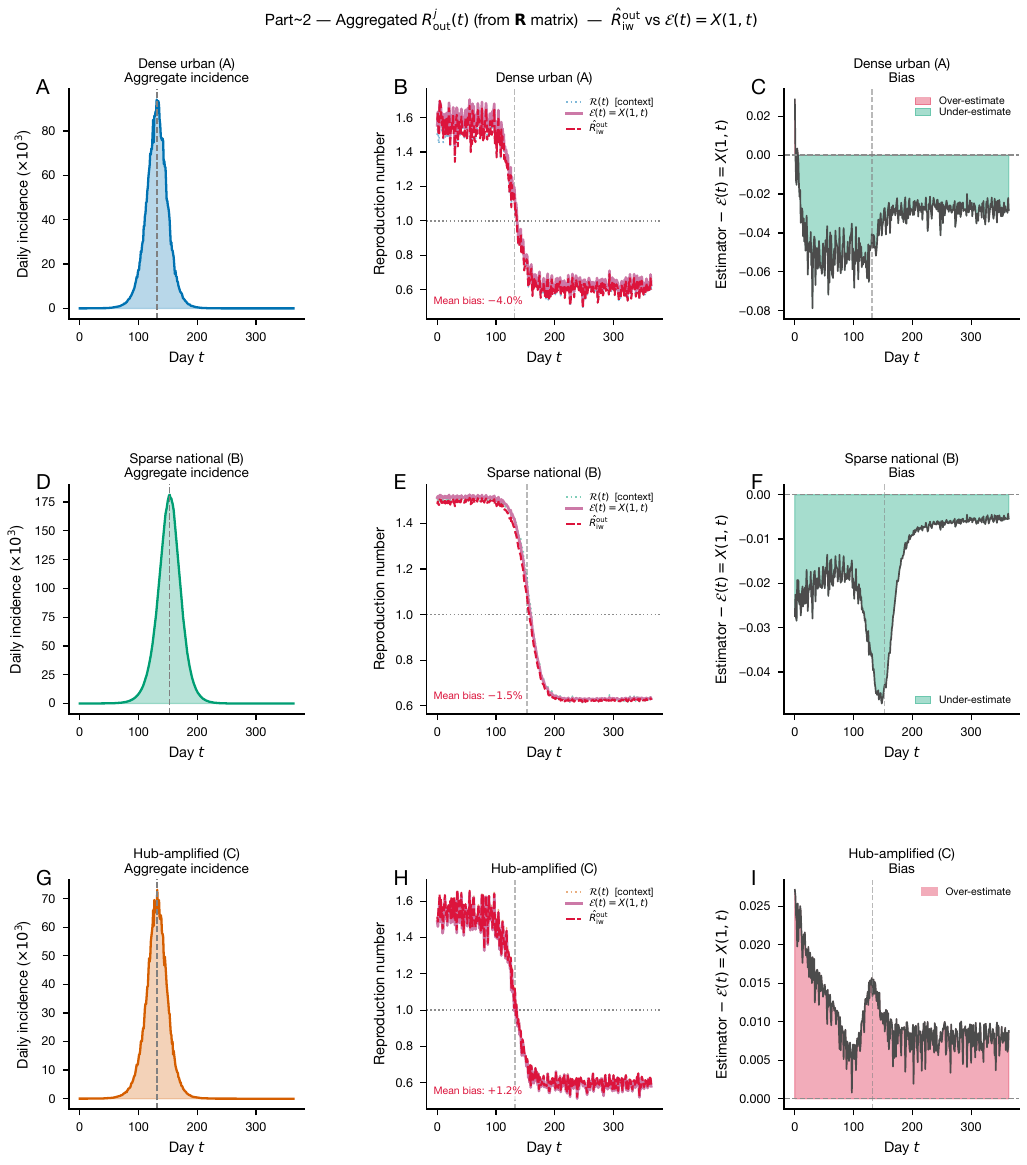}
\caption{\label{fig:naive_out_iw_vs_Et} \footnotesize \textbf{Incidence-weighted $R_{\text{out}}(t)$ versus $\mathcal{E}(t)$} As in Figure \ref{fig:naive_out_iw_vs_Rt}, but comparing the incidence-weighted mean of outward reproduction numbers $R_{\text{out}}^j(t)$ against the spatial risk-averse reproduction number $\mathcal{E}(t)$ for the dense-urban setting (Scenario A).}
\end{figure}
\newpage
\subsubsection{Arithmetic mean of mobility-informed outward reproduction numbers}
\begin{figure}[H]
\centering
\includegraphics[scale = 0.88]{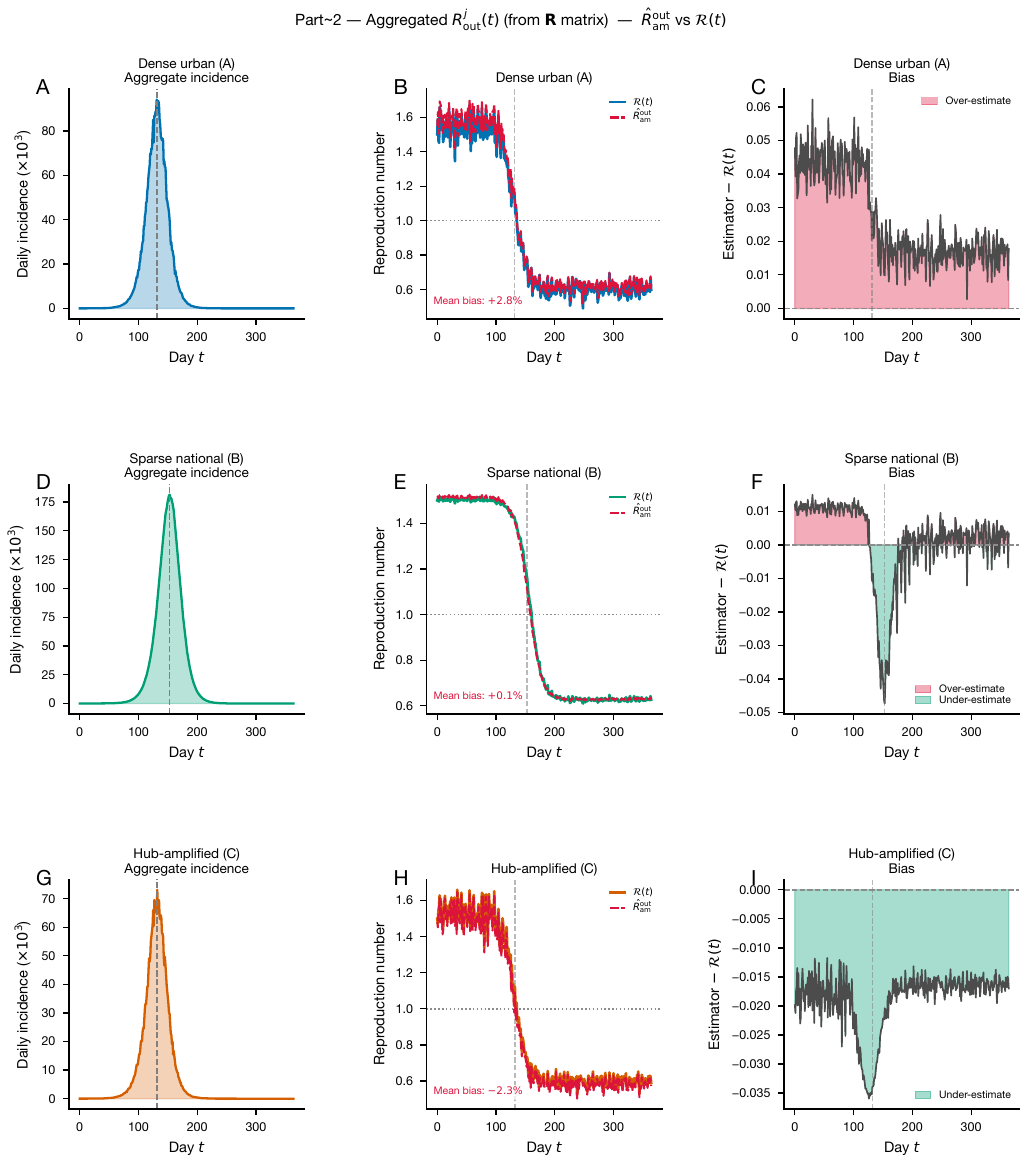}
\caption{\label{fig:naive_out_am_vs_Rt} \footnotesize \textbf{Arithmetic mean of $R_{\text{out}}(t)$ versus $\mathcal{R}(t)$} Comparison of the arithmetic mean across locations of the mobility-informed outward reproduction numbers $R_{\text{out}}^j(t)$ against our framework's network-level reproduction number $\mathcal{R}(t)$ for the dense-urban setting (Scenario A), sparse-national setting (Scenario B), and hub-amplified setting (Scenario C). }
\end{figure}
\FloatBarrier

\begin{figure}[H]
\centering
\includegraphics[scale = 0.88]{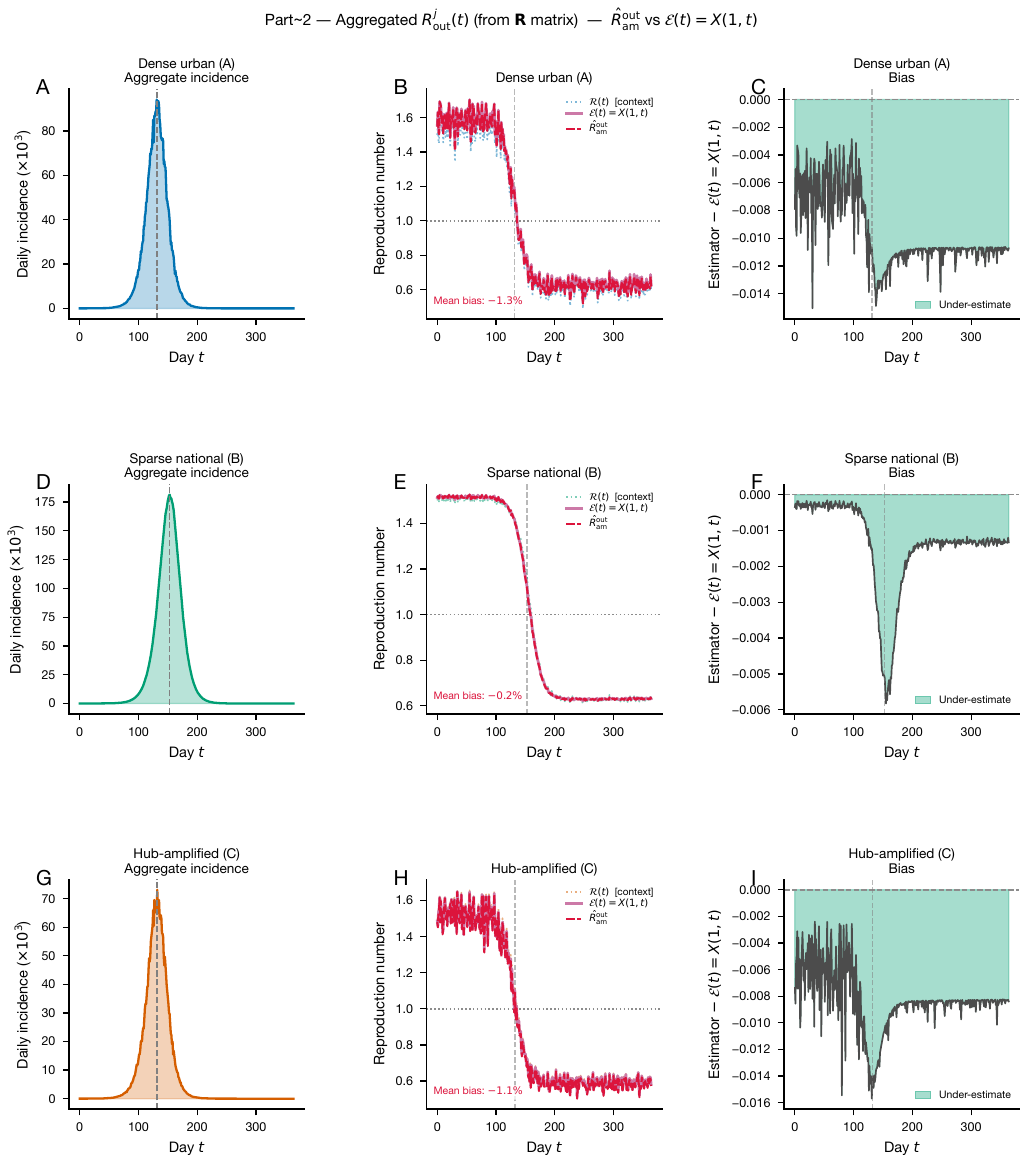}
\caption{\label{fig:naive_out_am_vs_Et} \footnotesize \textbf{Arithmetic mean of $R_{\text{out}}(t)$ versus $\mathcal{E}(t)$} As in Figure \ref{fig:naive_out_am_vs_Rt}, but comparing the arithmetic mean of outward reproduction numbers $R_{\text{out}}^j(t)$ against the spatial risk-averse reproduction number $\mathcal{E}(t)$ for the dense-urban setting (Scenario A), sparse-national setting (Scenario B), and hub-amplified setting (Scenario C).}
\end{figure}
\FloatBarrier

\subsubsection{Relative transmissibility weighting of independent, closed-population reproduction numbers}

\begin{figure}[H]
\centering
\includegraphics[scale = 0.88]{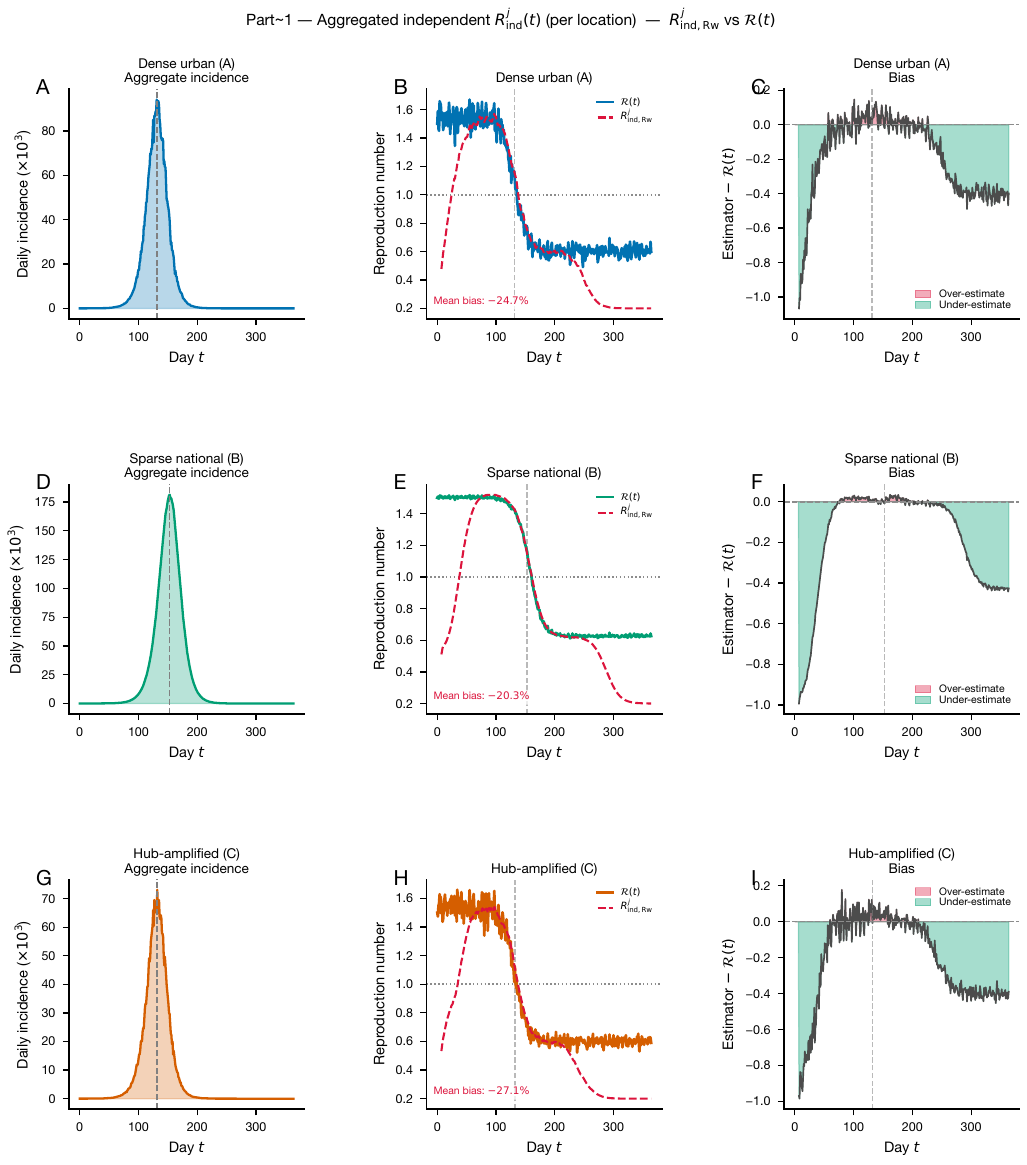}
\caption{\label{fig:naive_ind_relW_vs_Rt} \footnotesize \textbf{Relative transmissibility weighting of $R_{\text{ind}}(t)$ versus $\mathcal{R}(t)$} Comparison of the relative-transmissibility-weighted mean across locations of the independent, closed-population reproduction numbers $R_{\text{ind}}^j(t)$ (each weighted by $R_{\text{ind}}^j(t)/\sum_k R_{\text{ind}}^k(t)$) against our framework's network-level reproduction number $\mathcal{R}(t)$ for the dense-urban setting (Scenario A), sparse-national setting (Scenario B), and hub-amplified setting (Scenario C). }
\end{figure}
\FloatBarrier

\newpage
\begin{figure}[H]
\centering
\includegraphics[scale = 0.88]{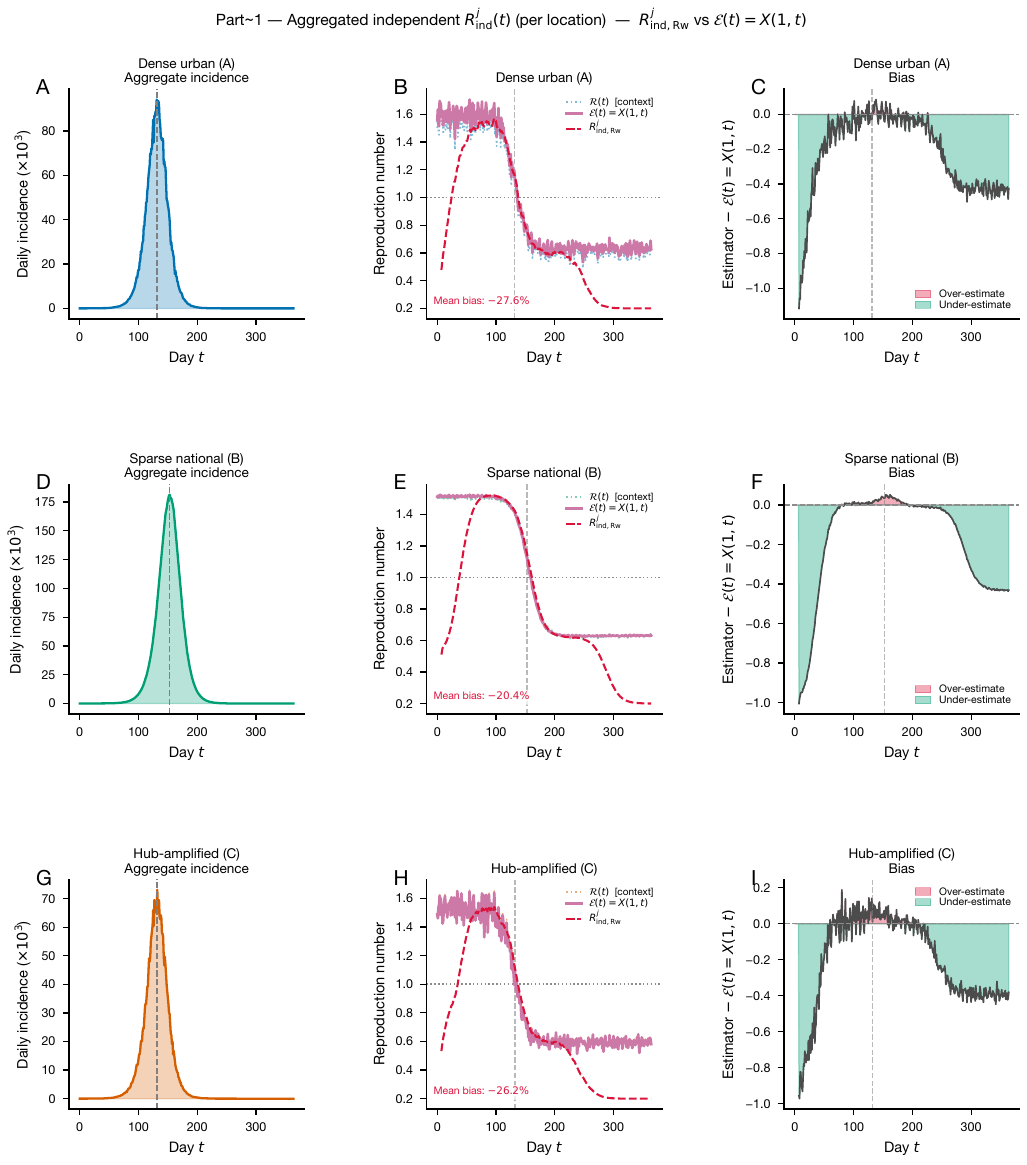}
\caption{\label{fig:naive_ind_relW_vs_Et} \footnotesize \textbf{Relative transmissibility weighting of $R_{\text{ind}}(t)$ versus $\mathcal{E}(t)$} As in Figure \ref{fig:naive_ind_relW_vs_Rt}, but comparing the relative-transmissibility-weighted mean of independent reproduction numbers $R_{\text{ind}}^j(t)$ against the spatial risk-averse reproduction number $\mathcal{E}(t)$ for the dense-urban setting (Scenario A), sparse-national setting (Scenario B), and hub-amplified setting (Scenario C).}
\end{figure}
\FloatBarrier

\subsubsection{Relative transmissibility weighting of mobility-informed outward reproduction numbers}

\begin{figure}[H]
\centering
\includegraphics[scale = 0.88]{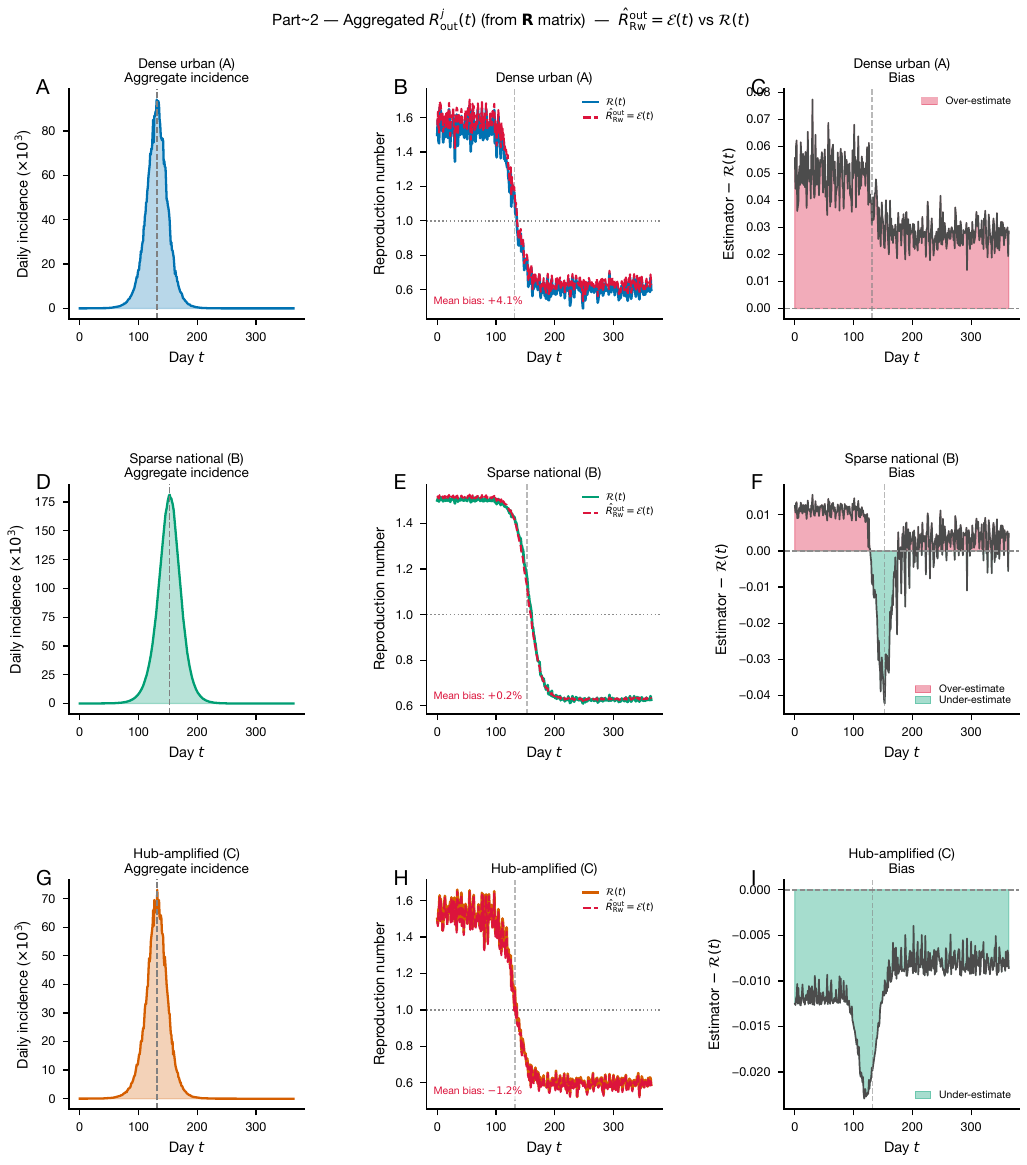}
\caption{\label{fig:naive_out_relW_vs_Rt} \footnotesize \textbf{Relative transmissibility weighting of $R_{\text{out}}(t)$, i.e. $\mathcal{E}(t)$, versus $\mathcal{R}(t)$} Comparison of the relative-transmissibility-weighted mean across locations of the mobility-informed outward reproduction numbers $R_{\text{out}}^j(t)$ (which yields the spatial risk-averse reproduction number $\mathcal{E}(t)$) against our framework's network-level reproduction number $\mathcal{R}(t)$ for the dense-urban setting (Scenario A), sparse-national setting (Scenario B), and hub-amplified setting (Scenario C). }
\end{figure}
\newpage
\subsubsection{Per-location reproduction numbers comparison}
\newpage
\begin{figure}[H]
\centering
\includegraphics[scale = 0.88]{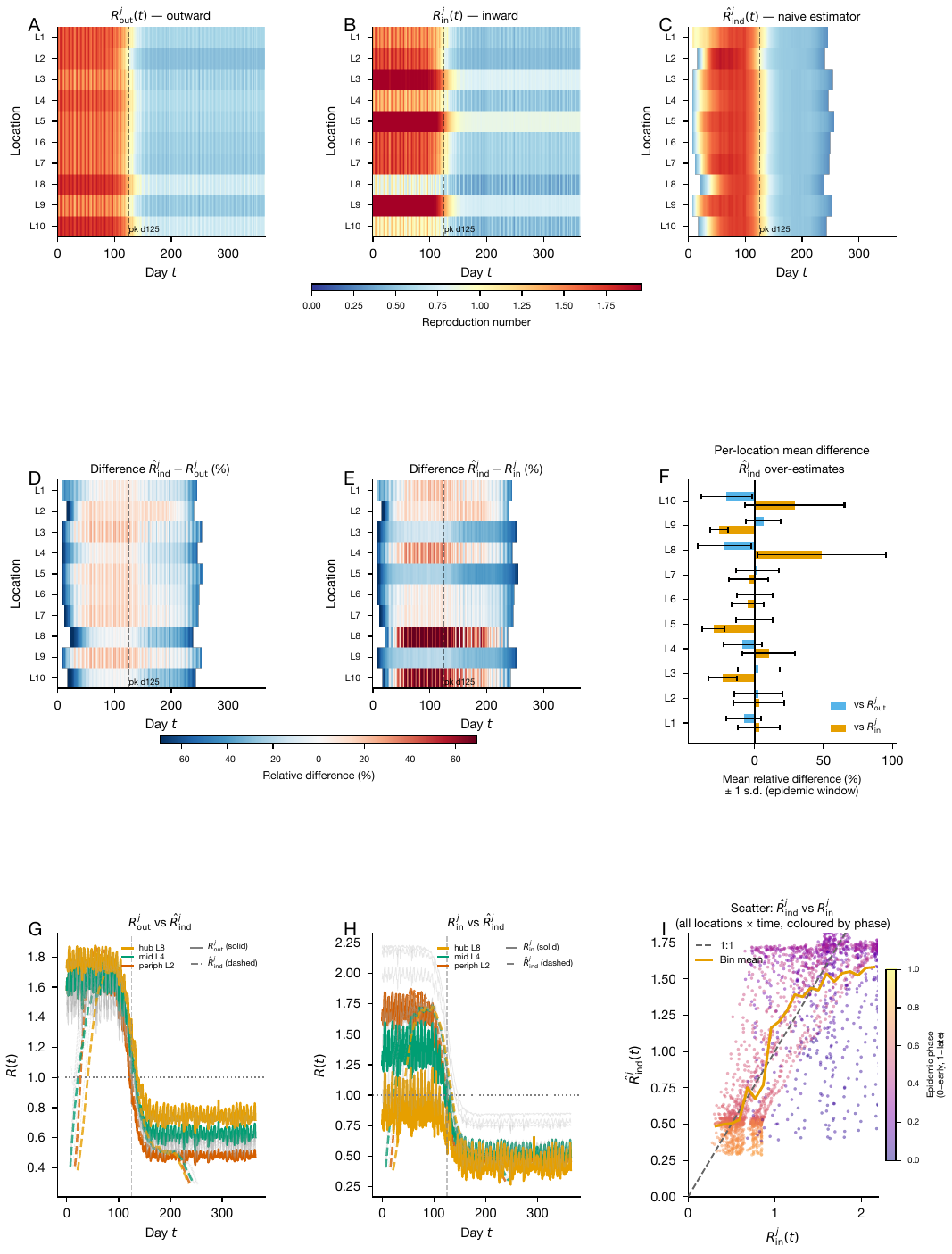}
\caption{\label{fig:SI_R_comparison} \footnotesize \textbf{Biasing effects of independent $R(t)$ estimators compared to inward and outward reproduction numbers:} \textbf{A)}, \textbf{B)}, and \textbf{C)} visualise the outward, inward, and independent estimates ($R_{\text{out}}^j(t), R_{\text{in}}^j(t),$ and $R_{\text{ind}}^j(t)$ respectively) over time for Scenario A. \textbf{D)}, and \textbf{E)} visualise the differences between $R_{\text{out}}^j(t)$ and $R_{\text{ind}}^j(t)$ and between $R_{\text{in}}^j(t)$ and $R_{\text{ind}}^j(t)$ respectively over time, with \textbf{F)} providing a per-location, aggregate summary of these differences over all time points. \textbf{G)}, and \textbf{H)} visualises, for three different classes of locations/nodes, $R_{\text{out}}^j(t)$ alongside $R_{\text{ind}}^j(t)$ and $R_{\text{in}}^j(t)$ alongside $R_{\text{ind}}^j(t)$ respectively over time. \textbf{L)} visualises values of $R_{\text{ind}}^j(t)$ (y-axis) against $R_{\text{in}}^j(t)$ (x-axes). See Tables \ref{tab:framework_outputs} -- \ref{tab:R_operational} for guidance on definitions, interpretations, and usage of our framework's quantities.  }
\end{figure}

\FloatBarrier


\subsection{Meeting location reproduction numbers}
\newpage
\begin{figure}[H]
\centering
\includegraphics[scale = 0.88]{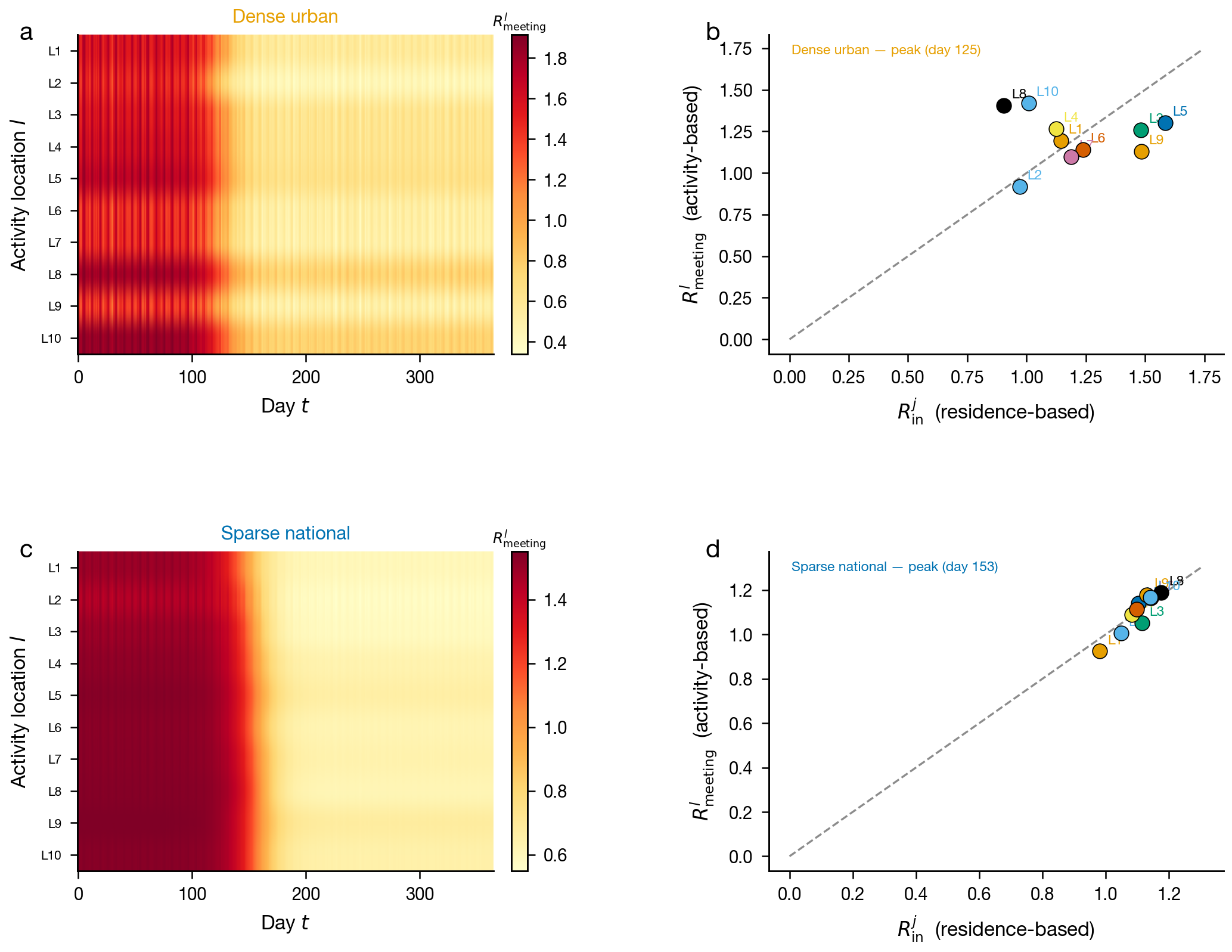}
\caption{\label{fig:SI5_meeting_combined} \footnotesize \textbf{Epidemic dynamics at meeting locations for Scenarios A (top) and B (bottom):} \textbf{A)} and \textbf{C)} visualise the meeting location reproduction number $R_{\text{meeting}}^l(t)$ for each location $l$ in Scenarios A and B respectively over time $t$. \textbf{B)} and \textbf{D)} compare $R_{\text{meeting}}^l(t)$ against $R_{\text{in}}^l(t)$ for each location at the peak of each epidemic in Scenarios A and B respectively.}
\end{figure}

\newpage


\subsection{Type reproduction numbers}
\begin{figure}[H]
\centering
\includegraphics[scale = 0.55]{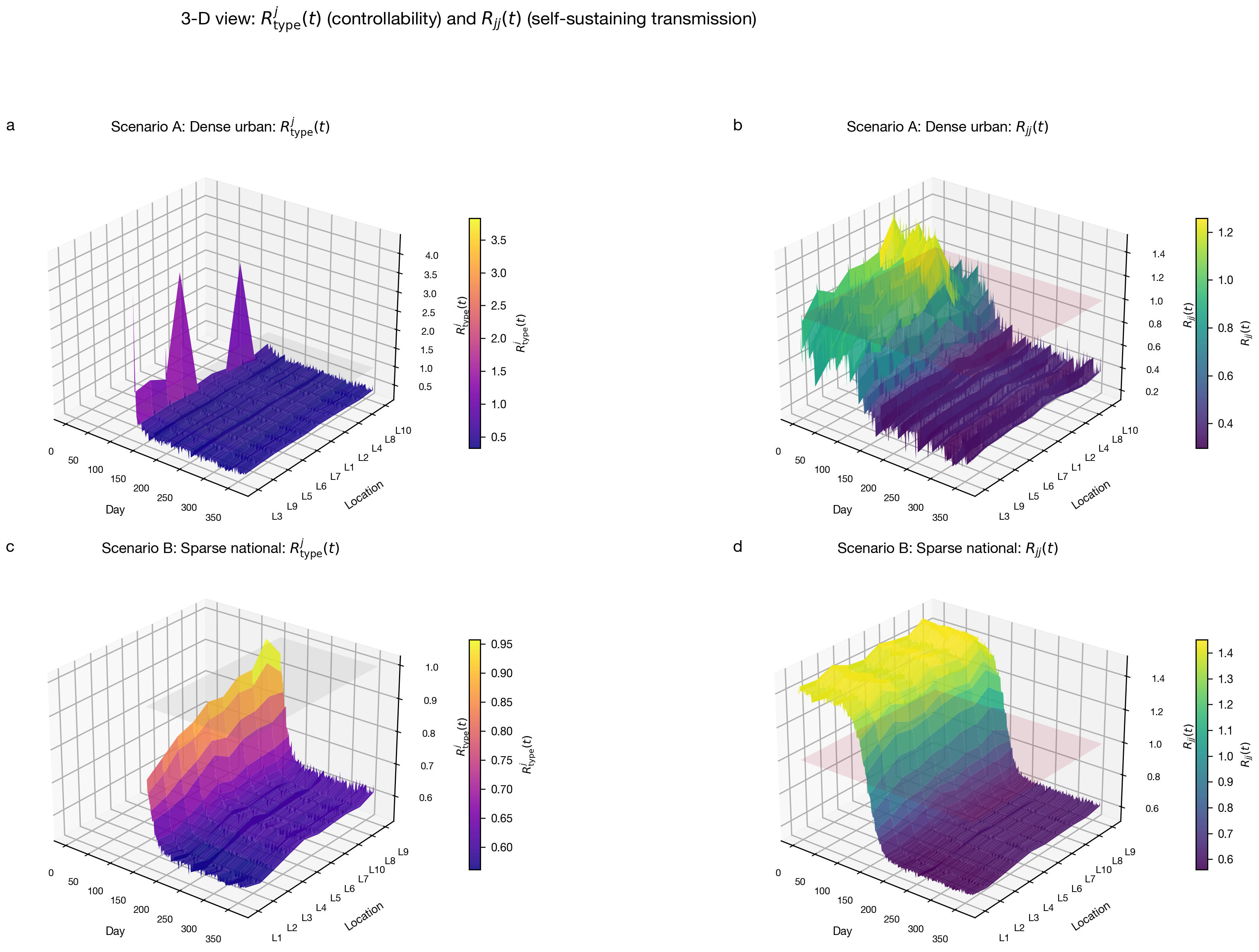}
\caption{\label{fig:type_surfaces} \footnotesize \textbf{Surface plots for type reproduction numbers and local, within-location reproduction numbers for Scenarios A (top) and B (below) :} \textbf{A)} and \textbf{C)} visualises the type reproduction numbers $R_{\text{type}^j}(t)$ (when defined) over time for each location in Scenarios A and B respectively. \textbf{B)} and \textbf{D)} visualises the within-location/local reproduction numbers $R_{jj}(t)$ over time for each location in Scenarios A and B respectively.}
\end{figure}
\FloatBarrier
\newpage
\begin{figure}[H]
\centering
\includegraphics[scale = 0.4]{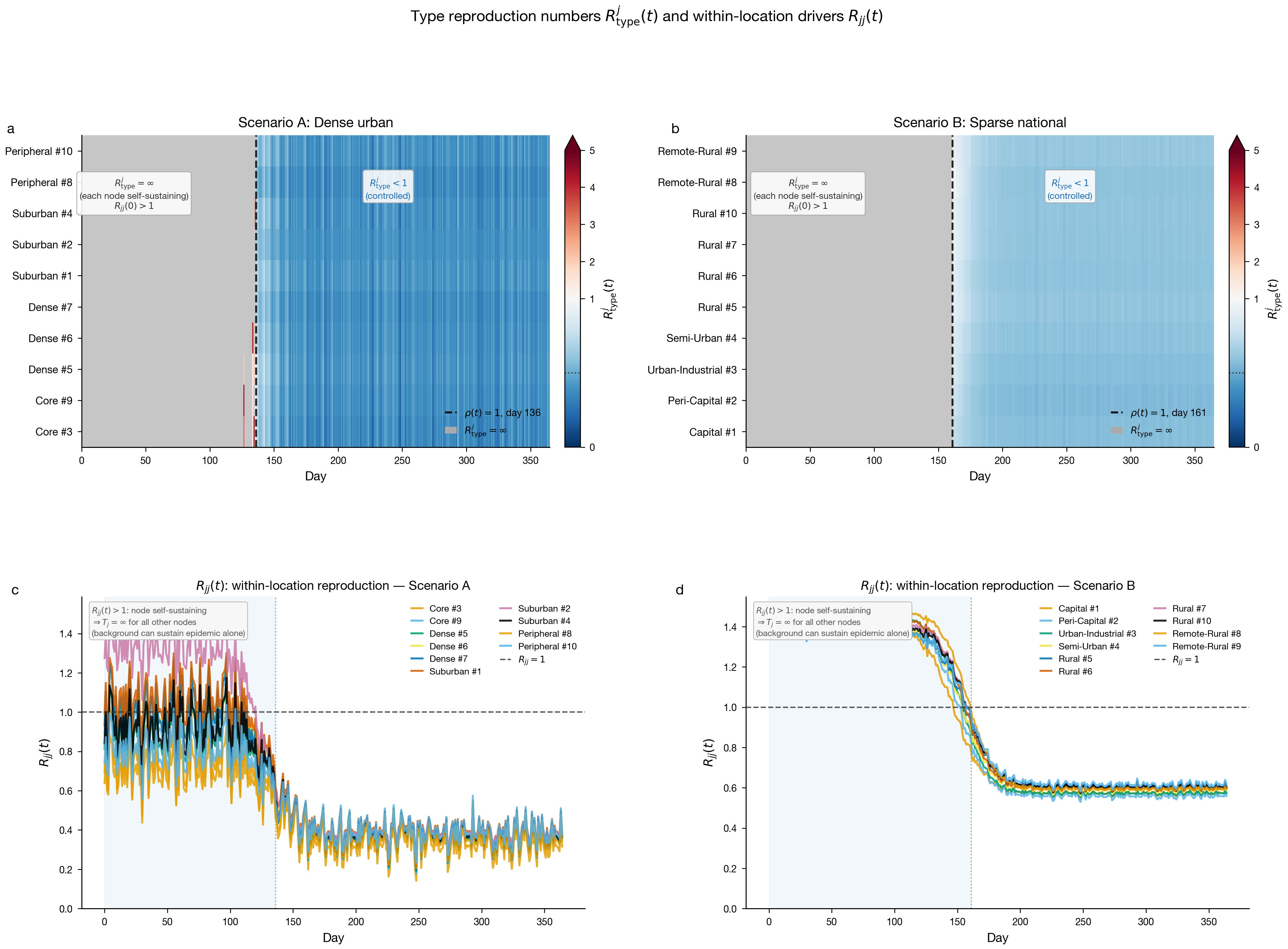}
\caption{\label{fig:type_heatmaps} \footnotesize \textbf{Heatmaps and line plots for type reproduction numbers and local, within-location reproduction numbers for Scenarios A (left) and B (right):} textbf{A)} and \textbf{B)} visualises the type reproduction numbers $R_{\text{type}^j}(t)$ (when defined) over time for each location in Scenarios A and B respectively. \textbf{C)} and \textbf{D)} visualises the within-location/local reproduction numbers $R_{jj}(t)$ over time for each location in Scenarios A and B respectively.}
\end{figure}
\FloatBarrier

\newpage
\begin{figure}[H]
\centering
\includegraphics[scale = 0.4]{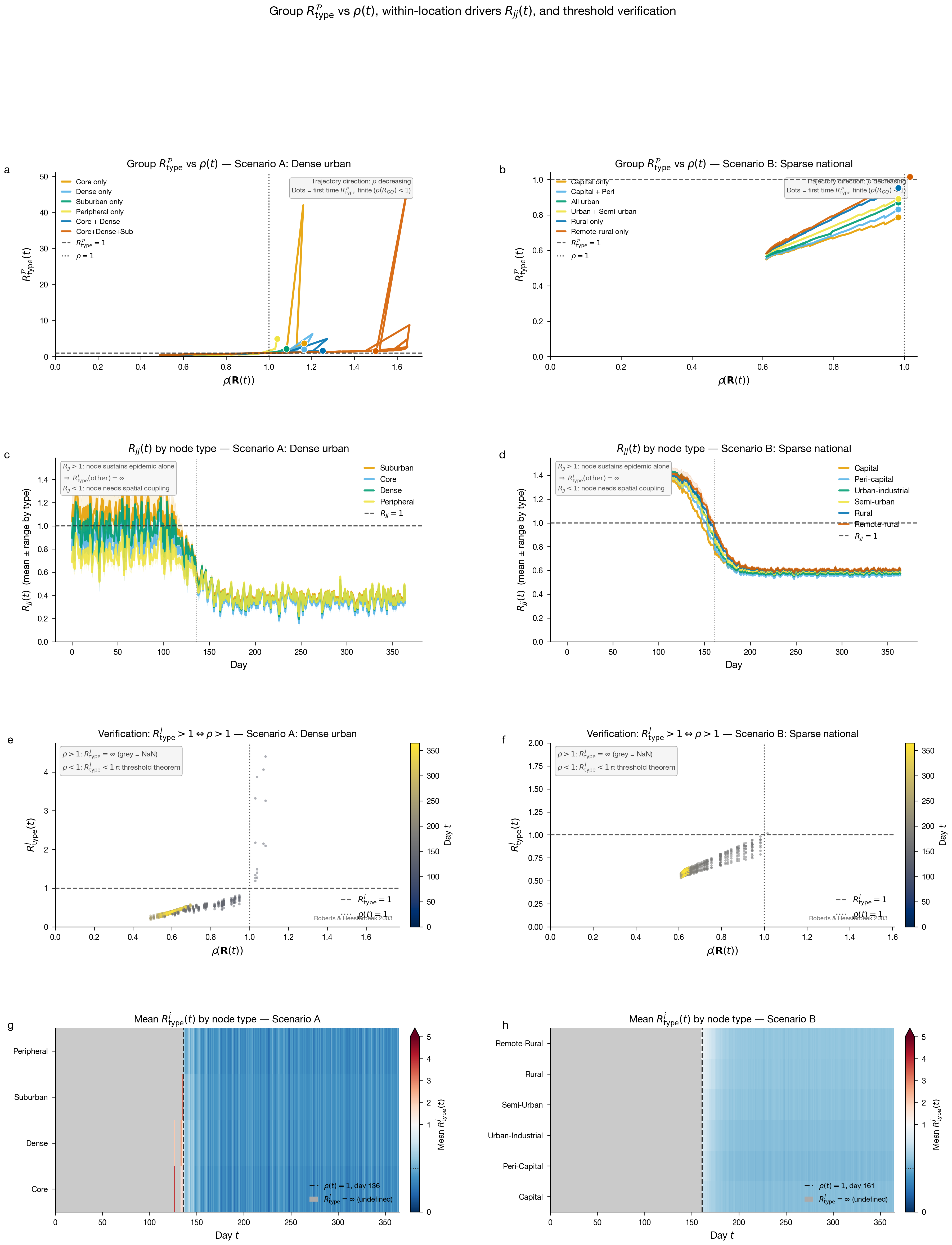}
\caption{\label{fig:type_groups} \footnotesize \textbf{Type reproduction numbers for groups of locations:} The group type reproduction number $R_{\text{type}}^{\mathcal{P}}(t)$ over time for a group of locations $\mathcal{P}$ (here, the core/densely populated locations) in Scenario A, treated jointly as a single type. The group type reproduction number quantifies the minimal control effort required across the group of locations to bring network-level epidemic growth under control, and can be defined at epidemic stages where the individual-location type reproduction numbers $R_{\text{type}}^{j}(t)$ are undefined.}
\end{figure}
\FloatBarrier
\newpage

\subsection{Coarse-national results (Scenario B)}
\begin{figure}[H]
\centering
\includegraphics[scale = 0.88]{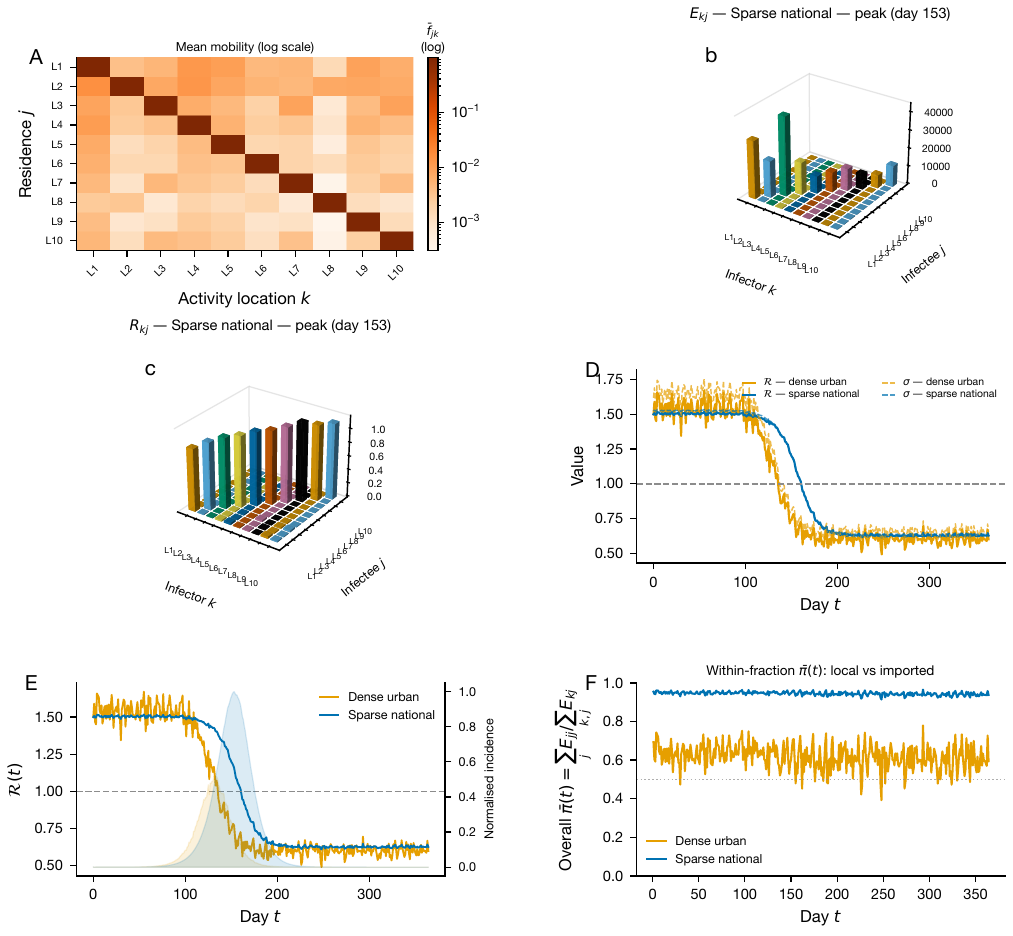}
\caption{\label{fig:05_comparison} \footnotesize \textbf{Comparison analysis of dynamics across different human movement patterns:} A) As a counterfactual analysis, sparse national movement dynamics are displayed (on a logarithmic scale) using the mean probabilities that resident of location $j$ are in location $k$. B) and C) display the between-location incidence $E_{kj}$ and reproduction numbers $R^{kj}$. D) Reactivity $\sigma(t)$ and network-level $\mathcal{R}(t)$ over time for both dense urban and sparse national settings. E) Network-level $\mathcal{R}(t)$ and incidence (normaliseD) over time for both dense urban and sparse national settings. F) Proportion of all incident infections that are generated locally between  residents of the same location over time for both dense urban and sparse national settings. See Tables \ref{tab:framework_outputs} -- \ref{tab:R_operational} for guidance on definitions, interpretations, and usage.}
\end{figure}
\newpage

\subsection*{Results in hub-amplified mobility network (Scenario C)}

\begin{figure}[H]
\centering
\includegraphics[scale = 0.88]{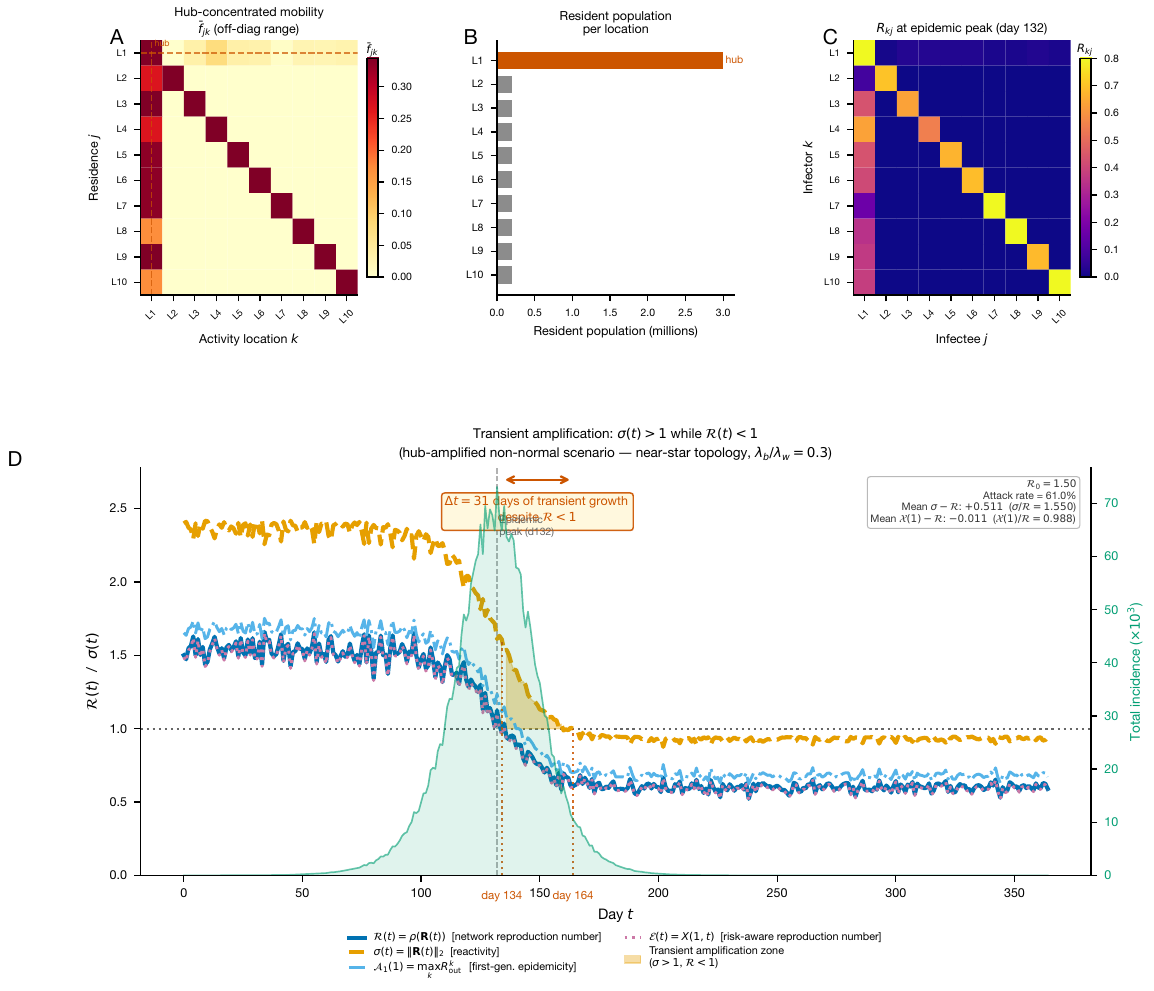}
\caption{\label{fig:scenario_C_transience} \footnotesize \textbf{Results for highly non-normal/heterogeneous (Scenario C) human movement patterns in a hub-amplified setting} \textbf{A)} visualises the time-averaged probability that residents of location $j$ are in location $k$, with row sums equal to 1.  \textbf{B)} captures the populations resident in each location. \textbf{C)} captures the between-location reproduction numbers $R^{kj}(t)$ at the epidemic incidence peak (day 132). \textbf{D)} visualises the infection incidence $E(t, 0)$, network-level reproduction number $\mathcal{R}(t)$, spatial risk-averse reproduction number $\mathcal{E}(t) = X(\alpha = 1, t)$, reactivity $\sigma(t)$, and first-generation epidemicity $A_1(1) = \max_{k} R_{\text{out}}^k(t)$. See Tables \ref{tab:framework_outputs} -- \ref{tab:R_operational} for guidance on definitions, interpretations, and usage.}
\end{figure}
\FloatBarrier

\newpage
\begin{figure}[H]
\centering
\includegraphics[scale = 0.88]{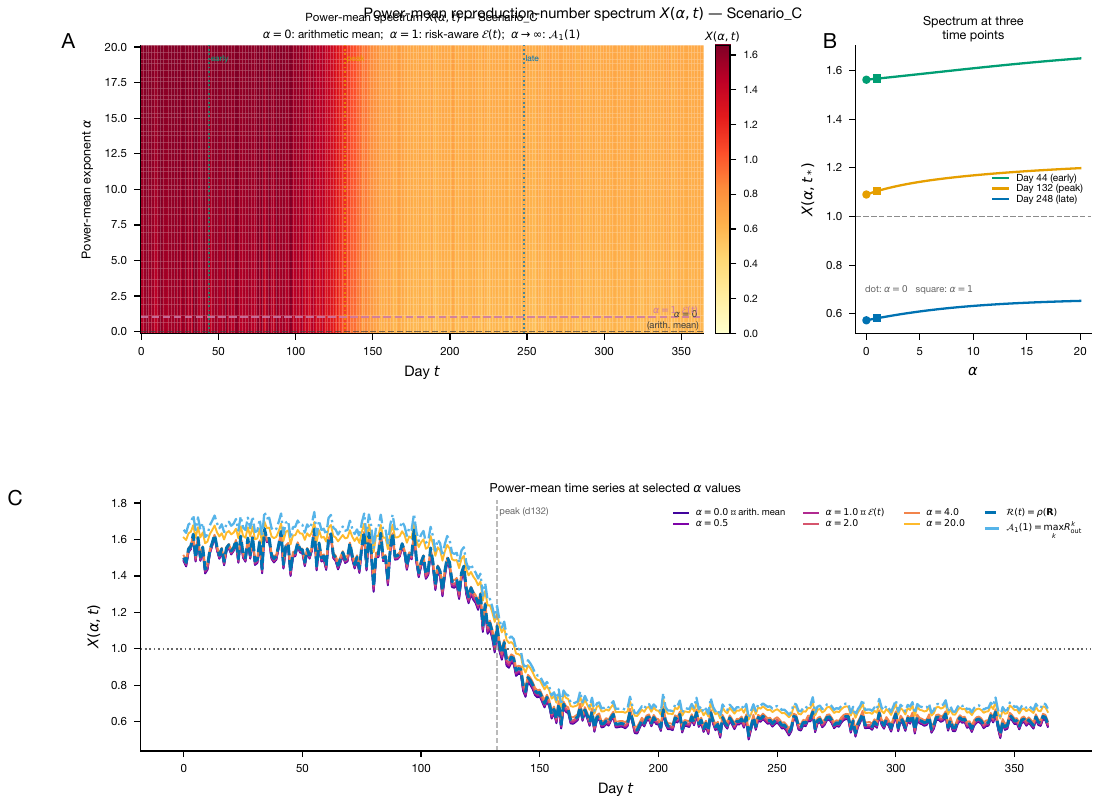}
\caption{\label{fig_counterfactual_nonnormal} \footnotesize \textbf{Results for increased non-normality/heterogeneity in human movement patterns (Scenario C):} Additional results for the hub-amplified setting (Scenario C). We show the Lehmer-mean spectrum $X(\alpha, t)$ across the continuum of weighting exponents $\alpha$ alongside the associated network-level transmissibility indicators, illustrating how they traverse from the arithmetic mean of the outward reproduction numbers ($\alpha = 0$) to the largest outward reproduction number $A_1(1)$ ($\alpha \rightarrow \infty$), with the spatial risk-averse reproduction number $\mathcal{E}(t) = X(1, t)$ and network-level $\mathcal{R}(t)$ shown for comparison. The strongly non-normal, hub-amplified movement patterns produce pronounced transience and slower mixing on the network.}
\end{figure}
\FloatBarrier


\end{appendix}
\end{document}